\magnification=1200
\hoffset=.0cm
\voffset=.0cm
\baselineskip=.53cm plus .53mm minus .53mm

%
%
%
%
\def\ref#1{\lbrack#1\rbrack}
%
%
%
%
\input amssym.def
\input amssym.tex
%
%
\font\teneusm=eusm10                    
\font\seveneusm=eusm7                   
\font\fiveeusm=eusm5                 
%
%
\font\sf=cmss10

%
%

%
%
\newfam\eusmfam
\textfont\eusmfam=\teneusm
\scriptfont\eusmfam=\seveneusm
\scriptscriptfont\eusmfam=\fiveeusm
\def\sans#1{\hbox{\sf #1}}

\def\proclaim #1. #2\par{\medbreak{\bf #1.\enspace}{\it #2}\par\medbreak}
%
%
%
%
%

\def\deg{{\rm deg}\hskip 1pt}

\def\ker{{\rm ker}\hskip 1pt}

\def\ad{{\rm ad}\hskip 1pt}

\def\GL{{\rm GL}\hskip 1pt}

\def\Hom{{\rm Hom}\hskip 1pt}

\def\Conn{{\rm Conn}\hskip 1pt}

\def\hst1{\hskip 1pt}
\def\square{\,\vbox{\hrule \hbox{\vrule height 0.25 cm 
\hskip 0.25 cm \vrule height 0.25 cm}\hrule}\,}
%
%
%
%
%

\hrule\vskip.5cm
\hbox to 16.5 truecm{April 1998  \hfil DFUB 98--7}
\hbox to 16.5 truecm{Version 1  \hfil hep-th/9804043}
\vskip.5cm\hrule
\vskip.9cm
\centerline{\bf BASIC AND EQUIVARIANT COHOMOLOGY IN}   
\centerline{\bf BALANCED TOPOLOGICAL FIELD THEORY}   
\vskip.4cm
\centerline{by}
\vskip.4cm
\centerline{\bf Roberto Zucchini}
\centerline{\it Dipartimento di Fisica, Universit\`a degli Studi di Bologna}
\centerline{\it V. Irnerio 46, I-40126 Bologna, Italy}
\centerline{\it and }
\centerline{\it INFN, sezione di Bologna}
\vskip.9cm
\hrule
\vskip.6cm
\centerline{\bf Abstract} 
\vskip.4cm
\par\noindent
We present a detailed algebraic study of the $N=2$ cohomological set--up 
describing the balanced topological field theory of Dijkgraaf 
and Moore. We emphasize the role of $N=2$ topological supersymmetry 
and $\goth s\goth l(2,\Bbb R)$ internal symmetry by a 
systematic use of superfield techniques and of an $\goth s\goth l(2,\Bbb R)$
covariant formalism. We provide a definition of $N=2$ basic and equivariant 
cohomology, generalizing Dijkgraaf's and Moore's, and of $N=2$ connection. 
For a general manifold with a group action, we show that: $i$) the $N=2$ 
basic cohomology is isomorphic to the tensor product of the ordinary 
$N=1$ basic cohomology and a universal $\goth s\goth l(2,\Bbb R)$ group 
theoretic factor: $ii$) the affine spaces of $N=2$ and $N=1$ connections 
are isomorphic.
\par\noindent
PACS no.: 0240, 0460, 1110. Keywords: Topological Field Theory, Cohomology.
\vfill\eject

\par\vskip .6cm
{\bf 0. Introduction}
\vskip .4cm
Topological quantum field theories are complicated often fully interacting 
local renormalizable field theories, yet they can be solved exactly and the 
solution is highly non trivial. Expectation values of topological observables 
provide topological invariants of the manifolds on which the fields propagate.
These invariants are independent from the couplings and to a large extent 
from the interactions between the fields. At the same time, topological field 
theories are often topological sectors of ordinary field theories. In this 
way, they are convenient testing grounds for subtle non perturbative field 
theoretic phenomena. See f. i. refs. \ref{1--3} for an updated comprehensive 
review on the subject and complete referencing.

$N=1$ cohomological topological field theories have been the object
of intense and exhaustive study. They can be understood in the framework
of equivariant cohomology of infinite dimensional vector bundles \ref{4--9}
and realized as Mathai-Quillen integral representations of Euler classes 
\ref{10--13}. The resulting formalism is elegant and general and covers 
the important case where the quotient by the action of a gauge symmetry 
group is required. Each of these models describes the differential topology 
of a certain moduli space, depending on the model considered: the field 
theoretic correlation functions of topological observables correspond to 
intersection numbers on the moduli space. 

$N=2$ cohomological topological field theories were discovered quite early
\ref{14--17}, but they did not arouse much interest until recently
when it became clear that they might provide important clues in the analysis
of $S$ duality in supersymmetric Yang--Mills theory and in the study
of the world volume theories of $D$--branes in string theory. 

In ref. \ref{18}, Vafa and Witten performed an exact strong coupling test of 
$S$ duality of $N=4$ supersymmetric 4--dimensional Yang--Mills theory by 
studying a topologically twist of the model yielding an $N=2$ cohomological
field theory. They showed that the partition function is 
$Z(\tau)=\sum_k a_k\exp(2\pi i\tau k)$, where $a_k$ is the Euler 
characteristic of the moduli space of $k$ instantons, and tested $S$ duality
by analyzing the modularity properties of $Z(\tau)$. Their work, 
inspired by the original work of Yamron \ref{14}, was soon developed and 
refined in a series of papers \ref{19--24}. 
In ref. \ref{25}, Bershadsky, Sadov and Vafa showed that the three 
$N=2$ cohomological topological field theories obtained by the  
non topological twistings of $N=4$ supersymmetric 4--dimensional 
Yang--Mills theory arose from curved $3$--branes embedded in Calabi-Yau 
manifolds and manifolds with exceptional holonomy groups. 
Their analysis was continued and further developed in refs. 
\ref{21,26--28}, where the connection with higher dimensional 
instantons was elucidated. In ref. \ref{29}, Park constructed a family of 
Yang-Mills instantons from $D$--instantons in topological twisted 
$N=4$ supersymmetric 4--dimensional Yang--Mills theory. 
In ref. \ref{30}, Hofman and Park worked out a 2--dimensional $N=2$ 
cohomological topological field theory as a candidate 
for covariant second quantized RNS superstrings,  
which they conjectured to be a formulation of $M$ theory.

All the endeavors mentioned above, and many other related ones, which
we cannot mention for lack of space, show that $N=2$ cohomological 
topological field theories are relevant in a variety of physical and 
mathematical issues. In spite of that, the body of literature devoted to the 
study of the geometry of such models is comparatively small. In ref. \ref{17},
Blau and Thompson worked out a Riemannian formulation of $N=2$ topological 
gauge theory using $N=2$ topological superfield techniques. In ref. \ref{31}, 
Dijkgraaf and Moore showed that all known $N=2$ topological models were 
examples of ``balanced topological field theories'' and developed a 
cohomological framework suitable for their study. In ref.  
\ref{21}, Blau and Thompson proved the equivalence of their earlier 
formulation and Dijkgraaf's and Moore's. These studies show that the 
partition function of every $N=2$ topological model calculates the Euler 
characteristic of some moduli space of vanishing virtual dimension. 
They also indicate that the appropriate cohomological scheme is provided 
by $N=2$ basic or equivariant cohomology. The present paper aims at a 
systematic study of the latter developing the ideas of \ref{31}. 

In general, a cohomological topological field theory is characterized by 
a symmetry Lie algebra $\goth g$, a graded algebra of fields $\sans f$ and 
a set of graded derivations on $\sans f$ generating a Lie algebra $\sans t$.
In turn, the topological algebra $\sans t$ provides the algebraic
and geometric framework for the definition of the topological observables 
\ref{1}.

As is well--known,
in $N=1$ cohomological topological field theory, $\sans t$ is 
generated by four derivations $k$, $d$, $j(\xi)$, $l(\xi)$, 
$\xi\in\goth g$, of degrees $0$, $1$, $-1$, $0$, respectively, obeying  
the graded commutation relations (2.1.14), (2.1.15a)--(2.1.15c), (2.1.16), 
(2.1.17a)--(2.1.17b) and (2.1.18a)--(2.1.18c) below.
$k$ is the ghost number operator. $d$ is the nilpotent topological charge.  
$j(\xi)$, $l(\xi)$ describe the action of the symmetry Lie algebra $\goth g$ 
on fields. The elements $\alpha\in\sans f$ are classified into the 
eigenspaces $\sans f^p$, $p\in\Bbb Z$, of $k$. The $N=1$ basic degree 
$p$ cohomology of $\sans f$ is defined by 
$$
\eqalignno{
&j(\xi)\alpha=0,\quad l(\xi)\alpha=0,\quad\xi\in\goth g,&(0.1)\cr
&d\alpha=0,\quad \alpha\equiv\alpha+d\beta,
\quad \beta\in\sans f^{p-1}, \quad \alpha\in\sans f^p.&(0.2)\cr}
$$

The $N=1$ Weil algebra $\sans w$, an essential element of the definition 
of the $N=1$ equivariant cohomology of $\sans f$, is generated by two 
$\goth g$ valued fields $\omega$, $\phi$ of degrees $1$, $2$, respectively. 
$\sans t$ acts on $\sans w$ according to (3.1.10a), (3.1.10b), (3.1.11a), 
(3.1.11b), (3.1.12a)--(3.1.12d) below. 

$k$, $d$, $j(\xi)$, $l(\xi)$ can 
be organized into two $N=1$ topological superderivation 
$$
\eqalignno{
H&=k-\theta d,&(0.3)\cr
I(\xi)&=j(\xi)+\theta l(\xi),\quad\xi\in\goth g.&(0.4)\cr}
$$
The Lie algebra structure of $\sans t$ is compatible with the underlying
$N=1$ topological supersymmetry, since the relevant commutation relations 
can be written in terms of the superderivations $H$, $I(\xi)$.
Similarly, $\omega$, $\phi$ can be organized into the $\goth g$ valued 
superfield
$$
W=\omega+\theta\Big(\phi-\hbox{$1\over 2$}[\omega,\omega]\Big).
\eqno(0.5)
$$
The action of $\sans t$ on $\sans w$ can be written 
in terms of the superderivations $H$, $I(\xi)$ and the superfield $W$
in a manifestly $N=1$ supersymmetric way.

Analogously, in $N=2$ cohomological topological field theory, 
$\sans t$ is generated by seven graded derivations $u_A$, $A=1,2$, 
$t_{AB}$, $A,B=1,2$, symmetric in $A,B$, $k$, $d_A$, $A=1,2$, $j(\xi)$, 
$j_A(\xi)$, $A=1,2$, $l(\xi)$, $\xi\in\goth g$, of degrees $-1$, $0$, $0$, 
$1$, $-2$, $-1$, $0$, respectively, obeying the graded commutation relations 
(2.2.16a)--(2.2.16c), (2.2.17a)--(2.2.17j), (2.2.18a)--(2.2.18c), 
(2.2.19a)--(2.2.19f), (2.2.20a)--(2.2.20f) below.  The $u_A$ are a sort of  
homotopy operators and constrain the cohomology of $\sans f$, defined shortly,
to an important extent. The $t_{AB}$ and $k$ are the generators of the 
internal $\goth s\goth l(2,\Bbb R)\oplus\Bbb R$ symmetry Lie algebra 
of $\sans t$. The $d_A$ are the nilpotent topological charges. $j(\xi)$, 
$j_A(\xi)$, $l(\xi)$ describe the action of the symmetry Lie algebra 
$\goth g$ on fields. The elements $\alpha\in\sans f$ are classified into 
the eigenspaces $\sans f^{n,p}$, $n\in\Bbb N$, $p\in\Bbb Z$, of the 
invariants $c$, $k$ of the internal algebra $\goth s\goth l(2,\Bbb R)
\oplus\Bbb R$. The $N=2$ basic type $n,p$ cohomology of $\sans f$ is 
defined by 
$$
\eqalignno{
&j(\xi)\alpha=0,\quad j_A(\xi)\alpha=0,\quad
l(\xi)\alpha=0, \quad\xi\in\goth g,&(0.6)\cr
&d_A\alpha=0,
\quad\alpha\equiv\alpha+\hbox{$1\over 2$}\epsilon^{KL}d_Kd_L\beta,
\quad\beta\in\sans f^{n,p-2},\quad\alpha\in\sans f^{n,p}. 
&(0.7)\cr}
$$
It is possible to show, using the basic relation 
$[d_A,u_B]=\hbox{$1\over 2$}(t_{AB}+\epsilon_{AB}k)$,
that this cohomology is trivial for $p\not=\pm n+1$.

The $N=2$ Weil algebra $\sans w$, entering the definition of $N=2$
equivariant cohomology, is generated by four $\goth g$ valued fields 
$\omega _A$, $A=1,2$, $\phi_{AB}$, $A,B=1,2$, symmetric in $A,B$, 
$\gamma$, $\rho _A$, $A=1,2$, of degrees $1$, $2$, $2$, $3$, 
respectively. $\sans t$ acts on $\sans w$ according to
(3.2.10a)--(3.2.10h), (3.2.11a)--(3.2.11h), (3.2.12a)--(3.2.12l).

$u_A$, $t_{AB}$, $k$, $d_A$, $j(\xi)$, $j_A(\xi)$, $l(\xi)$ 
can be organized into two $N=2$ topological superderivation 
$$
\eqalignno{
H_A&=u_A+\hbox{$1\over 2$}\theta^K(t_{AK}-\epsilon_{AK}k)
-\hbox{$1\over 2$}\epsilon_{KL}\theta^K\theta^L d_A,&(0.9)\cr
I(\xi)&=j(\xi)+\theta^Kj_K(\xi)
+\hbox{$1\over 2$}\epsilon_{KL}\theta^K\theta^Ll(\xi), 
\quad \xi\in\goth g.&(0.10)\cr}
$$
The Lie algebra structure of $\sans t$ is compatible with the underlying
$N=2$ topological supersymmetry, since the relevant commutation relations 
can be written in terms of the superderivations $H_A$, $I(\xi)$.
Similarly, $\omega_A$, $\phi_{AB}$, $\gamma$, $\rho_A$ 
can be organized into the $\goth g$ valued superfield
$$
\eqalignno{
W&=\omega_A+\theta^K\Big(\phi_{AK}+\epsilon_{AK}\gamma
-\hbox{$1\over 2$}[\omega_A,\omega_K]\Big)
+\hbox{$1\over 2$}\epsilon_{MN}\theta^M\theta^N
\Big(-2\rho_A-\epsilon^{KL}[\omega_K,\phi_{AL}]&\cr
&+[\omega_A,\gamma]+\hbox{$1\over 6$}
\epsilon^{KL}[\omega_K,[\omega_L,\omega_A]]\Big).&(0.11)\cr}
$$
The action of $\sans t$ on $\sans w$ can be written 
in terms of the superderivations $H_A$, $I(\xi)$ and the superfield $W$
in a manifestly $N=2$ supersymmetric way.

In the first part of this paper, we study the topological algebra 
$\sans t$ and the Weil algebra $\sans w$ abstractly both in the $N=1$ and 
in the $N=2$ case. We show that their structure is essentially dictated by 
rather general requirements of closure and topological supersymmetry, 
called stability conditions, which can be defined for any value of $N$. In 
this way, we show that all these algebraic structures can be derived from a 
single universal model independent notion and also suggest an obvious 
method for generalizing them to higher values of $N$. In the second part of 
the paper, we define basic and equivariant cohomology, abstract
connections and the Weil homomorphism both in the $N=1$ and in the $N=2$ case 
and study some of their properties. Finally, in the third part of the paper,
we study the cohomology of manifolds carrying a right group action and show 
that, in this important case, the $N=2$ type $(k,k+1)$ basic cohomology is 
isomorphic to the tensor product of the $N=1$ degree $k$ basic cohomology 
and the completely symmetric tensor space $\bigvee^{k-1}\Bbb R^2$
and that the affine spaces of $N=2$ and $N=1$ connections are isomorphic. 

Throughout the paper, we stress the role of topological supersymmetry, also
because we feel that, on this score, confusing claims have appeared in the
literature. This has allowed us to discover the derivations $u_A$ and $k$
introduced above, which are not mentioned in ref. \ref{31}, but which are  
required by $N=2$ topological supersymmetry and constrain structurally
the $N=2$ cohomology. 

The definition of $N=2$ basic cohomology given above is more general
than that used in ref. \ref{31}, which is limited to the important
case where $n=1$. In our judgement, this definition is more appropriate, 
yielding the aforementioned fundamental relation between the $N=1$ and 
$N=2$ basic cohomologies of manifolds with a right group action. 

This paper is organized as follows. In sect. 1, we introduce and precisely
define the crucial notion of stability. In sect. 2, we show that both the 
$N=1$ and $N=2$ topological algebras can be derived in a model independent 
way by imposing non degeneracy and stability. In sect. 3, we show similarly
that non degeneracy and stability allow to obtain in abstract fashion
both the $N=1$ and $N=2$ Weil algebra. In sect. 4, we define the relevant
notions of $N=1$ and $N=2$ basic cohomology, highlighting their similarities 
and differences. In sect. 5, we study the Weil superoperation and 
its cohomology. In sect. 6, we define $N=1$ and $N=2$ abstract connections,
equivariant cohomology and the related Weil homomorphism. In sect. 7, we 
apply our algebraic set up to study the cohomology of manifolds carrying
a right group action and work out the relation between $N=1$ and $N=2$ 
cohomology. Finally, sect. 8 outlines future lines of inquiry.
\par\vskip .6cm
{\bf 1. Superalgebras, supermodules and stability}
\vskip .4cm

\vskip.2cm {\it 1.1. $\Bbb Z$ graded algebras and the corresponding 
superalgebras}

All the vector spaces, algebras and modules considered in this section are 
real and $\Bbb Z$ graded. If $\sans s$ is such a space, 
we denote by $\sans s^k$ the subspace of $\sans s$ of degree $k\in\Bbb Z$. 
If $\sans s=\sans s^0$, $\sans s$ is called ungraded.

Let $N\in\Bbb N$. Let $\theta^A$, $A=1,\cdots,N$, be a $N$--uple of 
Grassmann odd generators which are conventionally assigned degree $-1$:
$$
\theta^A\theta^B+\theta^B\theta^A=0, \quad A,B=1,\cdots,N;\quad
\deg\theta^A=-1,\quad A=1,\cdots,N.
\eqno(1.1.1)
$$
The $\theta^A$ generate a Grassmann algebra $\Lambda_N[\theta]$.

Let $\sans a$ be a $\Bbb Z$ graded algebra. 
The $N$ superalgebra $\sans A_N$ associated to $\sans a$
is the graded tensor product algebra
$$
\sans A_N=\Lambda_N[\theta]\hat\otimes\sans a
\eqno(1.1.2)
$$
with the canonical $\Bbb Z$ grading.

\vskip.2cm {\it 1.2. Stability}

In this subsection, we introduce the notion of stability, which 
will play an important role throughout this paper.

Let $\sans u$ and $\sans a$ be an ungraded real vector space and a 
$\Bbb Z$ graded real algebra, respectively. Let ${\cal T}
=\{T_r|r\in R\}$ be a finite subset of $\Hom(\sans u,\sans A_N)$.

\proclaim Definition 1.2.1. $\cal T$ is called stable if:           
\item{$i$)} for $r\in R$, there is $p^{\cal T}{}_r\in\Bbb Z$ such that, 
for all $x\in\sans u$, $T_r(x)\in\sans A_N^{p^{\cal T}{}_r}$;
\item{$ii$)} for $x\in\sans u$, $x\not=0$,
the $T_r(x)$ are linearly independent;   
\item{$iii$)} there are constants $k^{\cal T}{}_{Ar}{}^s\in\Bbb R$, 
$r,s\in R$, $A=1,\cdots,N$, such that, for $x\in\sans u$,
$$
\partial_AT_r(x)=k^{\cal T}{}_{Ar}{}^sT_s(x),
\quad\quad r\in R,~A=1,\cdots,N,
\eqno(1.2.1)
$$
\item{\hphantom{$ii$)}}where $\partial_A=\partial/\partial\theta^A$.
\item{If $\sans u$} is an algebra, it is also required that:
\item{$iv$)} there are constants $g^{\cal T}{}_{rs}{}^t\in\Bbb R$, 
$r,s,t\in R$, such that, for $x,y\in\sans u$,
$$
T_r(x)T_s(y)=g^{\cal T}{}_{rs}{}^tT_t(xy), 
\quad\quad r,s\in R.
\eqno(1.2.2)
$$

For $x\in\sans u$, denote by $\sans T(x)$ the linear 
span of the $T_r(x)$. Then, by the above, $\sans T(x)$
is stable under $\partial_A$. Further, when $\sans u$ is an 
algebra, $\sans T(x)\sans T(y)\subseteq \sans T(xy)$.

The following remarks are in order.

\item{$a$)} By $ii$ above, if $\sans u\not=0$, the constants
$k^{\cal T}{}_{Ar}{}^s$ are uniquely determined; further, when $\sans u$
is an algebra and $\sans u\sans u\not=0$,
the $g^{\cal T}{}_{rs}{}^t$, also, are uniquely determined.

\item{$b$)} By grading reasons, when $\sans u\not=0$,
$$
k^{\cal T}{}_{Ar}{}^s=0, 
\quad\hbox{unless $p^{\cal T}{}_r-p^{\cal T}{}_s+1=0$};
\eqno(1.2.3)
$$
further, when $\sans u$ is an algebra such that $\sans u\sans u\not=0$,
$$
g^{\cal T}{}_{rs}{}^t=0, 
\quad\hbox{unless $p^{\cal T}{}_r+p^{\cal T}{}_s-p^{\cal T}{}_t=0$}.
\eqno(1.2.4)
$$

\item{$c$)} As $\partial_A\partial_B+\partial_B\partial_A=0$, one has 
$$
k^{\cal T}{}_{Ar}{}^tk^{\cal T}{}_{Bt}{}^s
+k^{\cal T}{}_{Br}{}^tk^{\cal T}{}_{At}{}^s=0,
\eqno(1.2.5)
$$
when $\sans u\not=0$;  further, if $\sans u$ is an algebra, being
$\partial_AT_r(x)T_s(y)
+(-1)^{p^{\cal T}{}_r}T_r(x)\partial_AT_s(y)
=g^{\cal T}{}_{rs}{}^t\partial_AT_t(xy)$ by (1.2.2), one has further
$$
k^{\cal T}{}_{Ar}{}^ug^{\cal T}{}_{us}{}^t
+(-1)^{p^{\cal T}{}_r}k^{\cal T}{}_{As}{}^ug^{\cal T}{}_{ru}{}^t
=g^{\cal T}{}_{rs}{}^uk^{\cal T}{}_{Au}{}^t,
\eqno(1.2.6)
$$
when $\sans u\sans u\not=0$.

\item{$d$)} If $\sans u$ is an algebra, there are further constraints
on the $g^{\cal T}{}_{rs}{}^t$ coming from such properties as
associativity, Jacobi relations, etc., when they hold.

Usually, the constraints (1.2.3)--(1.2.6), are so stringent that,
given some extra non singularity assumptions, allow the determination of  
the $k^{\cal T}{}_{Ar}{}^s$ and $g^{\cal T}{}_{rs}{}^t$, up to the natural 
equivalence relation associated to the non singular linear redefinition of 
the $T_r$.

Let $\sans a$ be $\Bbb Z$ graded real algebra.
Let ${\cal H}=\{H_r|r\in R\}$ be a finite subset of $\sans A_N$.

\proclaim Definition 1.2.2. $\cal H$ is called stable, if, defining
$$
T^{\cal H}{}_r(x)=xH_r,\quad\quad x\in\Bbb R,
\eqno(1.2.7)
$$
${\cal T}^{\cal H}=\{T^{\cal H}{}_r|r\in R\}$ is a stable subset of 
$\Hom(\Bbb R,\sans A_N)$ as defined above. 

We denote by $\sans H$ the linear span of the $H_r$. 
Then, $\sans H$ is a subalgebra of $\sans A_N$.

Let $\sans u$, $\sans a$, $\sans e$, $\sans q$ be an ungraded real algebra, 
a $\Bbb Z$ graded real algebra, an ungraded real left $\sans u$ module and 
a $\Bbb Z$ graded real left $\sans a$ module algebra, respectively.
Let ${\cal T}=\{T_r|r\in R\}$, ${\cal M}=\{M_i|i\in I\}$ be stable 
subsets of $\Hom(\sans u,\sans A_N)$ and $\Hom(\sans e,\sans Q_N)$,
respectively. Let $C\in\sans Q_N^0$ such that $\partial_A C=0$.

\proclaim Definition 1.2.3. 
We say that $\cal T$ acts $C$--stably on $\cal M$ on the left if:
\item{$i$)} if $C\not=0$, then for $m\in\sans e$, $C$ is linearly 
independent from the $M_i(m)$;
\item{$ii$)} for $x\in\sans u$,
$$
T_r(x)C=0;
\eqno(1.2.8)
$$
\item{$iii$)} 
there are constants $c^{\cal TM}{}_{ri}{}^j\in\Bbb R$, $r\in R$, $i,j\in I$, 
and bilinear maps $\varphi^{\cal TM}{}_{ri}:\sans u\times\sans e\mapsto\Bbb R$,
$r\in R$, $i\in I$ such that, for any $x\in\sans u$ and $m\in\sans e$,
$$
T_r(x)M_i(m)=c^{\cal TM}{}_{ri}{}^jM_j(xm)+\varphi^{\cal TM}{}_{ri}(x,m)C, 
\quad\quad r\in R,~i\in I.
\eqno(1.2.9)
$$
When $C=0$ above, we say that  $\cal T$ acts stably on $\cal M$ 
on the left.

So, for any $x\in\sans u$ and $m\in\sans e$,
$\sans T(x)\sans M(m)\subseteq\sans M(xm)\oplus\Bbb RC$.

Note the following.

\item{$a$)} By the above definition, if $\sans u\sans e\not=0$, the constants
$c^{\cal MT}{}_{ri}{}^j$ are uniquely determined; also, if $C\not=0$,
the maps $\varphi^{\cal TM}{}_{ri}$ are uniquely determined. 

\item{$b$)} By grading reasons, when $\sans u\sans e\not=0$,
$$
c^{\cal TM}{}_{ri}{}^j=0, 
\quad\hbox{unless $p^{\cal T}{}_r+p^{\cal M}{}_i-p^{\cal M}{}_j=0$};
\eqno(1.2.10)
$$
similarly, when $C\not=0$,
$$
\varphi^{\cal TM}{}_{ri}(x,m)=0,
\quad\hbox{unless $p^{\cal T}{}_r+p^{\cal M}{}_i=0$}.
\eqno(1.2.11)
$$

\item{$c$)} As, by (1.2.9), $\partial_A T_r(x)M_i(m)
+(-1)^{p^{\cal T}{}_r}T_r(x)\partial_A M_i(m)
=c^{\cal TM}{}_{ri}{}^j\partial_A M_j(xm)$, one has
$$
k^{\cal T}{}_{Ar}{}^sc^{\cal TM}{}_{si}{}^j
+(-1)^{p^{\cal T}{}_r}k^{\cal M}{}_{Ai}{}^kc^{\cal TM}{}_{rk}{}^j
=c^{\cal TM}{}_{ri}{}^kk^{\cal M}{}_{Ak}{}^j,
\eqno(1.2.12)
$$
when $\sans u\sans e\not=0$; further, as $(T_r(x)T_s(y))M_i(m)
=T_r(x)(T_s(y)M_i(m))$, 
$$
g^{\cal T}{}_{rs}{}^tc^{\cal TM}{}_{ti}{}^j
-c^{\cal TM}{}_{si}{}^kc^{\cal TM}{}_{rk}{}^j=0,
\eqno(1.2.13)
$$
$$
g^{\cal T}{}_{rs}{}^t\varphi^{\cal TM}{}_{ti}(xy,m)
-c^{\cal TM}{}_{si}{}^j\varphi^{\cal TM}{}_{rj}(x,ym)=0,
\eqno(1.2.14)
$$
when $\sans u\sans e\not=0$, $C\not=0$, respectively.

Usually, the constraints (1.2.10)--(1.2.14), are so stringent that,
given some extra non singularity assumptions, allow the determination of  
the $c^{\cal TM}{}_{ri}{}^j$ and $\varphi^{\cal TM}{}_{ri}$, up to the natural 
equivalence relation associated to the non singular linear redefinition 
of the $T_r$ and $M_i$.

Let $\sans a$, $\sans e$, $\sans q$ be an ungraded real algebra, an 
ungraded real vector space and a $\Bbb Z$ graded real left 
$\sans a$ module algebra, respectively.
Let ${\cal H}=\{H_r|r\in R\}$, ${\cal M}=\{M_i|i\in I\}$ be stable subsets 
of $\sans A_N$ and $\Hom(\sans e,\sans Q_N)$, respectively. 

\proclaim Definition 1.2.4. We say that $\cal H$ acts stably on
$\cal M$, if ${\cal T}^{\cal H}$ (cfr. def. 1.2.2) acts stably on
$\cal M$ as defined above, where $\sans e$ has the structure of $\Bbb R$
module corresponding to that of real vector space.

\vskip.2cm {\it 1.2. The N=1,2 cases}

In this paper, we concentrate on the cases $N=1,2$. In this subsection, 
we introduce notation suitable for these special $N$ values.

Let $\sans a$ be a $\Bbb Z$ graded real algebra. 

Let $N=1$. In this case, one can set $\theta^1=\theta$ for simplicity.
If $X\in \sans A_1^p$ for some $p\in\Bbb Z$, then $X$ is of the form
$$
X=x+\theta\tilde x
\eqno(1.3.1)
$$
with $x\in \sans a^p$ and $\tilde x\in \sans a^{p+1}$. Note that
$$
x=X\big|_{\theta=0}.
\eqno(1.3.2)
$$
Denoting $\partial=\partial/\partial\theta$, we define
$$
\tilde X=\partial X.
\eqno(1.3.3)
$$
Clearly, $\tilde X\in\sans A_1^{p+1}$. Indeed,
$$
\tilde X=\tilde x.
\eqno(1.3.4)
$$

Let $N=2$.
If $X\in \sans A_2^p$ for some $p\in\Bbb Z$, then $X$ is of the form
$$
X=x+\theta^Ax_{,A}+\hbox{$1\over 2$}\epsilon_{KL}\theta^K\theta^L\tilde x
\eqno(1.3.5)
$$
with $x\in \sans a^p$, $x_{,A}\in \sans a^{p+1}$ and 
$\tilde x\in \sans a^{p+2}$
\footnote{}{}
\footnote{${}^1$}{The totally antisymmetric symbols $\epsilon_{AB}$,
$\epsilon^{AB}$ are normalized so that $|\epsilon_{12}|
=|\epsilon^{12}|=1$ and $\epsilon^{AK}\epsilon_{KB}
=\epsilon_{BK}\epsilon^{KA}=\delta^A_B$.}. Note that
$$
x=X\big|_{\theta=0}.
\eqno(1.3.6)
$$
Denoting $\partial_A=\partial/\partial\theta^A$, we define
$$
X_{,A}=\partial_A X.
\eqno(1.3.7)
$$
Clearly, $X_{,A}\in\sans A_2^{p+1}$. Indeed,
$$
X_{,A}=x_{,A}+\epsilon_{AK}\theta^K\tilde x.
\eqno(1.3.8)
$$
So,
$$
x_{,A}=X_{,A}\big|_{\theta=0}.
\eqno(1.3.9)
$$
Finally, we set
$$
\tilde X=\hbox{$1\over 2$}\epsilon^{KL}\partial_K\partial_LX.
\eqno(1.3.10)
$$
Clearly, $\tilde X\in\sans A_2^{p+2}$, as
$$
\tilde X=\tilde x.
\eqno(1.3.11)
$$

The analysis of stability simplify considerably in the $N=1,2$ cases. 
In the $N=1$ case, stability is an almost trivial notion. For $N=2$, this 
is no longer so. The notion of stability is nevertheless useful, since it
shows quite clearly in what sense several $N=2$ algebraic structures 
generalize their $N=1$ counterparts.
\par\vskip .6cm
{\bf 2. Fundamental superstructures}         
\vskip .4cm
Let $\sans d$ be a $\Bbb Z$ graded real Lie algebra, i.e. a $\Bbb Z$
graded real algebra whose product is graded antisymmetric.
Let $\goth g$ be a real ungraded Lie algebra.

\vskip.2cm {\it 2.1. The fundamental N=1 superstructure}

\proclaim Proposition 2.1.1.
Let $H\in\sans D_1^0$. Assume that the set $\{H,\tilde H\}$ is stable 
(cfr. def. 1.2.2). 
Let $\sans H$ the linear span of this set. Assume that
$$
[\sans H^0,\sans H^1]=\sans H^1.
\eqno(2.1.1)
$$
Then, perhaps after substituting $H$ by $\kappa H$ for some suitable 
$\kappa\in\Bbb R_\times$, one has 
$$
\matrix{~&~\cr
[H,H]=0, \hfill &[H,\tilde H]=\tilde H, \hfill\cr
~&~\cr 
[\tilde H,\tilde H]= 0.&\hfill&~\cr
~&~\cr}
\eqno(2.1.2a)-(2.1.2c)
$$

{\it Proof}. It is straightforward to check that (2.1.2a)--(2.1.2c)
are compatible with the graded antisymmetry and Jacobi relations of 
$\sans D_1$. By stability and grading reasons, 
one has $[\sans H^i,\sans H^j]\subseteq
\sans H^{i+j}$, where $\sans H^0=\Bbb R H$, $\sans H^1=\Bbb R \tilde H$ 
and $\sans H^i=0$ else. So, (2.1.2a) holds. Further, by stability
(cfr. eq. (1.2.2)),
$$
[H,\tilde H]=a\tilde H,
\eqno(2.1.3)
$$
where $a\in\Bbb R$.  Owing to (2.1.1), $a\not=0$. Renormalize $H$ into 
$a^{-1}H$. Then, one can set $a=1$ in (2.1.3), getting (2.1.2b). Applying 
$\partial$ to (2.1.2b), one obtains (2.1.2c). \hfill $\square$

\proclaim Proposition 2.1.2.
Let $I:\goth g\mapsto \sans D_1$ be a homomorphism such that $I(\goth g)
\subseteq\sans D_1^{-1}$. Assume that the set $\{I,\tilde I\}$ is stable
(cfr. def. 1.2.1).
For $\xi\in\goth g$, let $\sans I(\xi)$ the linear span of the set
$\{I(\xi),\tilde I(\xi)\}$. Assume that
$$
[\sans I^{-1}(\xi),\sans I^0(\eta)]=\sans I^{-1}([\xi,\eta]),
\quad\quad \xi,\eta\in\goth g.
\eqno(2.1.4)
$$
Then, perhaps after substituting $I$ by $\lambda I$ for some suitable 
$\lambda\in\Bbb R_\times$, one has
$$
\matrix{~&~\cr
[I(\xi),I(\eta)]=0, \hfill&[I(\xi),\tilde I(\eta)]=I([\xi,\eta]),\hfill\cr
~&~\cr
[\tilde I(\xi),\tilde I(\eta)]=\tilde I([\xi,\eta]),\hfill
&\hphantom{[I(\xi),\tilde I(\eta)]=I([\xi,\eta]),}
\quad\quad \xi,\eta\in\goth g.\hfill\cr
~&~\cr}
\eqno(2.1.5a)-(2.1.5c)
$$

{\it Proof}. It is straightforward to check that (2.1.5a)--(2.1.5c)
are compatible with the graded antisymmetry and Jacobi relations of 
$\sans D_1$. By stability and grading reasons, one has 
$[\sans I^i(\xi),\sans I^j(\eta)]\subseteq\sans I^{i+j}([\xi,\eta])$, 
where $\sans I^{-1}(\xi)=\Bbb R I(\xi)$,  $\sans I^0(\xi)
=\Bbb R \tilde I(\xi)$ and $\sans I^i(\xi)=0$ else.
So, (2.1.5a) holds. Further, by stability (cfr. eq. (1.2.2)),
$$
[I(\xi),\tilde I(\eta)]=aI([\xi,\eta]),
\eqno(2.1.6)
$$
where $a\in \Bbb R$. Owing to (2.1.4), $a\not=0$, 
if $[\goth g,\goth g]\not=0$, or one can choose $a\not=0$,
if $[\goth g,\goth g]=0$. Renormalize $I$ into $a^{-1}I$. Then, 
one can set $a=1$ in (2.1.6), getting (2.1.5b). Applying 
$\partial$ to this latter relation, one obtains (2.1.5c). \hfill $\square$

\proclaim Proposition 2.1.3.
Assume that $[\goth g,\goth g]\not=0$. Assume that $H$ 
and $I$ satisfy the hypotheses of props. 2.1.1 and 2.1.2, 
respectively. Assume further that $\{H,\tilde H\}$ 
acts stably on $\{I,\tilde I\}$ (cfr. def. 1.2.4). 
Assume finally that
$$
[\sans H^0,\sans I^{-1}(\xi)]=\sans I^{-1}(\xi),
\quad\quad\xi\in\goth g.
\eqno(2.1.7)
$$
Then, one has 
$$
\matrix{~&~\cr
[H,I(\xi)]=-I(\xi),\hfill&[H,\tilde I(\xi)]=0,\hfill\cr
~&~\cr
[\tilde H,I(\xi)]=-\tilde I(\xi),\hfill&[\tilde H,\tilde I(\xi)]=0,
\quad\quad\xi\in\goth g.\hfill\cr
~&~\cr}
\eqno(2.1.8a)-(2.1.8d)
$$

{\it Proof}. It is straightforward to check that (2.1.8a)--(2.1.8d)
are compatible with the graded antisymmetry and Jacobi relations of 
$\sans D_1$ and with relations (2.1.2a)--(2.1.2c), (2.1.5a)--(2.1.5c). 
By stability and grading reasons, one has $[\sans H^i,\sans I^j(\xi)]
\subseteq\sans I^{i+j}(\xi)$, where $\sans H^i$, $\sans I^i(\xi)$ have 
been defined above (2.1.3) and (2.1.6), respectively. Taking stability into 
account (cfr. eq. (1.2.9)), one has
$$
[H,I(\xi)]=aI(\xi), \quad [H,\tilde I(\xi)]=b\tilde I(\xi),
\eqno(2.1.9),(2.1.10)
$$
where $a, b\in\Bbb R$. By applying $\partial$ to (2.1.9) and using 
(2.1.10), one gets
$$
[\tilde H,I(\xi)]=(a-b)\tilde I(\xi).
\eqno(2.1.11)
$$
Owing to (2.1.7), $a\not=0$. From (2.1.5b) and the Jacobi identity, one has  
$[H,I([\xi,\eta])]-[[H,I(\xi)],\tilde I(\eta)]-[I(\xi),[H,\tilde I(\eta)]]=0$,
which, upon using (2.1.9), (2.1.10), gives the relation $-bI([\xi,\eta])=0$. 
As $[\goth g,\goth g]\not=0$ and $I(\zeta)\not=0$ for $\zeta\not=0$ 
by stability, $b=0$. From (2.1.3b) and the Jacobi identity, $[\tilde H,I(\xi)]
-[H,[\tilde H, I(\xi)]]+[[H,I(\xi)],\tilde H]=0$, which, upon using
(2.1.9)--(2.1.11), gives $a(a+1)I(\xi)=0$. As $\goth g\not=0$ 
and $I(\zeta)\not=0$ for $\zeta\not=0$, $a=-1$. Substituting the found values 
of $a$ and $b$ into (2.1.9)--(2.1.11), one gets (2.1.8a)--(2.1.8c). (2.1.8d)
is obtained by applying $\partial$ to (2.1.8c). \hfill $\square$

For an arbitrary ungraded Lie algebra $\goth g$, relations (2.1.2a)--(2.1.2c),
(2.1.5a)--(2.1.5c), (2.1.8a)--(2.1.8d) define a $\Bbb Z$ graded real Lie 
algebra $\sans t_\theta$. As shown by the above three propositions, 
when $\goth g$ is not Abelian, $\sans t_\theta$ is the unique solution of a 
suitable set of stability and non degeneracy conditions. When $\goth g$ is 
Abelian, $\sans t_\theta$ is a particular case of a certain class of 
$\Bbb Z$ graded Lie algebras. We shall not investigate this issue any further. 

Note that $\sans t_\theta\simeq\sans t_0=:\sans t$. 
The $\Bbb Z$ graded real Lie algebra $\sans t$ is called fundamental 
$N=1$ superstructure of $\goth g$. It is defined in terms of the four
generators $h$, $\tilde h$, $i(\xi)$, $\tilde\imath(\xi)$, $\xi\in\goth g$,
satisfying relations (2.1.2a)--(2.1.2c), (2.1.5a)--(2.1.5c), 
(2.1.8a)--(2.1.8d). More customarily, one sets
$$
k=h,\quad d=-\tilde h,
\eqno(2.1.12a),(2.1.12b)
$$
$$
j(\xi)=i(\xi),\quad l(\xi)=\tilde\imath(\xi), 
\quad\quad \xi\in\goth g.
\eqno(2.1.13a),(2.1.13b)
$$
From (2.1.2a)--(2.1.2c), (2.1.5a)--(2.1.5c) and (2.1.8a)--(2.1.8d), one sees
that $k$, $d$, $j$ and $l$ satisfy the relations
$$
[k,k]=0,
\eqno(2.1.14)
$$
$$
\matrix{~&~\cr
[k,d]=d, \hfill &~\hfill\cr
~&~\cr 
[k,j(\xi)]=-j(\xi),\hfill&[k,l(\xi)]=0,\quad\quad\xi\in\goth g,\hfill\cr
~&~\cr}
\eqno(2.1.15a)-(2.1.15c)
$$
$$
[d,d]=0,
\eqno(2.1.16)
$$
$$
\matrix{~&~\cr
[d,j(\xi)]=l(\xi),\hfill&[d,l(\xi)]=0,
\quad\quad\xi\in\goth g,\hfill\cr
~&~\cr}
\eqno(2.1.17a),(2.1.17b)
$$
$$
\matrix{~&~\cr
[j(\xi),j(\eta)]=0, \hfill&[j(\xi),l(\eta)]=j([\xi,\eta]),\hfill\cr
~&~\cr
[l(\xi),l(\eta)]=l([\xi,\eta]),\hfill&
\hphantom{[j(\xi),l(\eta)]=j([\xi,\eta]),}
\quad\quad\xi,\eta\in\goth g.\hfill\cr
~&~\cr}
\eqno(2.1.18a)-(2.1.18c)
$$
Note that, by (2.1.14), $k$ generates an ungraded Lie subalgebra
$$ 
\sans i\simeq\Bbb R
\eqno(2.1.19)
$$
of $\sans t$. $\sans i$ is called the internal symmetry 
algebra of the fundamental $N=1$ superstructure $\sans t$.

\vskip.2cm {\it 2.2. The fundamental N=2 superstructure}

\proclaim Proposition 2.2.1.
Let $H_A\in\sans D_2^{-1}$, $A=1,2$. Assume that the set 
$\{H_A|A=1,2\}\cup\{H_{A,B}|A,B=1,2\}\cup\{\tilde H_A|A=1,2\}$ is stable 
(cfr. def. 1.2.2). 
Let $\sans H$ the linear span of this set. Assume that
$$
[\sans H^{-1},\sans H^0]=\sans H^{-1}.
\eqno(2.2.1)
$$
Then, perhaps after substituting $H_A$ by $\kappa_A{}^BH_B$ for a suitable 
matrix $\kappa\in\GL(2,\Bbb R)$, one has 
$$
\matrix{~&~\cr
[H_A,H_B]=0,\hfill&[H_A,H_{B,C}]=\epsilon_{AB}H_C,\hfill\cr
~&~\cr
[H_A,\tilde H_B]=-H_{A,B},\hfill&
[H_{A,C},H_{B,D}]=\epsilon_{AB}H_{C,D}-\epsilon_{DC}H_{B,A},\hfill\cr
~&~\cr
[H_{A,C},\tilde H_B]=-\epsilon_{BC}\tilde H_A,\hfill&
[\tilde H_A,\tilde H_B]=0.\hfill\cr
~&~\cr}
\eqno(2.2.2a)-(2.2.2f)
$$

{\it Proof}. It is straightforward to check that (2.2.2a)--(2.2.2f)
are compatible with the graded antisymmetry and Jacobi relations of 
$\sans D_2$. By stability and grading reasons, one has 
$[\sans H^i,\sans H^j]\subseteq
\sans H^{i+j}$, where $\sans H^{-1}=\bigoplus_{A=1,2}\Bbb R H_A$, 
$\sans H^0=\bigoplus_{A,B=1,2}\Bbb R H_{A,B}$,
$\sans H^1=\bigoplus_{A=1,2}\Bbb R \tilde H_A$ and $\sans H^i=0$ else.
So, (2.2.2a) holds. Further, by stability (cfr. eq. (1.2.2)),  
$$
[H_A,H_{B,C}]=a_{ABC}{}^KH_K,\quad
[H_A,\tilde H_B]=b_{AB}{}^{KL}H_{K,L},
\eqno(2.2.3),(2.2.4)
$$
where $a_{ABC}{}^D,b_{AB}{}^{CD}\in\Bbb R$ are certain constants. 
Applying $\partial_C$ to (2.2.2a), one gets 
$[H_{A,C},H_B]+[H_{B,C},H_A]=0$. Substituting (2.2.3) into this relation 
and exploiting the linear independence of the $H_A$, one has 
$a_{ABC}{}^D+a_{BAC}{}^D=0$, which entails
$a_{ABC}{}^D=\epsilon_{AB}a_C{}^D$ for certain constants 
$a_A{}^B\in\Bbb R$. Now, on account of (2.2.1) and (2.2.3), the matrix 
$a=(a_A{}^B)$ must be invertible. Redefine $H_A$ into $(\det a)^{-1}
a_A{}^KH_K$. The new $H_A$ satisfy obviously the same assumptions as 
the old ones. Further, the new $H_A$ satisfy relations (2.2.3), where, now, 
by construction, $a_{ABC}{}^D=\epsilon_{AB}\delta_C{}^D$. In this way, 
we get (2.2.2b). Next, applying $\partial_D$ to (2.2.2b), one gets 
$[H_{A,C},H_{B,D}]+\epsilon_{CD}[\tilde H_A,H_B]=\epsilon_{AB}H_{C,D}$. 
From this relation, using (2.2.2a), (2.2.2b) and the Jacobi identity,
one obtains $[H_{A,C},[H_{B,D},H_E]]-[H_{B,D},[H_{A,C},H_E]]
-\epsilon_{CD}[[\tilde H_A,H_E],H_B]-\epsilon_{AB}[H_{C,D},H_E]=0$. 
Substituting (2.2.2b) and (2.2.4)
into this identity and exploiting the linear independence of the $H_A$,
one finds that  $b_{AB}{}^{CD}=-\delta_A{}^C\delta_B{}^D$. Substituting this
expression back into (2.2.4), one obtains (2.2.2c). Next, substituting 
(2.2.2c) into the relation $[H_{A,C},H_{B,D}]=\epsilon_{AB}H_{C,D}+
\epsilon_{DC}[\tilde H_A,H_B]$ obtained earlier, one gets (2.2.2d).
(2.2.2e) follows from applying $\partial_C$ to (2.2.2c). Finally,
(2.2.2f) follows from applying $\partial_D$ to (2.2.2e). \hfill $\square$

\proclaim Proposition 2.2.2.
Let $I:\goth g\mapsto \sans D_2$ be a linear map such that $I(\goth g)
\subseteq\sans D_2^{-2}$. Assume that the set 
$\{I,\tilde I\}\cup\{I_{,A}|A=1,2\}$ is stable 
(cfr. subsect. 1.3). For $\xi\in\goth g$, let $\sans I(\xi)$ be the linear 
span of $\{I(\xi),\tilde I(\xi)\}\cup\{I_{,A}(\xi)|A=1,2\}$. Assume that
$$
[\sans I^{-2}(\xi),\sans I^0(\eta)]=\sans I^{-2}([\xi,\eta]),
\quad\quad\xi,\eta\in\goth g.
\eqno(2.2.5)
$$
Then, perhaps after substituting $I$ by $\lambda I$ for a suitable 
$\lambda\in\Bbb R_\times$, one has 
$$
\matrix{~&~\cr
[I(\xi),I(\eta)]=0,\hfill&[I(\xi),I_{,A}(\eta)]=0,\hfill\cr
~&~\cr
[I(\xi),\tilde I(\eta)]=I([\xi,\eta]),\hfill&
[I_{,A}(\xi),I_{,B}(\eta)]=\epsilon_{AB}I([\xi,\eta]),\hfill\cr
~&~\cr
[I_{,A}(\xi),\tilde I(\eta)]=I_{,A}([\xi,\eta]),\hfill&
[\tilde I(\xi),\tilde I(\eta)]=\tilde I([\xi,\eta]),
\quad\quad\xi,\eta\in\goth g.\hfill\cr
~&~\cr}
\eqno(2.2.6a)-(2.2.6f)
$$

{\it Proof}. It is straightforward to check that (2.2.6a)--(2.2.6f)
are compatible with the graded antisymmetry and Jacobi relations of 
$\sans D_2$. By stability and grading reasons,
one has $[\sans I^i(\xi),\sans I^j(\eta)]\subseteq\sans 
I^{i+j}([\xi,\eta])$, where $\sans I^{-2}(\xi)=\Bbb R I(\xi)$, 
$\sans I^{-1}(\xi)=\bigoplus_{A=1,2}\Bbb R I_{,A}(\xi)$,
$\sans I^0(\xi)=\Bbb R \tilde I(\xi)$ and $\sans I^i(\xi)=0$ else.
So, (2.2.6a), (2.2.6b) hold. Further, by stability (cfr. eq (1.2.2)), 
one has 
$$
[I(\xi),\tilde I(\eta)]=aI([\xi,\eta]),
\eqno(2.2.7)
$$
where $a\in\Bbb R$. Owing to (2.2.5), $a\not=0$, if $[\goth g,\goth g]\not=0$,
or one can choose $a\not=0$, if $[\goth g,\goth g]=0$. 
Renormalize, $I$ into $a^{-1}I$. Then, one can set $a=1$ in (2.2.7),
getting (2.2.6c). Applying $\partial_B$ to (2.2.6b) and using (2.2.6c), 
one gets (2.2.6d) readily. Applying successively $\partial_A$ and 
$\partial_A\partial_B$ to (2.2.6c), one gets (2.2.6e), (2.2.6f).
\hfill $\square$

\proclaim Proposition 2.2.3.
Assume that $[\goth g,\goth g]\not=0$. Assume that $H$ and $I$ satisfy 
the hypotheses of props. 2.2.1 and 2.2.2, respectively. 
Assume further that $\{H_A|A=1,2\}\cup\{H_{A,B}|A,B=1,2\}\cup
\{\tilde H_A|A=1,2\}$ acts stably on $\{I,\tilde I\}
\cup\{I_{,A}|A=1,2\}$ (cfr. def. 1.2.4). Assume finally that
$$
[\sans H^0,\sans I^{-2}(\xi)]=\sans I^{-2}(\xi),
\quad\quad\xi\in\goth g.
\eqno(2.2.8)
$$
Then, one has 
$$
\matrix{~&~\cr
[H_A,I(\xi)]=0,\hfill&   
[H_A,I_{,B}(\xi)]=\epsilon_{AB}I(\xi),\hfill\cr
~&~\cr
[H_A,\tilde I(\xi)]=0,\hfill&   
[H_{A,B},I(\xi)]=\epsilon_{AB}I(\xi),\hfill\cr
~&~\cr
[H_{A,C},I_{,B}(\xi)]=\epsilon_{AB}I_{,C}(\xi),\hfill&   
[H_{A,B},\tilde I(\xi)]=0,\hfill\cr
~&~\cr
[\tilde H_A,I(\xi)]=-I_{,A}(\xi),\hfill&   
[\tilde H_A,I_{,B}(\xi)]=\epsilon_{AB}\tilde I(\xi),\hfill\cr
~&~\cr
[\tilde H_A,\tilde I(\xi)]=0.\hfill&   
\hphantom{[H_{A,B},I(\xi)]=\epsilon_{AB}I(\xi),}
\quad\quad\xi\in\goth g.\hfill\cr
~&~\cr}
\eqno(2.2.9a)-(2.2.9i)
$$

{\it Proof}. It is straightforward to check that (2.2.9a)--(2.2.9i)
are compatible with the graded antisymmetry and Jacobi relations of 
$\sans D_2$ and with (2.2.2a)--(2.2.2f), (2.2.6a)--(2.2.6f). 
By stability and grading reasons, $[\sans H^i,\sans I^j(\xi)]
\subseteq\sans I^{i+j}(\xi)$, where $\sans H^i$ and $\sans I^i(\xi)$ have 
been defined above (2.2.3) and (2.2.7), respectively. So, (2.2.9a) holds. 
Further, by stability (cfr. eq. (1.2.9)), 
$$
[H_A,I_{,B}(\xi)]=a_A{}^K\epsilon_{KB}I(\xi)
\quad [H_A,\tilde I(\xi)]=b_A{}^KI_{,K}(\xi),
\eqno(2.2.10),(2.2.11)
$$
where $a_A{}^B, b_A{}^B\in\Bbb R$. Applying $\partial_C$ to (2.2.9a) and 
(2.2.10), and using (2.2.10), (2.2.11) to cast the relations so obtained, 
one gets further 
$$
[H_{A,B},I(\xi)]=a_A{}^K\epsilon_{KB}I(\xi),\quad
[H_{A,C},I_{,B}(\xi)]=a_A{}^K\epsilon_{KB}I_{,C}(\xi)
+\epsilon_{BC}b_A{}^KI_{,K}(\xi).
\eqno(2.2.12),(2.2.13)
$$
Now, from (2.2.6e) and the Jacobi identity, 
$[[H_A,I_{,B}(\xi)],\tilde I(\eta)]-[I_{,B}(\xi),[H_A,\tilde I(\eta)]]
-[H_A,I_{,B}([\xi,\eta])]=0$. Substituting (2.2.10), (2.2.11) into this 
relation and using (2.2.6e), one obtains $b_A{}^K\epsilon_{KB}I([\xi,\eta])
=0$. As $[\goth g,\goth g]\not=0$ and $I(\zeta)\not=0$ for $\zeta\not=0$
by stability, one has $b_A{}^B=0$. Next, from (2.2.2b) and the Jacobi 
identity, one has $[H_{A,C},[H_D,I_{,B}(\xi)]]-[H_D,[H_{A,C},I_{,B}(\xi)]]
-\epsilon_{AD}[H_C,I_{,B}(\xi)]=0$. Substituting (2.2.10), (2.2.12), (2.2.13)
into this relation, one gets
$-\epsilon_{AD}\epsilon^{BK}[-a_C{}^La_L{}^K+a_L{}^La_C{}^K-a_C{}^K]I(\xi)=0$. 
Since $\goth g\not=0$ and $I(\zeta)\not=0$ for $\zeta\not=0$, 
$-a_A{}^Ka_K{}^B+a_K{}^Ka_A{}^B-a_A{}^B=0$. It is a simple algebraic exercise 
to show that there are only two 2 by 2 real matrices satisfying this relation:
$a_A{}^B=0$ and $a_A{}^B=\delta_A{}^B$. The first solution is not admissible,
because, by (2.2.12), it would violate (2.2.8). So, the second holds. 
By substituting the values of $a_A{}^B$ and $b_A{}^B$ thus found in 
(2.2.10)--(2.2.13), one gets (2.2.9b)--(2.2.9e). (2.2.9f) follows from 
applying $\partial_B$ to (2.2.9c). (2.2.9g) follows from applying 
$\partial_C$ to (2.2.9d) and using (2.2.9e) to express the result.
Applying successively $\partial_B$ and $\partial_B\partial_C$ to (2.2.9g), 
one gets (2.2.9h), (2.2.9i). \hfill $\square$

For an arbitrary ungraded Lie algebra $\goth g$, relations (2.2.2a)--(2.2.2f), 
(2.2.6a)--(2.2.6f), (2.2.9a)--(2.2.9i) define a $\Bbb Z$ graded real Lie 
algebra $\sans t_\theta$. As shown by the above three propositions, when 
$\goth g$ is not Abelian, $\sans t_\theta$ is the unique solution of a 
suitable set of stability and non degeneracy conditions. When $\goth g$ is 
Abelian, $\sans t_\theta$ is a particular case of a certain class of 
$\Bbb Z$ graded Lie algebras. We shall not investigate this matter any 
further. Note the analogy to the $N=1$ case. 

Note that $\sans t_\theta\simeq\sans t_0=:\sans t$. 
The $\Bbb Z$ graded real Lie algebra $\sans t$ is called fundamental 
$N=2$ superstructure of $\goth g$. It is defined in terms
of the six generators $h_A$, $h_{A,B}$, $\tilde h_A$, $i(\xi)$, 
$i_{,A}(\xi)$, $\tilde\imath(\xi)$, $\xi\in\goth g$ satisfying relations 
(2.2.2a)--(2.2.2f), (2.2.6a)--(2.2.6f), (2.2.9a)--(2.2.9i).
To make contact with ref. \ref{31}, one sets
$$
\matrix{~&~\cr
t_{AB}=h_{A,B}+h_{B,A},\hfill&k=\epsilon^{KL}h_{K,L},\hfill\cr
~&~\cr
u_A=h_A, \hfill&d_A=-\tilde h_A,\hfill\cr
~&~\cr}
\eqno(2.2.14a)-(2.2.14d)
$$
$$
j(\xi)=i(\xi),\quad j_A(\xi)=i_{,A}(\xi),\quad l(\xi)=\tilde\imath(\xi), 
\quad\quad \xi\in\goth g.
\eqno(2.2.15a)-(2.2.15c)
$$
From (2.2.2a)--(2.2.2f), (2.2.6a)--(2.2.6f) and (2.2.9a)--(2.2.9i), one sees
that $t_{AB}$, $k$, $u_A$, $d_A$, $j$, $j_A$ and $l$ satisfy the relations
$$
\matrix{
[t_{AC},t_{BD}]= \epsilon_{AB}t_{CD}+\epsilon_{CB}t_{AD}
+\epsilon_{AD}t_{BC}+\epsilon_{CD}t_{BA},\hfill\cr 
~\cr 
\matrix{[k,t_{AB}]=0,\hfill &[k,k]=0, \hfill\cr}
~\cr}
\eqno(2.2.16a)-(2.2.16c)
$$
$$
\matrix{~&~\cr
[t_{AC},u_B]=\epsilon_{AB}u_C+\epsilon_{CB}u_A, \hfill&[k,u_A]=-u_A ,\hfill\cr
~&~\cr
[t_{AC},d_B]=\epsilon_{AB}d_C+\epsilon_{CB}d_A, \hfill&[k,d_A]=d_A ,\hfill\cr
~&~\cr
[t_{AB},j(\xi)]=0, \hfill&[k,j(\xi)]=-2j(\xi),\hfill\cr
~&~\cr
[t_{AC},j_B(\xi)]=\epsilon_{AB}j_C(\xi)+\epsilon_{CB}j_A(\xi), 
\hfill&[k,j_A(\xi)]=-j_A(\xi) ,\hfill\cr
~&~\cr
[t_{AB},l(\xi)]=0, \hfill&[k,l(\xi)]=0,
\quad\quad \xi\in\goth g,\hfill\cr
~&~\cr}
\eqno(2.2.17a)-(2.2.17j)
$$
$$
\matrix{~&~\cr
[u_A,u_B]=0,\hfill&[d_A,u_B]={1\over 2}(t_{AB}+\epsilon_{AB}k), 
\hfill\cr
~&~\cr
[d_A,d_B]=0,\hfill&~\cr
~&~\cr}
\eqno(2.2.18a)-(2.2.18c)
$$
$$
\matrix{~&~\cr
[u_A,j(\xi)]=0,\hfill&   
[u_A,j_B(\xi)]=\epsilon_{AB}j(\xi),\hfill\cr
~&~\cr
[u_A,l(\xi)]=0,\hfill& 
[d_A,j(\xi)]=j_A(\xi),\hfill\cr  
~&~\cr
[d_A,j_B(\xi)]=-\epsilon_{AB}l(\xi),\hfill&
[d_A,l(\xi)]=0,\quad\quad \xi\in\goth g,\hfill\cr
~&~\cr}
\eqno(2.2.19a)-(2.2.19f)
$$
$$
\matrix{~&~\cr
[j(\xi),j(\eta)]=0,\hfill&[j(\xi),j_A(\eta)]=0,\hfill\cr
~&~\cr
[j(\xi),l(\eta)]=j([\xi,\eta]),\hfill&
[j_A(\xi),j_B(\eta)]=\epsilon_{AB}j([\xi,\eta]),\hfill\cr
~&~\cr
[j_A(\xi),l(\eta)]=j_A([\xi,\eta]),\hfill&
[l(\xi),l(\eta)]=l([\xi,\eta]),
\quad\quad \xi,\eta\in\goth g.\hfill\cr
~&~\cr}
\eqno(2.2.20a)-(2.2.20f)
$$
Note that, from (2.2.16a)--(2.2.16c), $t_{AB}$, $k$ generate an ungraded
Lie subalgebra
$$
\sans i\simeq\goth s\goth l(2,\Bbb R)\oplus\Bbb R
\eqno(2.2.21)
$$
of $\sans t$. $\sans i$ is called the internal symmetry algebra of 
the $N=2$ fundamental superstructure $\sans t$ and plays an important role. 
\par\vskip .6cm
{\bf 3. The Weil algebra}
\vskip .4cm
Let $\sans d$ be a $\Bbb Z$ graded real Lie algebra.
Let $\sans z$ be a $\Bbb Z$ graded real left $\sans d$ module algebra 
with unity $1$, where the action of $\sans d$ on $\sans z$ is derivative,
i. e. it obeys the graded Leibniz rule. Finally, let $\goth g$ be 
an ungraded real Lie algebra.

\vskip.2cm {\it 3.1. The N=1 case}

\proclaim Proposition 3.1.1. 
Assume that $[\goth g,[\goth g,\goth g]]\not=0$. Assume that $H$ and $I$ 
satisfy the hypotheses of props. 2.1.1 and 2.1.2, respectively. Let 
$W:\goth g^\vee\mapsto \sans Z_1$ be a homomorphism 
such that $W(\goth g^\vee)\subseteq 
\sans Z_1^1$. Suppose $\{W,\tilde W\}$ is stable and that 
$\{H,\tilde H\}$ and $\{I,\tilde I\}$ act stably and $1$--stably on
$\{W,\tilde W\}$ (cfr. defs. 1.2.3, 1.2.4), respectively, the module 
action of $\goth g$ on $\goth g^\vee$ being the coadjoint one. 
For $\mu\in\goth g^\vee$, denote by $\sans W(\mu)$ the linear span 
of $\{W(\mu),\tilde W(\mu)\}$. Assume that
$$
\sans I^{-1}(\xi)\sans W^1(\mu)=\Bbb R1,
\eqno(3.1.1)
$$
for some $\xi\in\goth g$, for fixed $\mu\in\goth g^\vee$, $\mu\not=0$,
and for some $\mu\in\goth g^\vee$, for fixed $\xi\in\goth g$, $\xi\not=0$.
Finally, suppose that
$$
\sans I^0(\xi)\sans W^1(\mu)=\sans W^1(\ad^\vee\xi\mu),
\eqno(3.1.2)
$$
for $\xi\in\goth g$ and $\mu\in\goth g^\vee$. Then, after perhaps 
redefining $W$ into $W\circ f$ for some invertible linear map
$f:\goth g^\vee\mapsto\goth g^\vee$ and viewing $W$ as an element of 
$\sans Z_1\otimes \goth g$, one has
$$
\matrix{~&~\cr
HW=W,\hfill&H\tilde W=2\tilde W,\hfill\cr
~&~\cr
\tilde HW=-\tilde W,\hfill&\tilde H\tilde W=0,\hfill\cr
~&~\cr}
\eqno(3.1.3a)-(3.1.3d)
$$
$$
\matrix{~&~\cr
I(\xi)W=\xi,\hfill&I(\xi)\tilde W=-[\xi,W],\hfill\cr
~&~\cr
\tilde I(\xi)W=-[\xi,W],\hfill&\tilde I(\xi)\tilde W=-[\xi,\tilde W],
\quad\quad\xi\in\goth g.\hfill\cr
~&~\cr}
\eqno(3.1.4a)-(3.1.4d)
$$

{\it Proof}. It is straightforward to check that (3.1.3a)--(3.1.3d) and 
(3.1.4a)--(3.1.4d) are compatible with relations (2.1.2a)--(2.1.2c),
(2.1.5a)--(2.1.5c) and (2.1.8a)--(2.1.8d). By stability, $1$--stability and 
grading reasons, $\sans H^i\sans W^j(\mu)\subseteq\sans W^{i+j}(\mu)$ and
$\sans I^i(\xi)\sans W^j(\mu)\subseteq\sans W^{i+j}(\ad^\vee\xi\mu)
\oplus\Bbb R\delta_{i+j,0}1$, where $\sans H^i$ and $\sans I^i(\xi)$ are 
given above (2.1.3) and (2.1.6), respectively, and $\sans W^1(\mu)
=\Bbb RW(\mu)$, $\sans W^2(\mu)=\Bbb R\tilde W(\mu)$ and $\sans W^i(\mu)=0$, 
else. We note further that $H1=0$ and $I(\xi)1=0$ as $H$ and $I(\xi)$ act as 
derivations and $1^2=1$. By stability (cfr. eq. (1.2.9)), one has thus
$$
HW(\mu)=aW(\mu),\quad \tilde HW(\mu)=b\tilde W(\mu),
\eqno(3.1.5),(3.1.6)
$$
where $a,b\in\Bbb R$. Further, by $1$--stability (cfr. eq. (1.2.9)), one has 
$$
I(\xi)W(\mu)=\varphi(\xi,\mu), \quad\tilde I(\xi)W(\mu)=cW(\ad^\vee\xi\mu),
\eqno(3.1.7),(3.1.8)
$$
where $\varphi:\goth g\times\goth g^\vee\mapsto\Bbb R$ is a bilinear map 
and $c\in\Bbb R$. Next, from (2.1.8a), one has the relation 
$([H,I(\xi)]+I(\xi))W(\mu)=0$, which, upon using (3.1.5), (3.1.7), 
gives $(1-a)\varphi(\xi,\mu)=0$. As $\varphi\not=0$, by (3.1.1), (3.1.7), 
$a=1$. Next, by (3.1.2), (3.1.8), $c\not=0$. From (2.1.5c), one has the
relation $([\tilde I(\xi),\tilde I(\eta)]-\tilde I([\xi,\eta]))W(\mu)=0$, 
which, using (3.1.8), yields $c(c+1)W(\ad^\vee[\xi,\eta]\mu)=0$. 
As, $c\not=0$ and $[\goth g,[\goth g,\goth g]]\not=0$ and 
$W(\nu)\not=0$ for $\nu\not=0$ by stability, one has $c=-1$. Next, we write
$\varphi(\xi,\mu)=\langle\mu,\phi(\xi)\rangle$, where $\phi:\goth g\mapsto
\goth g$ is a linear map, which on account of (3.1.1), (3.1.7) is invertible. 
From (2.1.5b), one has the relation $([I(\xi),\tilde I(\eta)]
-I([\xi,\eta]))W(\mu)=0$, which, using (3.1.7), (3.1.8), gives the 
relation  $\langle\mu,[\phi(\xi),\eta]-\phi([\xi,\eta])\rangle=0$. So,
$[\phi,\ad\eta]=0$. Redefine $W$ into $W\circ\phi^{\vee-1}$. 
After doing so, (3.1.7) holds with $\varphi(\xi,\mu)=\langle\mu,\xi\rangle$,
while (3.1.8) is formally invariant. In this way, one obtains 
(3.1.4a), (3.1.4c). (3.1.4b) follows from applying $\partial$ to (3.1.4a)
and using (3.1.4c). (3.1.4d) follows from applying $\partial$ to (3.1.4c).
Next, by (2.1.8c), one has the relation $[[\tilde H,I(\xi)]
+\tilde I(\xi)]W(\mu)=0$, which, by  using (3.1.6), (3.1.4a)--(3.1.4c) gives 
$-(b+1)W(\ad^\vee\xi\mu)=0$. As $[\goth g,\goth g]\not=0$ and $W(\nu)\not=0$ 
for $\nu\not=0$, $b=-1$. By substituting the values of $a$ and $b$ thus found
into (3.1.5), (3.1.6), one gets (3.1.3a). (3.1.3c). (3.1.3b) follows from 
applying $\partial$ to (3.1.3a) and using (3.1.3c). (3.1.3d) follows from 
applying $\partial$ to (3.1.3c).  \hfill $\square$

For an arbitrary ungraded Lie algebra $\goth g$, relations (3.1.3a)--(3.1.3d), 
(3.1.4a)--(3.1.4d) define a $\Bbb Z$ graded real left module algebra 
$\sans w_\theta$ of the graded real Lie algebra $\sans t_\theta$ 
(cfr. subsect. 2.1). 
As shown by the above propositions, when $[\goth g,[\goth g,\goth g]]\not=0$, 
$\sans w_\theta$ is the unique solution of a suitable set of stability and non 
degeneracy conditions. Else, $\sans w_\theta$ is a particular case of a 
certain class of $\Bbb Z$ graded left module algebra of $\sans t_\theta$.

Note that $\sans w_\theta\simeq\sans w_0=:\sans w$. $\sans w$ is called 
$N=1$ Weil algebra of $\goth g$ and is a $\Bbb Z$ graded real left module 
algebra of the fundamental $N=1$ superstructure $\sans t$ of $\goth g$
(cfr. subsect. 2.1). 
It is defined in terms of the generators $1$, $w$, $\tilde w$ and the 
derivations $h$, $\tilde h$, $i(\xi)$, $\tilde\imath(\xi)$, $\xi\in\goth g$,
satisfying relations (3.1.3a)--(3.1.3d), (3.1.4a)--(3.1.4d). However,
in the standard treatment, $\sans w$ is usually presented as follows. Define
$$
\omega=w,\quad\phi=\tilde w+(1/2)[w,w].
\eqno(3.1.9)
$$
Then, one has
$$
k\omega=\omega,\quad k\phi=2\phi,
\eqno(3.1.10a),(3.1.10b)
$$
$$
d\omega=\phi-(1/2)[\omega,\omega],\quad d\phi=-[\omega,\phi],
\eqno(3.1.11a),(3.1.11b)
$$
$$
\matrix{~&~\cr
j(\xi)\omega=\xi,\hfill& j(\xi)\phi=0,\hfill\cr
~&~\cr
l(\xi)\omega=-[\xi,\omega],\hfill& l(\xi)\phi=-[\xi,\phi],
\quad\quad\xi\in\goth g,\hfill\cr
~&~\cr}
\eqno(3.1.12a)-(3.1.12d)
$$
where $k$, $d$, $j$, $l$ are given by (2.1.12a), (2.1.12b), (2.1.13a), 
(2.1.13b). Note that $\omega$ is just another name for $w$. 
$\phi$ is by construction `horizontal', i. e. satisfying (3.1.12b).

\vskip.2cm {\it 3.2. The N=2 case}

\proclaim Proposition 3.2.1. 
Assume that $[\goth g,[\goth g,\goth g]]\not=0$. Assume that $H_A$, $A=1,2$, 
and $I$ satisfy the hypotheses of props. 2.2.1 and 2.2.2, respectively. 
Let $W_A:\goth g^\vee\mapsto\sans Z_2$, $A=1,2$, be homomorphisms 
such that $W(\goth g^\vee)\subseteq\sans Z_2^1$. Suppose that 
$\{W_A|A=1,2\}\cup\{W_{A,B}|A,B=1,2\}\cup\{\tilde W_A|A=1,2\}$ is stable 
and that $\{H_A|A=1,2\}\cup\{H_{A,B}|A,B=1,2\}\cup\{\tilde H_A|A=1,2\}$ 
and $\{I,\tilde I\}\cup\{I_{,A}|A=1,2\}$ act stably and $1$--stably on
$\{W_A|A=1,2\}\cup\{W_{A,B}|A,B=1,2\}\cup\{\tilde W_A|A=1,2\}$
(cfr. defs. 1.2.3, 1.2.4), respectively, the module action of $\goth g$ on 
$\goth g^\vee$ being the coadjoint one. For $\mu\in\goth g^\vee$,
denote by $\sans W(\mu)$ the linear span of $\{W_A(\mu)|A=1,2\}\cup
\{W_{A,B}(\mu)|A,B=1,2\}\cup\{\tilde W_A(\mu)|A=1,2\}$. Assume that
$$
\sans I^{-1}(\xi)\sans W^1(\mu)=\Bbb R1,
\eqno(3.2.1)
$$
for some $\xi\in\goth g$, for fixed $\mu\in\goth g^\vee$, $\mu\not=0$,
and for some $\mu\in\goth g^\vee$, for fixed $\xi\in\goth g$, $\xi\not=0$.
Finally, suppose that
$$
\sans I^0(\xi)\sans W^1(\mu)=\sans W^1(\ad^\vee\xi\mu),
\eqno(3.2.2)
$$
for $\xi\in\goth g$ and $\mu\in\goth g^\vee$. Then, after perhaps 
redefining $W_A$ into $\lambda_A{}^BW_B\circ f$ for some matrix
$\lambda\in\GL(2,\Bbb R)$ and invertible linear map
$f:\goth g^\vee\mapsto\goth g^\vee$ and viewing $W$ as an element of 
$\sans Z_2\otimes \goth g$, one has
$$
\matrix{~&~\cr
H_AW_B=0,\hfill&H_AW_{B,C}=-\epsilon_{BC}W_A,\hfill\cr
~&~\cr
H_A\tilde W_B=-W_{A,B}-W_{B,A},\hfill&H_{A,C}W_B=-\epsilon_{BC}W_A,\hfill\cr
~&~\cr
H_{A,C}W_{B,D}=\epsilon_{CB}W_{A,D}-\epsilon_{DC}W_{B,A},\hfill
&H_{A,C}\tilde W_B=-\epsilon_{BC}\tilde W_A-\epsilon_{AC}\tilde W_B,
\hfill\cr
~&~\cr
\tilde H_AW_B=-W_{B,A},\hfill
&\tilde H_AW_{B,C}=\epsilon_{AC}\tilde W_B,\hfill\cr
~&~\cr
\tilde H_A\tilde W_B=0,\hfill&~\hfill\cr
~&~\cr}
\eqno(3.2.3a)-(3.2.3i)
$$
$$
\matrix{~&~\cr
I(\xi)W_A=0,\hfill&I(\xi)W_{A,B}=\epsilon_{AB}\xi,\hfill\cr
~&~\cr
I(\xi)\tilde W_A=-[\xi,W_A],\hfill&I_{,A}(\xi)W_B=\epsilon_{AB}\xi,\hfill\cr
~&~\cr
I_{,A}(\xi)W_{B,C}=-\epsilon_{AC}[\xi,W_B],\hfill
&I_{,A}(\xi)\tilde W_B=-[\xi,W_{B,A}],\hfill\cr
~&~\cr
\tilde I(\xi)W_A=-[\xi,W_A],\hfill
&\tilde I(\xi)W_{A,B}=-[\xi,W_{A,B}],\hfill\cr
~&~\cr
\tilde I(\xi)\tilde W_A=-[\xi,\tilde W_A],\hfill
&\hphantom{\tilde I(\xi)\tilde W_A=-[\xi,\tilde W_A],}
\quad\quad\xi\in\goth g.\hfill\cr
~&~\cr}
\eqno(3.2.4a)-(3.2.4i)
$$

{\it Proof}. It is straightforward to check that (3.2.3a)--(3.2.3i) and 
(3.2.4a)--(3.2.4i) are compatible with relations (2.2.2a)--(2.2.2f),
(2.2.6a)--(2.2.6f) and (2.2.9a)--(2.2.9i). By stability, $1$--stability and 
grading reasons, one has 
$\sans H^i\sans W^j(\mu)\subseteq\sans W^{i+j}(\mu)$ and
$\sans I^i(\xi)\sans W^j(\mu)\subseteq\sans W^{i+j}(\ad^\vee\xi\mu)
\oplus\Bbb R\delta_{i+j,0}1$, where $\sans H^i$ and $\sans I^i(\xi)$ are 
given above (2.2.3) and (2.2.7), respectively, and $\sans W^1(\mu)
=\bigoplus_{A=1,2}\Bbb RW_A(\mu)$, $\sans W^2(\mu)=
\bigoplus_{A,B=1,2}\Bbb RW_{A,B}(\mu)$, $\sans W^3(\mu)
=\bigoplus_{A=1,2}\Bbb R\tilde W_A(\mu)$ and $\sans W^i(\mu)=0$, else. 
We note further that $H_A1=0$ and $I(\xi)1=0$ as $H_A$ and $I(\xi)$ act as 
derivations and $1^2=1$. Thus, by stability, (3.2.3a) holds and one has 
$$
H_{A,C}W_B(\mu)=a_{ABC}{}^KW_K(\mu),
\quad \tilde H_AW_B(\mu)=b_{AB}{}^{KL}W_{K,L}(\mu),
\eqno(3.2.5),(3.2.6)
$$
where $a_{ABC}{}^D,b_{AB}{}^{CD}\in\Bbb R$. Further, by $1$--stability,
(3.2.4a) holds and one has 
$$
I_{,A}(\xi)W_B(\mu)=\varphi_{AB}(\xi,\mu), 
\quad\tilde I(\xi)W_A(\mu)=c_A{}^KW_K(\ad^\vee\xi\mu),
\eqno(3.2.7),(3.2.8)
$$
where $\varphi_{AB}:\goth g\times\goth g^\vee\mapsto\Bbb R$ is a bilinear map 
and $c_A{}^B\in\Bbb R$. Now, by (2.2.2d), 
$([H_{A,C},H_{B,D}]-\epsilon_{AB}H_{C,D}+\epsilon_{DC}H_{B,A})W_E(\mu)=0$.
Upon substituting (3.2.5) into this relation, one obtains 
$(a_{AEC}{}^Ka_{BKD}{}^L-a_{BED}{}^Ka_{AKC}{}^L
+\epsilon_{AB}a_{CED}{}^L-\epsilon_{DC}a_{BEA}{}^L)W_L(\mu)=0$. 
Since the $W_A(\nu)$ are linearly independent for $\nu\not=0$,  
$a_{AEC}{}^Ka_{BKD}{}^F-a_{BED}{}^Ka_{AKC}{}^F+\epsilon_{AB}a_{CED}{}^F
-\epsilon_{DC}a_{BEA}{}^F=0$. As is easy to see, this relation implies 
that the matrices $a_{AB}=(a_{ACB}{}^D)$ form a 2 dimensional
representation of the Lie algebra $\goth s\goth l(2,\Bbb R)\oplus \Bbb R$. 
A simple Lie algebraic analysis shows that either $i$) $a_{ABC}{}^D
=-\epsilon_{BC}\delta_A{}^D-\epsilon_{AC}a\delta_B{}^D$ or $ii$) 
$a_{ABC}{}^D=-\epsilon_{AC}a_B{}^D$ up to equivalence, 
where $a,a_A{}^B\in\Bbb R$. From (2.2.9e), one has the relation 
$([H_{A,C},I_{,B}(\xi)]-\epsilon_{AB}I_{,C}(\xi))W_D(\mu)=0$,
which upon using (3.2.5), (3.2.7), entails that 
$-a_{ADC}{}^K\varphi_{BK}(\xi,\mu)-\epsilon_{AB}\varphi_{CD}(\xi,\mu)=0$. 
Substituting the expression of $a_{ACB}{}^D$ found above into
this identity, one gets after some simple rearrangements
$a\varphi_{AB}(\xi,\mu)=0$ in case $i$ and $\varphi_{AB}(\xi,\mu)=0$ 
in case $ii$. As $\varphi_{AB}$ cannot vanish by (3.2.1), (3.2.7), 
only case $i$ with $a=0$ can occur. So, $a_{ABC}{}^D
=-\epsilon_{BC}\delta_A{}^D$. Using this expression, one sees then that 
$\epsilon_{CD}\varphi_{AB}(\xi,\mu)-\epsilon_{AB}\varphi_{CD}(\xi,\mu)=0$, 
which implies that $\varphi_{AB}(\xi,\mu)=\epsilon_{AB}\varphi(\xi,\mu)$ 
for some bilinear map $\varphi:\goth g\times\goth g^\vee\mapsto\Bbb R$.
Next, by (3.2.2), (3.2.8), the matrix $c=(c_A{}^B)$ is invertible. 
From (2.2.6f), one has the relation $[[\tilde I(\xi),\tilde I(\eta)]
-\tilde I([\xi,\eta])]W_A(\mu)=0$, which, using (3.2.8), yields 
$c_A{}^K(c_K{}^L+\delta_K{}^L)W_L(\ad^\vee[\xi,\eta]\mu)=0$. 
As, $c$ is invertible, $[\goth g,[\goth g,\goth g]]\not=0$ and 
the $W_A(\nu)$ are linearly independent for $\nu\not=0$, one has 
$c_A{}^B=-\delta_A{}^B$. Next, we write $\varphi(\xi,\mu)=
\langle\mu,\phi(\xi)\rangle$, where $\phi:\goth g\mapsto\goth g$ is a 
linear map, which on account of (3.2.1) is invertible. 
From (2.2.6e), one has the relation $([I_{,A}(\xi),\tilde I(\eta)]
-I_{,A}([\xi,\eta]))W_B(\mu)=0$, which, using (3.2.7), (3.2.8), gives the 
relation  $\epsilon_{AB}\langle\mu,[\phi(\xi),\eta]
-\phi([\xi,\eta])\rangle=0$. So, $[\phi,\ad\eta]=0$. 
Redefine $W$ into $W\circ\phi^{\vee-1}$. 
After doing so, (3.2.7) holds with $\varphi(\xi,\mu)=\langle\mu,\xi\rangle$,
while (3.2.8) is formally invariant. In this way, 
one obtains (3.2.4d), (3.2.4g). By applying $\partial_B$ to (3.2.4a)
and using (3.2.4d), one obtains (3.2.4b).
Applying $\partial_C$ to (3.2.4d) and using (3.2.4g), one obtains
(3.2.4e). Applying $\partial_C$ to (3.2.4b) and using (3.2.4e),
one gets (3.2.4c). Applying $\partial_B$, $\partial_C\partial_B$ to 
(3.2.4g), one gets (3.2.4h), (3.2.4i), respectively. Applying $\partial_D$ 
to (3.2.4e) and using (3.2.4h), one gets (3.2.4f). Next, by (2.2.9h), 
$([\tilde H_A,I_{,B}(\xi)]-\epsilon_{AB}\tilde I(\xi))W_C(\mu)=0$.
Using (3.2.6), (3.2.4d), (3.2.4e), (3.2.4g), this yields 
$(\epsilon_{AB}\delta_C{}^K+b_{AC}{}^{KL}\epsilon_{LB})
W_K(\ad^\vee\xi\mu)=0$. As $[\goth g,\goth g]\not=0$ and
the $W_A(\nu)$ are linearly independent for $\nu\not=0$, one has 
$b_{AB}{}^{CD}=-\delta_A{}^D\delta_B{}^C$. Substituting 
the expressions of $a_{ABC}{}^D$ and $b_{AB}{}^{CD}$ obtained  
into (3.2.5), (3.2.6), one obtains (3.2.3d), (3.2.3g). 
By applying $\partial_B$ to (3.2.3a) and using (3.2.3d), one obtains 
(3.2.4b). Applying $\partial_D$ to (3.2.3d) and using (3.2.3g), one obtains
(3.2.3e). Applying $\partial_D$ to (3.2.3b) and using (3.2.3e),
one gets (3.2.3c). Applying $\partial_C$, $\partial_D\partial_C$ to (3.2.3g), 
one gets (3.2.3h), (3.2.3i), respectively. Applying $\partial_E$ to
(3.2.3e) and using (3.2.3h), one gets (3.2.3f).\hfill $\square$

For an arbitrary ungraded Lie algebra $\goth g$, relations (3.2.3a)--(3.2.3i), 
(3.2.4a)--(3.2.4i) define a $\Bbb Z$ graded real left module algebra 
$\sans w_\theta$ of the graded real Lie algebra $\sans t_\theta$ 
(cfr. subsect. 2.2).
As shown by the above proposition, when $[\goth g,[\goth g,\goth g]]\not=0$, 
$\sans w_\theta$ is the unique solution of a suitable set of stability and 
non degeneracy conditions. Else, $\sans w_\theta$ is a particular case of a 
certain class of $\Bbb Z$ graded left module algebra of $\sans t_\theta$.
Note the analogy to the $N=1$ case

Note that $\sans w_\theta\simeq\sans w_0=:\sans w$. $\sans w$ is called 
$N=2$ Weil algebra of $\goth g$ and is a $\Bbb Z$ graded real left module 
algebra of the fundamental $N=2$ superstructure $\sans t$ of $\goth g$
(cfr. subsect. 2.2).
It is defined in terms of the generators $1$, $w_A$, $w_{A,B}$, $\tilde w_A$
of the derivations $h_A$, $h_{A,B}$, $\tilde h_A$, $i(\xi)$, 
$i_{,A}(\xi)$, $\tilde\imath(\xi)$, $\xi\in\goth g$ satisfying relations 
(3.2.3a)--(3.2.3i), (3.2.4a)--(3.2.4i).
To make contact with ref. \ref{31}, we shall present $\sans w$ as follows.
Define
$$
\matrix{~&~\cr
\omega_A=w_A,\hfill
&\phi_{AB}={1\over 2}(w_{A,B}+w_{B,A}+[w_A,w_B]),\hfill\cr
~&~\cr
\gamma=-{1\over 2}\epsilon^{KL}w_{K,L},\hfill
&\rho_A=-{1\over 2}\tilde w_A-{1\over 2}\epsilon^{KL}[w_K,w_{A,L}]
-{1\over 6}\epsilon^{KL}[w_K,[w_L,w_A]].\hfill\cr
~&~\cr}
\eqno(3.2.9a),(3.2.9b)
$$
Then, one has 
$$
\matrix{~&~\cr
t_{AC}\omega_B=\epsilon_{AB}\omega_C+\epsilon_{CB}\omega_A, 
\hfill&k\omega_A=\omega_A ,\hfill\cr
~&~\cr
t_{AC}\phi_{BD}=\epsilon_{AB}\phi_{CD}+\epsilon_{CB}\phi_{AD}
+\epsilon_{AD}\phi_{BC}+\epsilon_{CD}\phi_{BA}, 
\hfill&k\phi_{AB}=2\phi_{AB},\hfill\cr
~&~\cr
t_{AB}\gamma=0, \hfill&k\gamma=2\gamma,\hfill\cr
~&~\cr
t_{AC}\rho_B=\epsilon_{AB}\rho_C+\epsilon_{CB}\rho_A, 
\hfill&k\rho_A=3\rho_A ,\hfill\cr
~&~\cr}
\eqno(3.2.10a)-(3.2.10h)
$$
$$
\matrix{~&~\cr
u_A\omega_B=0,\hfill&u_A\phi_{BC}=0,\hfill\cr
~&~\cr
u_A\gamma=-\omega_A, \hfill&u_A\rho_B=\phi_{AB},\hfill\cr
~&~\cr
d_A\omega_B=-{1\over 2}[\omega_A,\omega_B]+\phi_{AB}-\epsilon_{AB}\gamma, 
\hfill
&d_A\phi_{BC}=-[\omega_A,\phi_{BC}]+\epsilon_{AB}\rho_C+\epsilon_{AC}\rho_B,
\hfill\cr
~&~\cr
d_A\gamma=-{1\over 2}[\omega_A,\gamma]+\rho_A
\hfill
&d_A\rho_B=-[\omega_A,\rho_B]
-{1\over 2}\epsilon^{KL}[\phi_{KA},\phi_{LB}],\hfill\cr
~&~\cr
\hphantom{d_A\gamma=}
+{1\over2}\epsilon^{KL}[\omega_K,\phi_{LA}-{1\over 6}[\omega_L,\omega_A]],
\hfill&~\hfill\cr
~&~\cr}
\eqno(3.2.11a)-(3.2.11h)
$$
$$
\matrix{~&~\cr
j(\xi)\omega_A=0,\hfill&j(\xi)\phi_{AB}=0,\hfill\cr
~&~\cr
j(\xi)\gamma=\xi, \hfill&j(\xi)\rho_A=0,\hfill\cr
~&~\cr
j_A(\xi)\omega_B=\epsilon_{AB}\xi,\hfill&j_A(\xi)\phi_{BC}=0,\hfill\cr
~&~\cr
j_A(\xi)\gamma=-{1\over 2}[\xi,\omega_A],\hfill&j_A(\xi)\rho_B=0,\hfill\cr
~&~\cr
l(\xi)\omega_A=-[\xi,\omega_A],\hfill&l(\xi)\phi_{AB}=-[\xi,\phi_{AB}],
\hfill\cr
~&~\cr
l(\xi)\gamma=-[\xi,\gamma], \hfill&l(\xi)\rho_A=-[\xi,\rho_A],
\quad\quad\xi\in\goth g,\hfill\cr
~&~\cr}
\eqno(3.2.12a)-(3.2.12l)
$$
where $t_{A,B}$, $k$, $u_A$, $d_A$, $j$, $j_A$, $l$
are given by by (2.2.14a)--(2.2.14d), (2.2.15a)--(2.2.15c). Note that
$\omega_A$ is just another name for $w_A$. $\gamma$ contains 
the information about $\tilde h_Aw_B$ not exhausted by $\phi_{AB}$.
By construction $\phi_{AB}$ and $\rho_A$ are `horizontal', i. e. satisfy
(3.2.12b), (3.2.12d), (3.2.12f), (3.2.12h). 
\par\vskip .6cm
{\bf 4. Superoperations and their cohomologies}
\vskip .4cm
Let $\goth g$ be an ungraded real Lie algebra. 

\vskip.2cm {\it 4.1. N=1 superoperations and their cohomologies}

\proclaim Definition 4.1.1. 
$\sans a$ is called an $N=1$ $\goth g$ superoperation if:
\item{$i$)} $\sans a$ is a $\Bbb Z$ graded real left module algebra 
of the fundamental $N=1$ superstructure $\sans t$ of $\goth g$
(cfr. subsect. 2.1);
\item{$ii)$} the action of $\sans t$ on $\sans a$ is derivative;
\item{$iii$)} $\sans a$ is completely reducible under the internal symmetry
algebra $\sans i$ of $\sans t$ (cfr. subsect. 2.1), 
\item{\hphantom{$iii$)}} the spectrum of the 
invariant $k$ of $\sans i$ is integer and the eigenspace $\sans a^p$
of $k$ of the eigen-
\item{\hphantom{$iii$)}}
value $p\in\Bbb Z$ is precisely the degree $p$ subspace of $\sans a$.

So, $\sans a$ is acted upon by four graded derivations $h$, $\tilde h$, 
$i(\xi)$, $\tilde\imath(\xi)$, $\xi\in\goth g$, of degree
$0$, $+1$, $-1$, $0$, respectively, satisfying relations
(2.1.2a)--(2.1.2c), (2.1.5a)--(2.1.5c), (2.1.8a)--(2.1.8d), or, 
equivalently, by four graded derivations $k$, $d$, $j(\xi)$, 
$l(\xi)$, $\xi\in\goth g$, of degree 
$0$, $+1$, $-1$, $0$, respectively, satisfying relations (2.1.14), 
(2.1.15a)--(2.1.15c), (2.1.16), (2.1.17a)--(2.1.17b), (2.1.18a)--(2.1.18c), 
the two sets of derivations being related as in (2.1.12a), (2.1.12b), 
(2.1.13a), (2.1.13b).

\proclaim Proposition 4.1.1. If $\sans a^{(r)}$, $r=1,2$, are two $N=1$ 
$\goth g$ superoperations, then their graded tensor product $\sans a
=\sans a^{(1)}\hat\otimes\sans a^{(2)}$ is also an $N=1$ $\goth g$ 
superoperation.

{\it Proof}. Indeed $\sans a$ satisfies the conditions stated 
in def. 4.1.1. \hfill $\square$

Let $\sans a$ be an $N=1$ $\goth g$ superoperation.

The pair $(\sans a, d)$ is an ordinary differential
complex, as the graded derivation $d$ has degree $+1$ and 
$[d,d]=0$. Its cohomology $H^*(\sans a)$, defined in the usual way by
$$
H^p(\sans a)=(\ker d\cap \sans a^p)/d\sans a^{p-1},\quad\quad p\in\Bbb Z,
\eqno(4.1.1)
$$
is the ordinary cohomology of the superoperation. Define 
$$
\sans a_{\rm basic}=\bigcap_{\xi\in\goth g}\ker j(\xi)\cap\ker l(\xi).
\eqno(4.1.2)
$$
By (2.1.17a), (2.1.17b), $\sans a_{\rm basic}$ is $d$ invariant. So,
$(\sans a_{\rm basic},d)$ is also a differential complex. 
Its cohomology $H^*_{\rm basic}(\sans a)$
$$
H^p_{\rm basic}(\sans a)=(\ker d\cap \sans a_{\rm basic}^p)/
d\sans a_{\rm basic}^{p-1},\quad\quad p\in\Bbb Z,
\eqno(4.1.3)
$$
is the basic cohomology of the superoperation. 

\proclaim Proposition 4.1.2. 
Each non zero (basic) cohomology class of degree $p$ defines a 1 dimensional
representation of the internal Lie algebra $\sans i$ of invariant $p$.

{\it Proof}. Set $k[x]=[kx]=p[x]$ for $[x]\in H^p(\sans a)$ 
($[x]\in H^p_{\rm basic}(\sans a)$) with arbitrary representative 
$x\in\sans a^{n,p}$ ($x\in\sans a^{n,p}_{\rm basic}$). 
\hfill $\square$

Though the above proposition is trivial, it is nevertheless interesting
because of its non trivial generalization to higher $N$. 

\vskip.2cm {\it 4.2. N=2 superoperations and their cohomologies}

\proclaim Definition 4.2.1. 
$\sans a$ is called an $N=2$ $\goth g$ superoperation if:
\item{$i$)} $\sans a$ is a $\Bbb Z$ graded real left module algebra 
of the fundamental $N=2$ superstructure $\sans t$ of $\goth g$
(cfr. subsect. 2.2);
\item{$ii$)} the action of $\sans t$ on $\sans a$ is derivative;
\item{$iii$)} $\sans a$ is completely reducible under the internal symmetry
algebra $\sans i$ of $\sans t$ (cfr. subsect. 2.2), 
\item{\hphantom{$iii$)}} the spectrum of the 
invariant $k$ of $\sans i$ is integer and the eigenspace $\sans a^p$
of $k$ of the eigen-
\item{\hphantom{$iii$)}}
value $p\in\Bbb Z$ is precisely the degree $p$ subspace of $\sans a$.

So, $\sans a$ is acted upon by six graded derivations $h_A$, $h_{A,B}$, 
$\tilde h_A$, $i(\xi)$, $i_{,A}(\xi)$, $\tilde\imath(\xi)$, $\xi\in\goth g$,
of degree $-1$, $0$, $+1$, $-2$, $-1$, $0$, respectively, satisfying relations 
(2.2.2a)--(2.2.2f), (2.2.6a)--(2.2.6f), (2.2.9a)--(2.2.9i), or, equivalently, 
by seven graded derivations $t_{AB}$, $k$, $u_A$, $d_A$, $j(\xi)$, 
$j_A(\xi)$, $l(\xi)$, $\xi\in\goth g$, of degree $0$, $0$, $-1$, $+1$, $-2$, 
$-1$, $0$, respectively, satisfying relations (2.2.16a)--(2.2.16c), 
(2.2.17a)--(2.2.17j), (2.2.18a)--(2.2.18c), (2.2.19a)--(2.2.19f), 
(2.2.20a)--(2.2.20f), the two sets of derivations being related as in 
(2.2.14a)--(2.2.14d), (2.2.15a)--(2.2.15c). 

Besides $k$, $\sans i$ possesses another invariant, namely
$$
c=-\hbox{$1\over 8$}\epsilon^{KL}\epsilon^{MN}t_{KM}t_{LN}.
\eqno(4.2.1)
$$
An irreducible representation of $\sans i$ is completely characterized 
up to equivalence by the values of $c$ and $k$, which we parametrize
as ${1\over 4}(n^2-1)$ and $p$, respectively, where $n\in\Bbb N$
and $p\in\Bbb Z$. $n$ is nothing but the dimension of the 
representation. Being completely reducible under $\sans i$,
$\sans a$ organizes into irreducible representations
of $\sans i$. We denote by $\sans a^{n,p}$ the eigenspace of
$c$, $k$ of eigenvalues ${1\over 4}(n^2-1)$, $p$, respectively.
It follows that $\sans a$ has a finer grading than the original one.

\proclaim Proposition 4.2.1. 
If $\sans a^{(r)}$, $r=1,2$, are two $N=2$ $\goth g$ superoperations,
then their graded tensor product $\sans a=\sans a^{(1)}\hat\otimes
\sans a^{(2)}$ is also an $N=2$ $\goth g$ superoperation.

{\it Proof}. Indeed $\sans a$ satisfies the conditions stated 
in def. 4.2.1. \hfill $\square$

Let $\sans a$ be an $N=2$ $ \goth g$ superoperation.

The graded derivations $d_A$ have degree $+1$ and satisfy $[d_A,d_B]=0$. So, 
one may define a double differential complex $(\sans a, d_A)$. We do not 
define cohomology in the usual way, as the standard definition would not 
be covariant with respect to $\sans i$.
Instead, we propose the following definition
generalizing that of ref. \ref{31}.
The ordinary cohomology $H^*(\sans a)$ is labelled by the values of the 
invariants $c$, $k$ of $\sans i$ and is defined as 
$$
H^{n,p}(\sans a)
=(\cap_{A=1,2}\ker d_A\cap \sans a^{n,p})
/\hbox{$1\over2$}\epsilon^{KL}d_Kd_L\sans a^{n,p-2},
\quad\quad (n,p)\in\Bbb N\times\Bbb Z.\vphantom{\bigg[}
\eqno(4.2.2)
$$
The basic subspace of $\sans a$ is defined as
$$
\sans a_{\rm basic}=\bigcap_{\xi\in\goth g}\ker j(\xi)\cap
\cap_{A=1,2}\ker j_A(\xi)\cap\ker l(\xi).
\eqno(4.2.3)
$$
Using (2.2.18d)--(2.2.18f), one can show that $\sans a_{\rm basic}$ 
is $d_A$ invariant. So, $(\sans a_{\rm basic},d_A)$ is also a double
differential complex. Its cohomology $H^*_{\rm basic}(\sans a)$
is defined
$$
H_{\rm basic}^{n,p}(\sans a)
=(\cap_{A=1,2}\ker d_A\cap \sans a_{\rm basic}^{n,p})
/\hbox{$1\over2$}\epsilon^{KL}d_Kd_L\sans a_{\rm basic}^{n,p-2},
\quad\quad (n,p)\in\Bbb N\times\Bbb Z, \vphantom{\bigg[}
\eqno(4.2.4)
$$
where $\sans a_{\rm basic}^{n,p}=\sans a^{n,p}\cap\sans a_{\rm basic}$,
and is the basic cohomology of the superoperation. 

The (basic) cohomology of any $N=2$ superoperation $\sans a$ is structurally
restricted, as indicated by the following.

\proclaim Proposition 4.2.2. One has
$$
H^{ n,p}(\sans a)=0,
\quad\quad\hbox{for $p\not=\pm n+1$}. \vphantom{\bigg[}
\eqno(4.2.5)
$$
Similarly,
$$
H_{\rm basic}^{ n,p}(\sans a)=0,
\quad\quad\hbox{for $p\not=\pm n+1$}. \vphantom{\bigg[}
\eqno(4.2.6)
$$

{\it Proof}. It is convenient for the time being to revert to the original
basis $h_A$, $h_{B,C}$, $\tilde h_D$ of $\sans t$, which allows for a more 
compact notation. Let $x\in\sans a$ such that $\tilde h_Ax=0$. Using 
(2.2.2b), (2.2.2c), it is easy to show that 
$$
[h_A+\epsilon^{KL}h_Kh_{L,A}]x
-\tilde h_A\hbox{$1\over 2$}\epsilon^{KL}h_Kh_Lx=0. \vphantom{\bigg[}
\eqno(4.2.7)
$$
Apply now $\tilde h_B$ to the left hand side of this equation and 
contract with $\epsilon^{BA}$. After a short calculation exploiting 
(2.2.2c), (2.2.2e), one gets
$$
\Big[-\hbox{$1\over 2$}\epsilon^{KL}\epsilon^{MN}h_{K,M}h_{L,N}
+\hbox{$1\over 2$}\epsilon^{KL}h_{K,L}\Big]x
-\hbox{$1\over 2$}\epsilon^{KL}\tilde h_K\tilde h_L
\hbox{$1\over 2$}\epsilon^{MN}h_Mh_Nx
=0.\vphantom{\bigg[}
\eqno(4.2.8)
$$
Using the relation $h_{A,B}={1\over 2}(t_{AB}-\epsilon_{AB}k)$,
following from (2.2.14a), (2.2.14b), and (2.2.14c), (2.2.14d) 
and (4.2.1) in (4.2.8), one gets finally
$$
\Big[c+\hbox{$1\over 4$}(1-(k-1)^2)\Big]x
-\hbox{$1\over 2$}\epsilon^{KL}d_Kd_L\hbox{$1\over 2$}\epsilon^{MN}u_Mu_Nx
=0.\vphantom{\bigg[}
\eqno(4.2.9)
$$
If $x\in\sans a^{n,p}$, (4.2.9)  yields
$$
\hbox{$1\over 4$}[n^2-(p-1)^2]x
-\hbox{$1\over 2$}\epsilon^{KL}d_Kd_L\hbox{$1\over 2$}\epsilon^{MN}u_Mu_Nx
=0.\vphantom{\bigg[}
\eqno(4.2.10)
$$
(4.2.10) yields (4.2.5) immediately. (4.2.6) follows also from (4.2.10)
upon checking that for $x\in\sans a_{\rm basic}$, 
${1\over 2}\epsilon^{MN}h_Mh_Nx\in\sans a_{\rm basic}$ as well, 
by (2.2.9a)--(2.2.9c). \hfill $\square$

\proclaim Proposition 4.2.3. The non trivial elements of $H^{n,p}(\sans a)$ 
($H^{n,p}_{\rm basic}(\sans a)$) fill irreducible representations of the 
internal symmetry algebra $\sans i$ of invariants $n,~p$. 

{\it Proof}. By (2.2.17c), (2.2.17d), if $x\in\sans a^{n,p}\cap\cap_{A=1,2}
\ker d_A$, then $t_{AB}x,~kx\in\sans a^{n,p}\cap\cap_{A=1,2}\ker d_A$ 
as well. Further, if $x\in{1\over 2}\epsilon^{KL}d_Kd_L\sans a^{n,p-2}$, 
$t_{AB}x,~kx \in{1\over 2}\epsilon^{KL}d_Kd_L\sans a^{n,p-2}$, also. 
One thus defines $t_{AB}[x]=[t_{AB}x]$ and $k[x]=[kx]$, for
any $[x]\in H^{n,p}(\sans a)$ with arbitrary representative $x\in\sans a^{n,p}
\cap\cap_{A=1,2}\ker d_A$. This yields the first part of the proposition.
The statement extends to basic cohomology, 
by noting that $t_{AB}x,~kx\in\sans a^{n,p}_{\rm basic}$ whenever
$x\in\sans a^{n,p}_{\rm basic}$, by (2.2.17e)--(2.2.17j). \hfill $\square$

Recall that the only irreducible $n$ dimensional module of $\sans i=\goth s
\goth l(2,\Bbb R)\oplus\Bbb R$ is the completely symmetric tensor space
$\bigvee^{n-1}\Bbb R^2$ up to equivalence. 
Hence, one has a tensor factorization of the form
$$
H^{n,p}(\sans a)
=K^{n,p}\otimes\hbox{$\bigvee^{n-1}$}\Bbb R^2,
\eqno(4.2.11)
$$
$$
H^{n,p}_{\rm basic}(\sans a)
=K^{n,p}_{\rm basic}\otimes\hbox{$\bigvee^{n-1}$}\Bbb R^2, 
\eqno(4.2.12)
$$
for certain real vector spaces $K^{n,p}$, $K^{n,p}_{\rm basic}$.
\par\vskip .6cm
{\bf 5. The Weil superoperation and its cohomologies}         
\vskip .4cm

Let $\goth g$ be an ungraded real Lie algebra.

\vskip.2cm {\it 5.1. The $N=1$ case}

Let $\sans w$ be the $N=1$ Weil algebra of $\goth g$ (cfr. subsect. 3.1). 
Then, $\sans w$ is an $N=1$ $\goth g$ superoperation (cfr. def. 4.1.1) 
called $N=1$ Weil superoperation.
Indeed, as shown in subsect. 3.1, $\sans w$ is a $\Bbb Z$ graded real left 
module algebra of the fundamental $N=1$ superstructure $\sans t$ of 
$\goth g$, the action of $\sans t$ on $\sans w$ is derivative
and $\sans w$ is obviously completely reducible under the internal 
symmetry algebra $\sans i$ with $k$ acting as the degree operator of 
$\sans w$ by (3.1.10a), (3.1.10b).

\proclaim Theorem 5.1.1. $H^p(\sans w)=0$ for $p\not=0$ and
$$
H^0(\sans w)\simeq\Bbb R.
\eqno(5.1.1) 
$$
Similarly, $H_{\rm basic}^p(\sans w)=0$, for $p\not=2s$ with $s\geq 0$, 
and 
$$
H_{\rm basic}^{2s}(\sans w)\simeq\big(\hbox{$\bigvee^s$}
\goth g^\vee\big)_{\ad^\vee\goth g},
\quad s\geq 0,
\eqno(5.1.2) 
$$
where $\big(\bigvee^s\goth g^\vee\big)_{\ad^\vee\goth g}$ denotes the subspace
of symmetrized tensor product
$\bigvee^s\goth g^\vee$ spanned by the elements which are invariant 
under the coadjoint action of $\goth g$.

{\it Proof}. Below, we shall use the following notation. 
Let $r\in\bigwedge^*\goth g^\vee\otimes\bigvee^*\goth g^\vee$. Let 
$\xi\in\Pi\goth g$, $\eta\in\goth g$, where $\Pi\goth g$ is the Grassmann odd 
partner of $\goth g$. We denote by $r(\xi,\eta)$ the evaluation of $r$ on 
$\sum_{p,q\geq 0}\xi^{\otimes p}\otimes\eta^{\otimes q}$.
Every element $z\in\sans w$ is of the form 
$z=r(w,\tilde w)$ for some $r\in\bigwedge^*\goth g^\vee
\otimes\bigvee^*\goth g^\vee$ uniquely determined by $z$. 
As $\deg w=1$, $\deg\tilde w=2$, 
$\sans w^p=0$ for $p<0$ and $\sans w^0=\Bbb R1$. Hence, $H^p(\sans w)=0$ 
for $p<0$ and $H^0(\sans w)\simeq\Bbb R$, trivially. Let $\sans w^{p>0}=
\bigoplus_{p>0}\sans w^p$. $\sans w^{p>0}$ is acted upon by
the graded derivations $h$, $\tilde h$ and two more graded derivations $i^*$, 
$\tilde\imath^*$ of degree $-1$, $0$, respectively,  defined by
$$
\matrix{~&~\cr
i^*w=0,\hfill&i^*\tilde w=w,\hfill\cr
~&~\cr
\tilde\imath^*w=w,\hfill&\tilde\imath^*\tilde w=\tilde w.\hfill\cr
~&~\cr}
\eqno(5.1.3a)-(5.1.3d)
$$
Identify $i^*$, $\tilde\imath^*$ with the linear maps 
$i^*(x)=x i^*$, $\tilde\imath^*(x)=x\tilde\imath^*$, $x\in\Bbb R$, 
Then, $h$, $\tilde h$, $i^*$, $\tilde\imath^*$ satisfy relations 
(2.1.2a)--(2.1.2c), (2.1.5a)--(2.1.5c), (2.1.8a)--(2.1.8d) with 
$\goth g=\Bbb R$. It follows that $\sans w^{p>0}$ is an $N=1$ $\Bbb R$
superoperation. Switch now to the derivations $k$, $d$, $j^*$, $l^*$
defined by (2.1.12a), (2.1.12b), (2.1.13a), (2.1.13b). 
By (2.1.17a), $j^*$ is a homotopy operator for $d$, for $l^*$ commutes with 
$j^*$ and $d$, by (2.1.17b), (2.1.18b), and  $l^*$ is invertible 
on $\sans w^{p>0}$, by (5.1.3c), (5.1.3d) and the definition of 
$\sans w^{p>0}$. Thus, the cohomology of $d$ is trivial on 
$\sans w^{p>0}$. This proves the first part of the theorem. 
Every element $z\in\sans w_{\rm basic}$ is of the form 
$z=r(\phi)$ for some $r\in\big(\bigvee^*\goth g^\vee\big)_{\ad^\vee\goth g}$
uniquely determined by $z$. 
Indeed, $z=r(\omega,\phi)$ for a unique $r\in\bigwedge^*\goth g^\vee
\otimes\bigvee^*\goth g^\vee$, by an argument similar to that employed
earlier, and, by (3.1.12a)--(3.1.12d), the basicity conditions 
$j(\xi)r(\omega,\phi)=0$, $l(\xi)r(\omega,\phi)=0$ imply that $r$ 
has polynomial degree $0$ in the first argument and is $\ad^\vee\goth g$ 
invariant. It follows that $\sans w_{\rm basic}^p=0$ for $p\not=2s$ with
$s\geq 0$, as $\deg\phi=2$. So, $H_{\rm basic}^p(\sans w)=0$ for 
$p\not=2s$ with $s\geq 0$. Let $s\geq 0$. If $z=r(\phi)$ with 
$r\in\big(\bigvee^s\goth g^\vee\big)_{\ad^\vee\goth g}$, then $dz=0$, by 
(3.1.11b) and the $\ad^\vee\goth g$ invariance of $r$. Hence, 
$\sans w_{\rm basic}^{2s}\cap\ker d=\sans w_{\rm basic}^{2s}$.
We thus have a linear injection $\mu:\sans w_{\rm basic}^{2s}\cap\ker d
\mapsto\big(\bigvee^s\goth g^\vee\big)_{\ad^\vee\goth g}$ given by 
$z\mapsto r$. As, $\sans w_{\rm basic}^{2s-1}=0$, $\mu$ induces a linear
bijection $\hat\mu:H_{\rm basic}^{2s}(\sans w)\mapsto
\big(\bigvee^s\goth g^\vee\big)_{\ad^\vee\goth g}$. \hfill $\square$

\vskip.2cm {\it 5.2. The $N=2$ case}

Let $\sans w$ be the $N=2$ Weil algebra of $\goth g$ (cfr. subsect. 3.2). 
Then, $\sans w$ is an $N=2$ $\goth g$ superoperation (cfr. def. 4.2.1) 
called $N=2$ Weil superoperation.
Indeed, as shown in subsect. 3.2, $\sans w$ is a $\Bbb Z$ graded real left 
module algebra of the fundamental $N=2$ superstructure $\sans t$ of 
$\goth g$, the action of $\sans t$ on $\sans w$ is derivative and 
$\sans w$ is obviously completely reducible under the internal 
symmetry algebra $\sans i$ with $k$ acting as the degree operator of 
$\sans w$ by (3.2.10a)--(3.2.10h).

\proclaim Theorem 5.2.1. $H^{n,p}(\sans w)=0$, for $(n,p)\not=(1,0)$,
and 
$$
H^{1,0}(\sans w)\simeq\Bbb R.
\eqno(5.2.1) 
$$
Similarly, $H_{\rm basic}^{n,p}(\sans w)=0$, for $(n,p)\not=(1,0)$,
$(2s,2s+1)$ with $s>0$, and
$$
H_{\rm basic}^{1,0}(\sans w)\simeq\Bbb R, \quad
H_{\rm basic}^{2s,2s+1}(\sans w)\simeq\big(\hbox{$\bigvee^s$}
\goth g^\vee\big)_{\ad^\vee\goth g}
\otimes\hbox{$\bigvee^{2s-1}$}\Bbb R^2,\quad s>0.
\eqno(5.2.2a), (5.2.2b)
$$

{\it Proof}. Below, we shall use the following notation. 
Let $r\in\bigwedge^*(\goth g^\vee\otimes\bigotimes^a\Bbb R^2)
\otimes\bigvee^*(\goth g^\vee\otimes\bigotimes^b\Bbb R^2)$. 
Let $\xi\in\Pi\goth g\otimes\bigotimes^a\Bbb R^{2\vee}$,
$\eta\in\goth g\bigotimes^b\Bbb R^{2\vee}$. We denote by
$r(\xi,\eta)$ the evaluation of $r$ on $\sum_{p,q\geq 0}\xi^{\otimes p}
\otimes\eta^{\otimes q}$. The above notation can be straightforwardly
generalized to the case where there are several $\xi$ and $\eta$.
Every element $z\in\sans w$ is of the form
$z=r(w,w_,,\tilde w)$ for some $r\in\bigwedge^*(\goth g^\vee\otimes\Bbb R^2)
\otimes\bigvee^*(\goth g^\vee\otimes\bigotimes^2\Bbb R^2)\otimes
\bigwedge^*(\goth g^\vee\otimes\Bbb R^2)$ uniquely determined by $z$. 
As $\deg w_A=1$, $\deg w_{A,B}=2$,
$\deg\tilde w_A=3$, $\sans w^{n,p}=0$, for $p<0$, and $\sans w^{n,0}
=\Bbb R\delta_{n,1}1$. So, $H^{n,p}(\sans w)=0$, for $p<0$, and 
$H^{n,0}(\sans w)\simeq\delta_{n,1}\Bbb R$, trivially.
Let $\sans w^{p>0}=\bigoplus_{n\in\Bbb N,p>0}\sans w^{n,p}$. 
$\sans w^{p>0}$ is acted upon by the graded derivations $h_A$, $h_{A,B}$,
$\tilde h_A$ and three more graded derivations $i$, $i_{,A}$,
$\tilde\imath$ of degree $-2$, $-1$, $0$, respectively, defined by
$$
\matrix{~&~&\cr
i^*w_A=0,\hfill&i^*w_{A,B}=0,\hfill&i^*\tilde w_A=w_A,\hfill\cr
~&~&~\cr
i^*{}_{,A}w_B=0,\hfill&i^*{}_{,A}w_{B,C}=-\epsilon_{CA}w_B,\hfill
&i^*{}_{,A}\tilde w_B=w_{B,A},\hfill\cr
~&~&~\cr
\tilde\imath^*w_A=w_A,\hfill&\tilde\imath^* w_{A,B}= w_{A,B},
\hfill&\tilde\imath^*\tilde w_A=\tilde w_A.\hfill\cr
~&~&\cr}
\eqno(5.2.3a)-(5.2.3i)
$$
Identify $i^*$, $i^*{}_{,A}$, $\tilde\imath^*$ with the linear maps 
$i^*(x)=xi^*$, $i^*{}_{,A}(x)=xi^*{}_{,A}$, $\tilde\imath^*(x)=x\tilde
\imath^*$, $x\in\Bbb R$. Then, $h_A$, $h_{A,B}$, $\tilde h_A$, $i$, $i_{,A}$, 
$\tilde\imath$ satisfy relations (2.2.2a)--(2.2.2f), (2.2.6a)--(2.2.6f),
(2.2.9a)--(2.2.9i) with $\goth g=\Bbb R$. From this fact, it is easy to 
see that $\sans w^{p>0}$ is an $N=2$ $\Bbb R$ superoperation.
Switch now to the derivations $t_{A,B}$, $k$, $u_A$, $d_A$,
$j^*$, $j^*_A$, $l^*$ defined by (2.2.14a)--(2.2.14d),
(2.2.15a)--(2.2.15c).
By (2.2.19e), $j^*{}_A$ is a homotopy operator for $d_A$, 
for $l^*$ commutes with $j^*{}_A$ and $d_A$, by (2.2.19f), (2.2.20e),
and $l^*$ is invertible on $\sans w^{p>0}$, by (5.2.3g)--(5.2.3i)
and the definition of $\sans w^{p>0}$. Indeed, using (2.2.19e),  
(2.2.19f), (2.2.20e), one can show that
$$
\Big[\hbox{$1\over 2$}\epsilon^{KL}d_Kd_L,
\hbox{$1\over 2$}\epsilon^{MN}j^*{}_Mj^*{}_N\Big]
=-l^*(l^*+\epsilon^{KL}j^*{}_Kd_L),
\eqno(5.2.4) 
$$
where, by (2.2.17g), (2.2.17h), ${1\over 2}\epsilon^{MN}j^*{}_Mj^*{}_N$ 
maps $\sans w^{n,p}$ into $\sans w^{n,p-2}$. Thus, the cohomology 
of $d_A$ is trivial on $\sans w^{p>0}$. This proves the first part of the 
theorem. Let us examine next the second part.
As $\sans w^{n,p}=0$, for $p<0$, and $\sans w^{n,0}=\Bbb R\delta_{n,1}1$, 
as shown earlier, and $1$ is obviously basic, $\sans w_{\rm basic}^{n,p}=0$, 
for $p<0$, and $\sans w_{\rm basic}^{n,0}=\Bbb R\delta_{n,1}1$.
Consequently, $H_{\rm basic}^{n,p}(\sans w)=0$ for $p<0$. 
and $H_{\rm basic}^{n,0}(\sans w)\simeq\delta_{n,1}\Bbb R$. On the other 
hand, by prop. 4.2.2, eq. (4.2.6), $H_{\rm basic}^{n,p}(\sans w)=0$ 
for $p\not=\pm n+1$. So, the only potentially non vanishing cohomology 
spaces which are left are $H_{\rm basic}^{n,n+1}(\sans w)$, $n\geq 1$, 
which we shall analyze next.
Every element $z\in\sans w_{\rm basic}$ is of the form
$z=r(\phi,\rho)$ for some $r\in\big(\bigvee^*(\goth g^\vee\otimes\bigvee^2
\Bbb R^2)\otimes\bigwedge^*(\goth g^\vee\otimes\Bbb R^2)
\big)_{\ad^\vee\goth g}$ uniquely determined by $z$. 
Indeed, $z=r(\omega,\gamma,\phi,\rho)$, 
for a unique $r\in\bigwedge^*(\goth g^\vee\otimes\Bbb R^2)\otimes
\bigvee^*\goth g^\vee\otimes\bigvee^*(\goth g^\vee
\otimes\bigvee^2\Bbb R^2)\otimes\bigwedge^*(\goth g^\vee\otimes\Bbb R^2)$, 
by an argument similar to that employed earlier in the proof, and,
by (3.2.12a)--(3.2.12l), the basicity conditions 
$j(\xi)r(\omega,\gamma,\phi,\rho)=0$, $j_A(\xi)r(\omega,\gamma,\phi,\rho)=0$,
$l(\xi)r(\omega,\gamma,\phi,\rho)=0$ imply that $r$ has polynomial degree 
$0$ in the first two arguments and is $\ad\goth g$ invariant.
Let $z=r(\phi,\rho)\in\sans w_{\rm basic}^{n,n+1}$.
From (3.2.10c), (3.2.10d), (3.2.10g), (3.2.10h) and 
the representation theory of $\sans i=\goth s\goth l(2,\Bbb R)
\oplus\Bbb R$, one knows that the total number of internal indices $A=1,2$ 
and the total degree carried by $\phi_{AB}$, $\rho_A$ in each monomial of
$r(\phi,\rho)$ must be $n-1+2\nu$ and $n+1$, respectively, where $2\nu$ is
the number of indices contracted by means of $\epsilon^{AB}$. 
Further, the $n-1$ uncontracted indices are totally symmetrized.
So, the numbers $m_\phi$, $m_\rho$ of occurences of $\phi_{AB}$, $\rho_A$ 
in a given monomial must satisfy the equations
$2m_\phi+1m_\rho=n-1+2\nu$, $2m_\phi+3m_\rho=n+1$.
Taking into account that $m_\phi$, $m_\rho$ are non negative integers,
one finds that $\nu=0$, $m_\phi=s-1$, $m_\rho=1$, for $n=2s$ with $s\geq 1$,
and $\nu=1$, $m_\phi=s$, $m_\rho=0$, for $n=2s-1$ with $s\geq 2$. 
Thus, the most general $z\in\sans w_{\rm basic}^{n,n+1}$ is of the form
$$
\eqalignno{
z&=u^{A_1\cdots A_{2s-1}}(\phi_{A_1A_s},\cdots,\phi_{A_{s-1}A_{2s-2}},
\rho_{A_{2s-1}}),\quad n=2s, ~s\geq 1,&(5.2.5a)\cr
z&=\hbox{$1\over 2$}\epsilon^{KL}
v^{A_1\cdots A_{2s-2}}(\phi_{A_1A_{s-1}},\cdots,\phi_{A_{s-2}A_{2s-4}}
\phi_{A_{2s-3}K}\phi_{A_{2s-2}L}),
\quad n=2s-1,~s\geq 2,&\cr
&&(5.2.5b)\cr}
$$
where $u^{A_1\cdots A_{2s-1}}\in\big(\bigvee^{s-1}\goth g^\vee\otimes
\goth g^\vee\big)_{\ad^\vee\goth g}$ totally symmetric in $A_1,\cdots,
A_{2s-1}$, $v^{A_1\cdots A_{2s-2}}\in\big(\bigvee^{s-2}\goth g^\vee\otimes
\bigwedge^2\goth g^\vee\big)_{\ad^\vee\goth g}$ totally symmetric in $A_1,
\cdots,A_{2s-2}$. Suppose now that $z\in\sans w_{\rm basic}^{n,n+1}
\cap\cap_{A=1,2}\ker d_A$ so that $z$, besides being of the form (5.2.5a), 
(5.2.5b), satisfies $d_Az=0$. Suppose first that $n=2s$. 
Using (5.2.5a), (3.2.11f), (3.2.11h), the symmetry properties 
and the $\ad\goth g$ invariance of 
$u^{A_1\cdots A_{2s-1}}$ and taking into account that terms with a 
different number of occurences of $\phi_{AB}$, $\rho_A$ are linearly 
independent, the condition $d_Az=0$ is equivalent to the equations 
$$
\eqalignno{
&\epsilon_{AA_{s-1}}
u^{A_1\cdots A_{2s-1}}(\phi_{A_1A_s},\cdots,\phi_{A_{s-2}A_{2s-3}},
\rho_{A_{2s-2}},\rho_{A_{2s-1}})=0,
&(5.2.6a)\cr
&\hbox{$1\over 2$}\epsilon^{KL}
u^{A_1\cdots A_{2s-1}}(\phi_{A_1A_s},\cdots,\phi_{A_{s-1}A_{2s-2}},
[\phi_{KA_{2s-1}},\phi_{LA}])=0.&(5.2.6b)\cr}
$$
As $\rho_A$ is odd and $u^{A_1\cdots A_{2s-1}}$ is totally symmetric in
$A_1,\cdots,A_{2s-1}$, (5.2.6a) entails that $u^{A_1\cdots A_{2s-1}}$ is 
totally symmetric in its $s$ arguments. Using this fact and the $\ad\goth g$ 
invariance of $u^{A_1\cdots A_{2s-1}}$, it is easy to see that (5.2.6b) is 
identically satisfied. Hence, $u^{A_1\cdots A_{2s-1}}\in
\big(\bigvee^s\goth g^\vee\big)_{\ad^\vee\goth g}$. Conversely, 
if this holds, then (5.2.6a), (5.2.6b) are fulfilled. The above analysis
shows that $\sans w_{\rm basic}^{2s,2s+1}\cap\cap_{A=1,2}\ker d_A$ is 
precisely the space of the $z$ of the form (5.2.5a) with 
$u^{A_1\cdots A_{2s-1}}\in\big(\bigvee^s\goth g^\vee\big)_{\ad^\vee\goth g}$ 
totally symmetric in $A_1,\cdots,A_{2s-1}$. Thus, we have a linear bijection
$\mu:\sans w_{\rm basic}^{2s,2s+1}\cap\cap_{A=1,2}\ker d_A\mapsto
\big(\bigvee^s\goth g^\vee\big)_{\ad^\vee\goth g}
\otimes\bigvee^{2s-1}\Bbb R^2$ defined by $z\mapsto
(u^{A_1\cdots A_{2s-1}})_{A_1,\cdots,A_{2s-1}=1,2}$. 
We note next that $\sans w_{\rm basic}^{2s,2s-1}=0$. 
Indeed, if $z=r(\phi,\rho)\in\sans w_{\rm basic}^{n,n-1}$,
the total number of internal indices $A=1,2$ and the total 
degree carried by $\phi_{AB}$, $\rho_A$ in each monomial of
$r(\phi,\rho)$ must be $n-1+2\nu$ and $n-1$, respectively, where $2\nu$ is
the number of indices contracted by means of $\epsilon^{AB}$. So, 
the numbers $m_\phi$, $m_\rho$ of occurences of $\phi_{AB}$, $\rho_A$ 
in a given monomial must satisfy the equations
$2m_\phi+1m_\rho=n-1+2\nu$, $2m_\phi+3m_\rho=n-1$.
Taking into account that $m_\phi$, $m_\rho$ are non negative integers,
one finds that there are no solutions for $n=2s$ with $s>0$, so
that $\sans w_{\rm basic}^{2s,2s-1}=0$ as announced.
Thus, the bijection $\mu$ above induces a bijection 
$\hat\mu:H_{\rm basic}^{2s,2s+1}(\sans w)\mapsto
\big(\bigvee^s\goth g^\vee\big)_{\ad^\vee\goth g}
\otimes\bigvee^{2s-1}\Bbb R^2$.
Suppose next that $n=2s-1$. Using (5.2.5b), (3.2.11f), (3.2.11h),
the (anti)symmetry properties and the $\ad\goth g$ invariance of 
$v^{A_1\cdots A_{2s-2}}$, the condition $d_Az=0$ is equivalent to the 
equation
$$
\eqalignno{
&(s-2)\epsilon^{KL}\epsilon_{AA_{s-2}}v^{A_1\cdots A_{2s-2}}
(\phi_{A_1A_{s-1}},\cdots,\phi_{A_{s-3}A_{2s-5}},\rho_{A_{2s-4}},
\phi_{A_{2s-3}K},\phi_{A_{2s-2}L})\vphantom{\big[}&\cr
&+v^{A_1\cdots A_{2s-2}}(\phi_{A_1A_{s-1}},\cdots,\phi_{A_{s-2}A_{2s-4}},
\rho_{A_{2s-3}},\phi_{A_{2s-2}A})\vphantom{\big[}&\cr
&+\epsilon^{KL}\epsilon_{AA_{2s-3}}
v^{A_1\cdots A_{2s-2}}(\phi_{A_1A_{s-1}},\cdots,\phi_{A_{s-2}A_{2s-4}},
\rho_K,\phi_{A_{2s-2}L})=0.\vphantom{\big[}&(5.2.7)\cr}
$$
Now, apply $u_B$ to this relation, using (3.2.11b), (3.2.11d), and then
contract with $\epsilon^{BA}$. One gets then ${1\over 2}\epsilon^{KL}
v^{A_1\cdots A_{2s-2}}(\phi_{A_1A_{s-1}},\cdots,\phi_{A_{s-2}A_{2s-4}}
\phi_{A_{2s-3}K}\phi_{A_{2s-2}L})=0$. So, $z=0$. We conclude that
$\sans w_{\rm basic}^{2s-1,2s}\cap\cap_{A=1,2}\ker d_A=0$.
Thus, $H_{\rm basic}^{2s-1,2s}(\sans w)=0$ as well. \hfill $\square$

\vskip.2cm {\it 5.3. The relation between the cohomologies of the 
$N=1$ and $N=2$ Weil superoperations}

Let $\sans w(n)$ denote the $N=n$ Weil superoperations, $n=1,~2$.

\proclaim Corollary 7.3.1. One has
$$
H^{n,\pm n+1}(\sans w(2))\simeq 
H^{\pm(n-{1\over 2})+{1\over 2}}(\sans w(1))\otimes
\hbox{$\bigvee^{n-1}$}\Bbb R^2,
\eqno(5.3.1)
$$
$$
H_{\rm basic}^{n,\pm n+1}(\sans w(2))\simeq 
H_{\rm basic}^{\pm(n-{1\over 2})+{1\over 2}}(\sans w(1))\otimes
\hbox{$\bigvee^{n-1}$}\Bbb R^2.
\eqno(5.3.2)
$$

{\it Proof}. Combine props. 5.1.1, 5.2.1. \hfill $\square$

Thus, the $N=1$ and $N=2$ cohomologies of $\sans w$ are intimately related.
\par\vskip .6cm
{\bf 6. Connections, equivariant cohomology and Weil homomorphism}         
\vskip .4cm

Let $\goth g$ be an ungraded real Lie algebra.

\vskip.2cm {\it 6.1. The $N=1$ case}

Let $\sans a$ be an $N=1$ $\goth g$ superoperation with unity, i. e. 
$\sans a$ as an algebra has a unity $1$.

\proclaim Definition 6.1.1. A connection $a$ on $\sans a$ is an element of 
$\sans a\otimes\goth g$ satisfying relations (3.1.10a), (3.1.12a), (3.1.12c) 
with $\omega$ substituted by $a$.

The curvature of $a$ is defined as usual as
$$
f=da+\hbox{$1\over 2$}[a,a].
\eqno(6.1.1)
$$
It is easy to see that $f$ satisfies relations (3.1.10b), 
(3.1.12b), (3.1.12d) with $\phi$ substituted by $f$.
In particular, being $j(\xi)f=0$ for any $\xi\in\goth g$, 
$f$ is horizontal. $a$, $f$ together fulfill (3.1.11a), (3.1.11b).

We denote by $\Conn(\sans a)$ the set of the connections of 
the $N=1$ $\goth g$ superoperation $\sans a$. $\Conn(\sans a)$ is an 
affine space modelled on $\sans a^1\otimes\goth g$.

\proclaim Proposition 6.1.1. Let $r\in\bigwedge^*\goth g^\vee
\otimes\bigvee^*\goth g^\vee$ be such that, for any connection 
$a\in\Conn(\sans a)$, $r(a,f)$ is a representative of some element of 
$H^p_{\rm basic}(\sans a)$ (see above eq. (5.1.3a) for the definition of 
the notation). Then, the basic cohomology class
$[r(a,f)]$ is independent from the choice of $a$.

{\it Proof}. We follow the methods of ref. \ref{32}. 
Consider the $N=1$ superoperation $\sans s$ 
generated by $s$, $\tilde s$ of degree $0$, $+1$, respectively, with
$$
\matrix{~&~\cr
h^ss=0, \hfill&
h^s\tilde s=\tilde s, \hfill\cr
~&~\cr
\tilde h^ss=-\tilde s, \hfill&
\tilde h^s\tilde s=0,\hfill\cr
~&~\cr}
\eqno(6.1.2a)-(6.1.2d)
$$
$$
i^s(\xi)=0, \quad\tilde\imath^s(\xi)=0,
\quad\quad\xi\in\goth g.
\eqno(6.1.3a),(6.1.3b)
$$
Next, we consider the graded tensor product superoperation
$\sans s\hat\otimes\sans a$ and the subalgebra $\sans c$ of 
$\sans s\hat\otimes\sans a$ generated by the elements of the form 
$a(s)$, $\tilde h^sa(s)$, $\tilde a(s)$, $\tilde h^s\tilde a(s)$,  
where $a:\Bbb R\mapsto\sans a\otimes\goth g$ is a polynomial such that, 
for fixed $\sigma\in\Bbb R$, $a(\sigma)$ is a connection on $\sans a$
and $\tilde a(\sigma)=-\tilde h a(\sigma)$. Next, we define a degree $0$ 
derivation $q$ on $\sans c$ by
$$
\matrix{~&~\cr
qa(s)=0, \hfill&
q\tilde a(s)=-\tilde h^sa(s), \hfill\cr
~&~\cr
q\tilde h^sa(s)=0, \hfill& 
q\tilde h^s\tilde a(s)=0. \hfill\cr
~&~\cr}
\eqno(6.1.4a)-(6.1.4d)
$$
Note that, for fixed $\sigma\in\Bbb R$, $a(\sigma)$, $\tilde a(\sigma)$
satisfy relations (3.1.3a)--(3.1.3d), (3.1.4a)--(3.1.4d) with $w$, $\tilde w$
replaced by $a(\sigma)$, $\tilde a(\sigma)$. Using this fact, one easily
checks that 
$$
[q,\tilde h]=\tilde h^s, \quad [q,\tilde h^s]=0,
\eqno(6.1.5a),(6.1.5b)
$$
$$
[q,h+h^s]=0,
\eqno(6.1.6)
$$
$$
[q,i(\xi)]=0,\quad [q,\tilde\imath(\xi)]=0,
\quad\quad\xi\in\goth g.
\eqno(6.1.7a),(6.1.7b)
$$
Let $r\in\bigwedge^*\goth g^\vee\otimes\bigvee^*\goth g^\vee$ be such 
that, for any connection $a$ on $\sans a$, $r[a]:=r(a,\tilde a)$ 
belongs to $\sans a_{\rm basic}\cap\ker\tilde h$. By (6.1.5a) and 
the fact that $\tilde hr[a]=0$, 
$$
\tilde h^sr[a(s)]=-\tilde hqr[a(s)].
\eqno(6.1.8)
$$
We note that, by (6.1.2c), (6.1.2d), (6.1.4a), (6.1.4b), $qr[a(s)]$ is 
necessarily of the form $qr[a(s)]=\tilde s\alpha(s|a)$, where
$\alpha(s|a)$ is a polynomial in $s$. From this expression and (6.1.2a), 
(6.1.2b), it follows that $h^sqr[a(s)]=qr[a(s)]$. By (6.1.6), one has then 
$$
hqr[a(s)]=q(h-1)r[a(s)].
\eqno(6.1.9)
$$
Further, from (6.1.7a), (6.1.7b) and the fact that 
$i(\xi)r[a]=0$, $\tilde\imath(\xi)r[a]=0$,
$$
i(\xi)qr[a(s)]=0,\quad
\tilde\imath(\xi)qr[a(s)]=0,\quad\quad\xi\in\goth g.
\eqno(6.1.10a),(6.1.10b)
$$
For any element $x$ of $\sans s\hat\otimes\sans a$ of the form 
$x=\tilde s\alpha(s)$ with $\alpha(s)$ a polynomials in $s$, we define 
$\int_{[0,1]}x=\int_0^1\alpha(\sigma)d\sigma$, where the right hand side is 
an ordinary Riemann integral. It is obvious that,
for any element of $f(s)$ of $\sans s\hat\otimes\sans a$
polynomial in $s$, $\tilde h^sf(s)$ is of the above form and 
$-\int_{[0,1]}\tilde h^sf(s)=f(1)-f(0)$. From (6.1.8), one has thus 
$$
r[a(1)]-r[a(0)]
=\tilde h\int_{[0,1]}qr[a(s)].
\eqno(6.1.11)
$$
By (2.1.12b), the right hand side of (6.1.11) belongs to $d\sans a$.
From (2.2.12a), (6.1.9), (6.1.10a), (6.1.10b), if $r[a]$ belongs to 
$\sans a^p_{\rm basic}$ for any connection $a$ on $\sans a$,
then $qr[a(\sigma)]$ belongs to $\sans a^{p-1}_{\rm basic}$ for $\sigma\in
\Bbb R$, so that $\int_{[0,1]}qr[a(s)]$ belongs to 
$\sans a^{p-1}_{\rm basic}$, too. \hfill $\square$

Consider the $N=1$ Weil $\goth g$ superoperation $\sans w$ (cfr. subsect.
5.1). Then $\omega$ is a connection on $\sans w$ with 
curvature $\phi$ (cfr. eq. (3.1.9)).

Given an $N=1$ $\goth g$ superoperation $\sans a$ with unity, one can 
define the graded tensor product $N=1$ $\goth g$ superoperation 
$\sans w\hat\otimes\sans a$ (cfr. subsect. 4.1). The latter is the equivariant 
$N=1$ superoperation associated to $\sans a$. The equivariant cohomology of 
$\sans a$ is, by definition the basic cohomology of 
$\sans w\hat\otimes\sans a$:
$$
H^p_{\rm equiv}(\sans a)=H^p_{\rm basic}(\sans w\hat\otimes\sans a),
\quad\quad p\in\Bbb Z.
\eqno(6.1.12)
$$

An equivariant cohomology class of $\sans a$ is represented by elements
of $\sans w\hat\otimes\sans a$ of the form $r(\omega,\phi)$, where
$r\in\bigwedge^*\goth g^\vee\otimes\bigvee^*\goth g^\vee\otimes\sans a$. 
The Weil generator $\omega$ constitutes a connection of 
$\sans w\hat\otimes\sans a$. If $a$ is a connection of 
$\sans a$, $a$ is a connection of $\sans w\hat\otimes\sans a$ as well. 
By prop. 6.1.1, $r(\omega,\phi)$ is equivalent to $r(a,f)$ in equivariant
cohomology. On the other hand, $r(a,f)$ is a representative of a basic 
cohomology class of $\sans a$, which, by prop. 6.1.1, is independent 
from $a$ in basic cohomology. Thus, there is a natural homomorphism 
of $H^*_{\rm equiv}(\sans a)$ into $H^p_{\rm basic}(\sans a)$, 
called $N=1$ Weil homomorphism.

\vskip.2cm {\it 6.2. The $N=2$ case}

Let $\sans a$ be an $N=2$ $\goth g$ superoperation with unity.

\proclaim Definition 6.2.1. A connection $(a_A)_{A=1,2}$, on $\sans a$ 
is a doublet of $\sans a\otimes\goth g$ satisfying relations 
(3.2.10a), (3.2.10b), (3.2.11a), (3.2.12a), (3.2.12e), (3.2.12i) with 
$\omega_A$ substituted by $a_A$.

The derived connection
$$
b=\hbox{$1\over 2$}\epsilon^{KL}d_Ka_L
\eqno(6.2.1)
$$
and the curvature and derived curvature
$$
\matrix{~\cr
f_{AB}=\hbox{$1\over 2$}(d_Aa_B+d_Ba_A+[a_A,a_B]),\hfill\cr
~\cr
g_A=-\hbox{$1\over 4$}\epsilon^{KL}d_Kd_La_A
-\hbox{$1\over 2$}\epsilon^{KL}[a_K,d_La_A]
-\hbox{$1\over 6$}\epsilon^{KL}[a_K,[a_L,a_A]]\hfill\cr
~\cr}
\eqno(6.2.2a),(6.2.2b)
$$
satisfy relations (3.2.10c)--(3.2.10h),  
(3.2.12b)--(3.2.12d), (3.2.12f)--(3.2.12h),
(3.2.12j)--(3. 2.12l) with $\gamma$, $\phi_{AB}$, $\rho_A$
substituted by $b$, $f_{AB}$, $g_A$, respectively.
In particular, being $j(\xi)f_{AB}=0$, $j_A(\xi)f_{BC}=0$, 
$j(\xi)g_A=0$, $j_A(\xi)g_B=0$ for any $\xi\in\goth g$, $f_{AB}$, 
$g_A$ are horizontal. $a_A$, $b$, $f_{AB}$, $g_A$ together satisfy 
(3.2.11b)--(3.2.11h).

We denote by $\Conn(\sans a)$ the set of the connections of 
the $N=2$ superoperation $\sans a$.
$\Conn(\sans a)$ is an affine space modelled on $\sans a^{2,1}\otimes\goth g$.

\proclaim Proposition 6.2.1. Let $r\in\bigwedge^*(\goth g^\vee\otimes
\Bbb R^2)\otimes\bigvee^*\goth g^\vee\otimes\bigvee^*(\goth g^\vee\otimes
\bigvee^2\Bbb R^2)\otimes\bigwedge^*(\goth g^\vee\otimes\Bbb R^2)$ 
be such that, for any connection $(a_A)_{A=1,2}\in\Conn(\sans a)$, 
$r(a,b,f,g)$ is a representative of some element of 
$H^{n,p}_{\rm basic}(\sans a)$ (see above eq. (5.2.3a) for the definition 
of the notation). Then, the basic cohomology class 
$[r(a,b,f,g)]$ is independent from the choice of $(a_A)_{A=1,2}$.

{\it Proof}. We generalize the methods of ref. \ref{32}.
Consider the $N=2$ superoperation $\sans s$ 
generated by $s$, $s_{,A}$, $\tilde s$ of degree $0$, $+1$, $+2$, 
respectively, with
$$
\matrix{~&~\cr
h^s{}_As=0, \hfill&
h^s{}_As_{,B}=0, \hfill\cr
~&~\cr
h^s{}_A\tilde s=-s_{,A}, \hfill&
h^s{}_{A,B}s=0, \hfill\cr
~&~\cr
h^s{}_{A,C}s_{,B}=-\epsilon_{BC}s_{,A}, \hfill& 
h^s{}_{A,B}\tilde s=-\epsilon_{AB}\tilde s, \hfill\cr
~&~\cr
\tilde h^s{}_As=-s_{,A}, \hfill&
\tilde h^s{}_As_{,B}=\epsilon_{AB}\tilde s,\hfill\cr
~&~\cr
\tilde h^s{}_A\tilde s=0,\hfill &
~\hfill\cr
~&~\cr}
\eqno(6.2.3a)-(6.2.3i)
$$
$$
i^s(\xi)=0, \quad i^s{}_{,A}(\xi)=0, \quad\tilde\imath^s(\xi)=0,
\quad\quad\xi\in\goth g.
\eqno(6.2.4a)-(6.2.4c)
$$
Next, we consider the graded tensor product superoperation
$\sans s\hat\otimes\sans a$ and the subalgebra $\sans c$ of 
$\sans s\hat\otimes\sans a$ generated by the elements of the form 
$a_A(s)$, $\tilde h^s{}_Aa_B(s)$, $\tilde h^s{}_A\tilde h^s{}_Ba_C(s)$, 
$a_{A,B}(s)$, $\tilde h^s{}_Aa_{B,C}(s)$, 
$\tilde h^s{}_A\tilde h^s{}_Ba_{C,D}(s)$, 
$\tilde a_A(s)$, $\tilde h^s{}_A\tilde a_B(s)$, 
$\tilde h^s{}_A\tilde h^s{}_B\tilde a_C(s)$, 
where $a_A:\Bbb R\mapsto\sans a\otimes\goth g$, $A=1,2$, is a
polynomial such that, for fixed $\sigma\in\Bbb R$, $a_A(\sigma)$ is a 
connection of $\sans a$ and $a_{A,B}(\sigma)=-\tilde h_Ba_A(\sigma)$,
$\tilde a_A(\sigma)={1\over 2}\epsilon^{KL}\tilde h_K\tilde h_La_A(\sigma)$.
Next, we define a degree $0$ derivation $q$ on $\sans c$ by
$$
\matrix{~&~\cr
qa_A(s)=0, \hfill&
qa_{A,B}(s)=-\tilde h^s{}_Ba_A(s), \hfill\cr
~&~\cr
q\tilde a_A(s)=-\epsilon^{KL}\tilde h^s{}_Ka_{A,L}(s), \hfill&
q\tilde h^s{}_Aa_B(s)=0, \hfill\cr
~&~\cr
q\tilde h^s{}_Aa_{B,C}(s)=\epsilon_{AC}{1\over 2}\epsilon^{KL}
\tilde h^s{}_K\tilde h^s{}_La_B(s), \hfill& 
q\tilde h^s{}_A\tilde a_B(s)
={1\over 2}\epsilon^{KL}\tilde h^s{}_K\tilde h^s{}_La_{B,A}(s),\hfill\cr
~&~\cr
q{1\over 2}\epsilon^{KL}\tilde h^s{}_K\tilde h^s{}_La_A(s)=0, \hfill& 
q{1\over 2}\epsilon^{KL}\tilde h^s{}_K\tilde h^s{}_La_{A,B}(s)=0, \hfill\cr
~&~\cr
q{1\over 2}\epsilon^{KL}\tilde h^s{}_K\tilde h^s{}_L\tilde a_A(s)=0. \hfill&
~\hfill\cr
~&~\cr}
\eqno(6.2.5a)-(6.2.5i)
$$
Note that, for fixed $\sigma\in\Bbb R$, $a(\sigma)$, $a_A(\sigma)$,
$\tilde a(\sigma)$ satisfy relations (3.2.3a)--(3.2.3i), (3.2.4a)--(3.2.4i) 
with $w$, $w_A$, $\tilde w$ replaced by $a(\sigma)$, $a_A(\sigma)$,
$\tilde a(\sigma)$. Using this fact, one easily checks that
$$
[q,\tilde h_A]=\tilde h^s{}_A, \quad
[q,\tilde h^s{}_A]=0,
\eqno(6.2.6a),(6.2.6b)
$$
$$
[q,h_{A,B}+h^s{}_{A,B}]=0,
\eqno(6.2.7)
$$
$$
[q,i(\xi)]=0,\quad [q,i_{,A}(\xi)]=0,\quad [q,\tilde\imath(\xi)]=0,
\quad\quad\xi\in\goth g.
\eqno(6.2.8a)-(6.2.8c)
$$
Using (6.2.6a), it is easy to show that
$$
\big[q,\hbox{$1\over 2$}\epsilon^{KL}\tilde h_K\tilde h_L\big]=
\epsilon^{KL}\tilde h^s{}_K\tilde h_L, \quad
\hbox{$1\over 2$}\big[q,\big[q,
\hbox{$1\over 2$}\epsilon^{KL}\tilde h_K\tilde h_L\big]\big]=
\hbox{$1\over 2$}\epsilon^{KL}\tilde h^s{}_K\tilde h^s{}_L.
\eqno(6.2.9a),(6.2.9b)
$$
Let $r\in\bigwedge^*(\goth g^\vee\otimes\Bbb R^2)
\otimes\bigvee^*(\goth g^\vee\otimes\bigotimes^2\Bbb R^2)\otimes
\bigwedge^*(\goth g^\vee\otimes\Bbb R^2)$ be such that, for any connection 
$a_A$, $A=1,2$, on $\sans a$, $r[a]:=r(a,a_,,\tilde a)$ belongs to
$\sans a_{\rm basic}\cap\cap_{A=1,2}\ker\tilde h_A$.
Using (6.2.9a), (6.2.9b) and the fact that $\tilde h_Ar[a]=0$, 
it is easy to see that
$$
\hbox{$1\over 2$}\epsilon^{KL}\tilde h^s{}_K\tilde h^s{}_Lr[a(s)]
=\hbox{$1\over 2$}\epsilon^{KL}\tilde h_K\tilde h_L
\hbox{$1\over 2$}q^2r[a(s)].
\eqno(6.2.10)
$$
We note that, by (6.2.3g)--(6.2.3i), (6.2.5a)--(6.2.5f), 
${1\over 2}q^2r[a(s)]$ is necessarily of the form 
${1\over 2}q^2r[a(s)]=\tilde s\alpha(s|a)+
{1\over 2}\epsilon^{KL}s_{,K}s_{,L}\beta(s|a)$, where $\alpha(s|a)$,
$\beta(s|a)$ are polynomials in $s$. From
this expression and (6.2.3d)--(6.2.3f), it follows that 
$h^s{}_{A,B}{1\over 2}q^2r[a(s)]=-\epsilon_{AB}{1\over 2}q^2r[a(s)]$. 
By (6.2.7), one has then 
$$
h_{A,B}\hbox{$1\over 2$}q^2r[a(s)]
=\hbox{$1\over 2$}q^2\big(h_{A,B}+\epsilon_{AB}\big)r[a(s)].
\eqno(6.2.11)
$$
Further, from (6.2.8a)--(6.2.8c) and the fact that
$i(\xi)r[a]=0$, $i_{,A}(\xi)r[a]=0$, $\tilde\imath(\xi)r[a]=0$, 
$$
i(\xi)\hbox{$1\over 2$}q^2r[a(s)]=0,\quad
i_{,A}(\xi)\hbox{$1\over 2$}q^2r[a(s)]=0,\quad
\tilde\imath(\xi)\hbox{$1\over 2$}q^2r[a(s)]=0,\quad\quad\xi\in\goth g.
\eqno(6.2.12a)-(6.2.12c)
$$
For any element $x$ of $\sans s\hat\otimes\sans a$ of the form 
$x=\tilde s\alpha(s)+{1\over 2}\epsilon^{KL}s_{,K}s_{,L}\beta(s)$
with $\alpha(s)$, $\beta(s)$ polynomials in $s$, we define 
$\int_{[0,1]}x=\int_0^1\alpha(\sigma)d\sigma$, where the right hand side is 
an ordinary Riemann integral. It is not difficult to show that,
for any element of $f(s)$ of $\sans s\hat\otimes\sans a$
polynomial in $s$, ${1\over 2}\epsilon^{KL}\tilde h^s{}_K\tilde h^s{}_Lf(s)$
is of the above form and 
$\int_{[0,1]}{1\over 2}\epsilon^{KL}\tilde h^s{}_K\tilde h^s{}_Lf(s)=
f(1)-f(0)$. From (6.2.10),
$$
r[a(1)]-r[a(0)]
=\hbox{$1\over 2$}\epsilon^{KL}\tilde h_K\tilde h_L
\int_{[0,1]}\hbox{$1\over 2$}q^2r[a(s)].
\eqno(6.2.13)
$$
By (2.2.14d), the right hand side of (6.2.13) belongs to
${1\over 2}\epsilon^{KL}d_Kd_L\sans a$.
From (2.2.14a), (2.2.14b), (6.2.11), (6.2.12a)--(6.2.12c), if  
$r[a]$ belongs to $\sans a^{n,p}_{\rm basic}$ for any connection $a_A$ on 
$\sans a$, then ${1\over 2}q^2r[a(\sigma)]$ belongs to 
$\sans a^{n,p-2}_{\rm basic}$ for $\sigma\in\Bbb R$, so that
$\int_{[0,1]}\hbox{$1\over 2$}q^2r[a(s)]$ belongs to 
$\sans a^{n,p-2}_{\rm basic}$, too. \hfill $\square$

Consider the $N=2$ Weil $\goth g$ superoperation $\sans w$ (cfr. subsect.
5.2). Then, $\omega_A$ is a connection of $\sans w$ with derived connection 
$\gamma$ and curvature and derived curvature $\phi_{AB}$, $\rho_A$ 
(cfr. eqs. (3.2.9a)--(3.2.9b)).

Given an $N=2$ $\goth g$ superoperation $\sans a$, one can 
define the graded tensor product $N=2$ $\goth g$ superoperation 
$\sans w\hat\otimes\sans a$ (cfr. subsect. 4.2). The latter is the 
equivariant $N=2$ superoperation associated to $\sans a$. The equivariant 
cohomology of $\sans a$ is by definition the basic cohomology of 
$\sans w\hat\otimes\sans a$:
$$
H^{n,p}_{\rm equiv}(\sans a)=H^{n,p}_{\rm basic}(\sans w\hat\otimes\sans a),
\quad\quad (n,p)\in\Bbb N\times\Bbb Z.
\eqno(6.2.14)
$$

An equivariant cohomology class of $\sans a$ is represented by elements of 
$\sans w\hat\otimes\sans a$ of the form $r(\omega,\gamma,\phi,\rho)$,
where $r\in\bigwedge^*(\goth g^\vee\otimes\Bbb R^2)
\otimes\bigvee^*\goth g^\vee\otimes\bigvee^*(\goth g^\vee\otimes
\bigvee^2\Bbb R^2)\otimes\bigwedge^*(\goth g^\vee\otimes\Bbb R^2)
\otimes\sans a$. The Weil generator $\omega_A$ constitutes a connection of 
$\sans w\hat\otimes\sans a$. If $a_A$ is a connection of 
$\sans a$, $a_A$ is a connection of $\sans w\hat\otimes\sans a$ as well. 
By prop. 6.2.1, $r(\omega,\gamma,\phi,\rho)$ is equivalent to 
$r(a,b,f,g)$ in equivariant cohomology. On the other hand, 
$r(a,b,f,g)$ is a representative of a basic cohomology class of 
$\sans a$, which, by prop. 6.2.1, is independent from $a_A$ in basic 
cohomology. Thus, there is a natural homomorphism of 
$H^{n,p}_{\rm equiv}(\sans a)$ into $H^{n,p}_{\rm basic}(\sans a)$, 
called $N=2$ Weil homomorphism.
\par\vskip .6cm
{\bf 7. Superoperations of a smooth manifold with a group action}         
\vskip .4cm
Let $M$ be a smooth $m$ dimensional real manifold. Thus, $M$ is 
endowed with a collection of smooth charts $(U_a,x_a)$, $a\in A$,
in the usual way. Let $M$ carry the right action of a real
Lie group $G$ with Lie algebra $\goth g$ (see ref. \ref{33} for 
an exhaustive treatment of the theory of manifolds with a group action).

Let $\sans s$ be a real Grassmann algebra such that $\sans s^0\simeq\Bbb R$.

\vskip.2cm {\it 7.1. N=1 differential geometry}

\proclaim Definition 7.1.1. An $N=1$ differential structure on $M$ is a 
collection $\{(U_a,X_a)|a\in A\}$, where
\item{$i)$} $\{U_a|a\in A\}$is an open covering of $M$;
\item{$ii)$} for each $a\in A$, $X_a:U_a\mapsto(\sans S_1^0)^m$ and 
$x_a=X_a|_{\theta=0}:U_a\mapsto\Bbb R^m$ is a coordinate of $M$;
\item{$iii)$} for $a,b\in A$ such that $U_a\cap U_b\not=\emptyset$,
$X_a=x_a\circ x_b{}^{-1}(X_b)$.

Below, we shall omit the chart indices $a, b,\ldots$ except when dealing 
with matching relations.

We write as usual
$$
X^i=x^i+\theta\tilde x^i,\quad \tilde X^i=\tilde x^i,
\eqno(7.1.1a),(7.1.1b)
$$
where $x^i:U\mapsto\Bbb R$, $\tilde x^i:U\mapsto\sans s^1$.

We introduce the $N=1$ covariant superderivatives
$$
D_i=\tilde\partial_{xi}+\theta\partial_{xi}, \quad
\tilde D_i=\partial_{xi},
\eqno(7.1.2a),(7.1.2b)
$$
where $\tilde\partial_{xi}=\partial/\partial\tilde x^i$. One has
relations 
$$
\matrix{~&~\cr
[D_i,D_j]=0,\hfill&[D_i,\tilde D_j]=0,\hfill\cr
~&~\cr
[\tilde D_i,\tilde D_j]=0.\hfill&~\hfill\cr
~&~\cr}
\eqno(7.1.3a)-(7.1.3c)
$$
Further, 
$$
\matrix{~&~\cr
D_iX^j=0,\hfill&D_i\tilde X^j=\delta_i^j,\hfill\cr
~&~\cr
\tilde D_iX^j=\delta_i^j,\hfill& \tilde D_i\tilde X^j=0.\hfill\cr
~&~\cr}
\eqno(7.1.4a)-(7.1.4d)
$$
Using (7.1.1a), (7.1.1b), it is straightforward to check that
relations (7.1.4a)--(7.1.4d) completely characterize $D_i$, $\tilde D_i$. 

The transformation properties of $X^i$ under chart changes, stated 
in def. 7.1.1, imply that
$$
\tilde X_a{}^i=\tilde X_b{}^j\tilde D_{bj}X_a{}^i.
\eqno(7.1.5)
$$
Using that (7.1.4a)--(7.1.4d) completely characterize $D_i$, $\tilde D_i$, 
one can show easily that they match as
$$
D_{ai}=\tilde D_{ai}X_b{}^jD_{bj}, \quad
\tilde D_{ai}=\tilde D_{ai}\tilde X_b{}^jD_{bj}
+\tilde D_{ai}X_b{}^j\tilde D_{bj}.
\eqno(7.1.6a),(7.1.6b)
$$

We denote by $\cal F$ the sheaf of germs of smooth $N=1$ functions on $M$ 
generated by $X^i$, $\tilde X^i$. By definition, a generic 
element $F\in{\cal F}(U)$ is a finite sum of the form $F=\sum_{p\geq 0}
f_{i_1\cdots i_p}\circ X\tilde X^{i_1}\cdots \tilde X^{i_p}$ for
certain smooth maps $f_{i_1\cdots i_p}:\Bbb R^m\mapsto\Bbb R$ 
antisymmetric in $i_1,\cdots,i_p$. It is easy to see that
$$
F=\sum_{p=0}^m\Big[F_{i_1\cdots i_p}
+\theta\partial_{xi_0}F_{i_i\cdots i_p}\tilde x^{i_0}\Big]
\tilde x^{i_1}\cdots \tilde x^{i_p},
\eqno(7.1.7)
$$
where $F_{i_1\cdots i_p}=f_{i_1\cdots i_p}\circ x$.
$\cal F$ has a natural grading corresponding to 
the total $\sans s$ degree of $\tilde x^i$.

We define on $U_a\cap U_b\not=\emptyset$,
$$
Z_{ab}{}^i{}_j=\tilde D_{bj}X_a{}^i.
\eqno(7.1.8)
$$
It is easy to see that $Z$ is a $\GL(m,{\cal F})$ 1--cocycle on $M$.
$Z$ is called the fundamental 1--cocycle of the $N=1$ differential 
structure. One can introduce in standard fashion the sheaf 
${\cal F}_{r,s}:={\cal F}(Z^{\otimes r}\otimes Z^{\vee\otimes s})$ 
of germs of smooth $N=1$ sections of $Z^{\otimes r}\otimes 
Z^{\vee\otimes s}$. We denote by $\sans f_{\theta r,s}$ the real
vector space of sections of ${\cal F}_{r,s}$ on $M$. 

Notice that $\sans f_{\theta r,s}\simeq\sans f_{0 r,s}$, 
since, by (7.1.7), any $F\in{\cal F}(U)$ is completely determined 
by $f=F|_{\theta=0}$.

$z=Z|_{\theta=0}$ is nothing but the tangent bundle 1--cocycle of $M$. 
By (7.1.1b), (7.1.5),  
$\sans f_{0 r,s}{}^p$, and thus also $\sans f_{\theta r,s}{}^p$ 
by the previous remark, can be identified with the space of degree $p$ 
smooth tensor fields of type $r,s$ on $M$. 

We are particularly interested in the space $\sans f_{\theta 0,0}$,
which is a graded algebra. 

We define
$$
H=\tilde X^iD_i,\quad \tilde H=-\tilde X^i\tilde D_i.
\eqno(7.1.9a), (7.1.9b)
$$
Using (7.1.5), (7.1.6a), (7.1.6b), it is easy to see that $H$, $\tilde H$ are 
globally defined derivations on $\sans f_{\theta 0,0}$.

Denoting by $c\xi$ the fundamental vector field on $M$ induced by
$\xi\in\goth g$, we define further
$$
I(\xi)=C^i\xi D_i,\quad \tilde I(\xi)=C^i\xi\tilde D_i
+\tilde X^j\tilde D_jC^i\xi D_i,
\eqno(7.1.10a),(7.1.10b)
$$
where $C\xi$ is the element of $\sans f_{\theta 1,0}{}^0$ corresponding to 
$c\xi$ given explicitly by 
$C^i\xi=c^i\xi+\theta\tilde x^j\partial_{xj}c^i\xi$.
By (7.1.6a), (7.1.8), $I(\xi)$, $\tilde I(\xi)$ are also globally 
defined derivations on $\sans f_{\theta 0,0}$.

Using the relation $D_iC^j\xi=0$, 
it is now straightforward to verify that $H$, $\tilde H$, $I$, 
$\tilde I$ satisfy relations (2.1.2a)--(2.1.2c), (2.1.5a)--(2.1.5c), 
(2.1.8a)--(2.1.8d). In this way, $\sans f_{\theta 0,0}$ becomes
a $\Bbb Z$ graded real left module algebra of the $\Bbb Z$ graded real Lie
algebra $\sans t_\theta$ (cfr. sect 2.1).

Thus, $\sans f:=\sans f_{0 0,0}$ acquires the structure of $N=1$ $\goth g$ 
superoperation (cfr. def. 4.1.1), the relevant graded derivations being  
$$
h=\tilde x^i\tilde \partial_{xi},\quad 
\tilde h=-\tilde x^i\partial_{xi}
\eqno(7.1.11a),(7.1.11b)
$$
$$
i(\xi)=c^i\xi\tilde \partial_{xi},\quad 
\tilde\imath(\xi)=c^i\xi\partial_{xi}
+\tilde x^j\partial_{xj}c^i\xi\tilde\partial_{xi}.
\eqno(7.1.12a),(7.1.12b)
$$
This superoperation is  canonically associated to the $N=1$ differential 
structure.

Now, from (7.1.7) and (7.1.5), it appears that the graded algebra
$\sans f$ is isomorphic to the graded algebra of ordinary 
differential forms on $M$. Under such an isomorphism, the derivations 
$k$, $d$, $j(\xi)$, $l(\xi)$, defined in (2.1.12a), (2.1.12b), 
(2.1.13a), (2.1.13b), correspond to the form degree $k_{\rm dR}$, 
the de Rham differential $d_{\rm dR}$, the contraction $j_{\rm dR}(\xi)$ 
and the Lie derivative $l_{\rm dR}(\xi)$, respectively. Therefore,
the above is nothing but a reformulation of the customary
theory of differential 
forms, so that, in particular, the (basic) cohomology of $\sans f$ 
is isomorphic to the (basic) de Rham cohomology.

\proclaim Theorem 7.1.1. There is an isomorphism of the $N=1$ (basic)
cohomology of $\sans f$ the de Rham (basic) cohomology of the
($G$) manifold $M$. Indeed, one has that $H^p(\sans f)=0$ 
($H_{\rm basic}^p(\sans f)=0$), except perhaps for $0\leq p\leq m$, 
and 
$$
H^p(\sans f)\simeq H_{\rm dR}^p(M),\quad 0\leq p\leq m,
\eqno(7.1.13)
$$
$$
H_{\rm basic}^p(\sans f)\simeq H_{\rm dR~basic}^p(M)
\quad 0\leq p\leq m.
\eqno(7.1.14)
$$

{\it Proof}. See the above remarks. \hfill $\square$

Recall that a connection $y$ on the $G$ space $M$ is a $\goth g$ valued
$1$ form satisfying  relations (3.1.10a), (3.1.12a), (3.1.12c) 
with  $j$, $l$, $\omega$ substituted by $j_{\rm dR}$, $l_{\rm dR}$, $y$
respectively \ref{33}. We denote by $\Conn(M)$ the affine space of the 
connections on $M$.

\proclaim Theorem 7.1.2. One has
$$
\Conn(\sans f)\simeq\Conn(M)
\eqno(7.1.15)
$$
(cfr. def. 6.1.1).

{\it Proof}. Any $a\in\sans f^1\otimes\goth g$ is locally 
of the form $a=a_i\tilde x^i$, where $a_i$ is a $\goth g$ valued 
smooth map. Define $\lambda(a)=a_id_{\rm dR}x^i$. Then, by the above remarks,
$\lambda(a)$ is a connection of $M$ if and only if $a$ is a connection 
of $\sans f$. The map $\lambda$ is obviously a bijection. 
\hfill $\square$

\vskip.2cm {\it 7.2. N=2 differential geometry}

\proclaim Definition 7.2.1. An $N=2$ differential structure on $M$ is a 
collection $\{(U_a,X_a)|a\in A\}$, where
\item{$i)$} $\{U_a|a\in A\}$is an open covering of $M$;
\item{$ii)$} for each $a\in A$, $X_a:U_a\mapsto(\sans S_2^0)^m$ and 
$x_a=X_a|_{\theta=0}:U_a\mapsto\Bbb R^m$ is a coordinate of $M$;
\item{$iii)$} for $a,b\in A$ such that $U_a\cap U_b\not=\emptyset$,
$X_a=x_a\circ x_b{}^{-1}(X_b)$.

Below, we shall omit the chart indices $a, b,\ldots$ except when dealing 
with matching relations.

We write as usual 
$$
\matrix{~&~\cr
X^i=x^i+\theta^Ax^i{}_{,A}+\hbox{$1\over 2$}\epsilon_{KL}
\theta^K\theta^L\tilde x^i,\hfill&
X^i{}_{,A}=x^i{}_{,A}+\epsilon_{AK}\theta^K\tilde x^i,\hfill\cr
~&~\cr
\tilde X^i=\tilde x^i,\hfill&~\cr
~&~\cr}
\eqno(7.2.1a)-(7.2.1c)
$$
where $x^i:U\mapsto\Bbb R$,  $x^i{}_{,A}:U\mapsto\sans s^1$,
$\tilde x^i:U\mapsto\sans s^2$. 

We introduce the $N=2$ covariant superderivatives
$$
\matrix{~&~\cr
D_i=\tilde\partial_{xi}+\epsilon_{KL}\theta^K\partial_{xi}{}^{,L}
+\hbox{$1\over 2$}\epsilon_{KL}\theta^K\theta^L\partial_{xi},
\hfill&
D_{i,A}=\epsilon_{AK}(\partial_{xi}{}^{,K}+\theta^K\partial_{xi}),
\hfill\cr
~&~\cr
\tilde D_i=\partial_{xi},\hfill&~\cr
~&~\cr}
\eqno(7.2.2a)-(7.2.2c)
$$
where $\partial_{xi}{}^{,A}=\partial/\partial x^i{}_{,A}$,
$\tilde\partial_{xi}=\partial/\partial\tilde x^i$. One has
$$
\matrix{~&~&~\cr
[D_i,D_j]=0,\hfill&[D_i,D_{j,A}]=0,\hfill&
[D_i,\tilde D_j]=0,\hfill\cr
~&~&~\cr
[D_{i,A},D_{j,B}]=0,\hfill&[D_{i,A},\tilde D_j]=0,\hfill&
[\tilde D_i,\tilde D_j]=0.\hfill\cr
~&~&~\cr}
\eqno(7.2.3a)-(7.2.3f)
$$
Further,
$$
\matrix{~&~&~\cr
D_iX^j=0,\hfill&D_iX^j{}_{,A}=0,\hfill&
D_i\tilde X^j=\delta_i^j,\hfill\cr
~&~&~\cr
D_{i,A}X^j=0,\hfill&D_{i,A}X^j{}_{,B}=\epsilon_{AB}\delta_i^j,\hfill&
D_{i,A}\tilde X^j=0,\hfill\cr
~&~&~\cr
\tilde D_iX^j=\delta_i^j,\hfill&\tilde D_iX^j{}_{,A}=0,\hfill&
\tilde D_i\tilde X^j=0.\hfill\cr
~&~&~\cr}
\eqno(7.2.4a)-(7.2.4i)
$$
By (7.2.1a)--(7.2.1c), relations (7.2.4a)--(7.2.4i) completely characterize 
$D_i$, $D_{i,A}$, $\tilde D_i$. 

The transformation properties of $X^i$ under chart changes, stated 
in def. 7.2.1, imply that
$$
X_a{}^i{}_{,A}=X_b{}^j{}_{,A}\tilde D_{bj}X_a{}^i,\quad
\tilde X_a{}^i=\tilde X_b{}^j\tilde D_{bj}X_a{}^i
+\hbox{$1\over 2$}\epsilon^{JK}X_b{}^j{}_{,J}X_b{}^k{}_{,K}
\tilde D_{bj}\tilde D_{bk}X_a{}^i.
\eqno(7.2.5a),(7.2.5b)
$$
Using that (7.2.4a)--(7.2.4i) completely characterize $D_i$, $D_{i,A}$, 
$\tilde D_i$, one can show easily that they match as
$$
\matrix{~\cr
\matrix{D_{ai}=\tilde D_{ai}X_b{}^jD_{bj},\hfill&
D_{ai,A}=\tilde D_{ai}X_b{}^j{}_{,A}D_{bj}
+\tilde D_{ai}X_b{}^jD_{bj,A},\hfill\cr}\hfill\cr
~\cr
\tilde D_{ai}=\tilde D_{ai}\tilde X_b{}^jD_{bj}
+\epsilon^{KL}\tilde D_{ai}X_b{}^k{}_{,K}D_{bk,L}
+\tilde D_{ai}X_b{}^j\tilde D_{bj}.\hfill\cr
~\cr}
\eqno(7.2.6a)-(7.2.6c)
$$

We denote by $\cal F$ the sheaf of germs of smooth $N=2$ functions on $M$  
generated by $X^i$, $X^i{}_{,A}$, $\tilde X^i$. By definition, a generic 
element $F\in{\cal F}(U)$ is a finite sum of the form $F=\sum_{p,q\geq 0}
f^{I_1\cdots I_p}_{i_1\cdots i_pi_{p+1}\cdots i_{p+q}}\circ X
X^{i_1}{}_{,I_1}\cdots X^{i_p}{}_{,I_p}
\tilde X^{i_{p+1}}\cdots \tilde X^{i_{p+q}}$ for
certain smooth maps $f^{I_1\cdots I_p}_{i_1\cdots i_p
i_{p+1}\cdots i_{p+q}}:\Bbb R^m\mapsto\Bbb R$ 
antisymmetric in the pairs $(i_1,I_1),\cdots,(i_p,I_p)$ and  
symmetric in $i_{p+1},\cdots,i_{p+q}$. It is straightforward though 
tedious to show that
$$
\eqalignno{
F&=\sum_{p=0}^{2m}\sum_{q=0}^{q_0}\Big\{
F^{I_1\cdots I_p}_{i_1\cdots i_pi_{p+1}\cdots i_{p+q}}
x^{i_1}{}_{,I_1}x^{i_2}{}_{,I_2}&\cr
&+\theta^K\Big[\delta_K^{I_0}\partial_{i_0}
F^{I_1\cdots I_p}_{i_1\cdots i_pi_{p+1}\cdots i_{p+q}}
x^{i_0}{}_{,I_0}x^{i_1}{}_{,I_1}x^{i_2}{}_{,I_2}
-p\epsilon_{KI_1}
F^{I_1\cdots I_p}_{i_1\cdots i_pi_{p+1}\cdots i_{p+q}}
x^{i_2}{}_{,I_2}\tilde x^{i_1}\Big]\vphantom{\sum_{p=0}^{2m}}&\cr
&+\hbox{$1\over 2$}\epsilon_{KL}\theta^K\theta^L\Big[
\hbox{$1\over 2$}\epsilon^{I_{-1}I_0}\partial_{i_{-1}}\partial_{i_0}
F^{I_1\cdots I_p}_{i_1\cdots i_pi_{p+1}\cdots i_{p+q}}
x^{i_{-1}}{}_{,I_{-1}}x^{i_0}{}_{,I_0}x^{i_1}{}_{,I_1}x^{i_2}{}_{,I_2}
\vphantom{\sum_{p=0}^{2m}}&\cr
&+\partial_{i_0}
F^{I_1\cdots I_p}_{i_1\cdots i_pi_{p+1}\cdots i_{p+q}}
x^{i_1}{}_{,I_1}x^{i_2}{}_{,I_2}\tilde x^{i_0}
-p\delta^{I_0}_{I_1}\partial_{i_0}
F^{I_1\cdots I_p}_{i_1\cdots i_pi_{p+1}\cdots i_{p+q}}
x^{i_0}{}_{,I_0}x^{i_2}{}_{,I_2}\tilde x^{i_1}
\vphantom{\sum_{p=0}^{2m}}&\cr
&+\hbox{$1\over 2$}p(p-1)\epsilon_{I_1I_2}
F^{I_1\cdots I_p}_{i_1\cdots i_pi_{p+1}\cdots i_{p+q}}
\tilde x^{i_1}\tilde x^{i_2}\Big]
\Big\}x^{i_3}{}_{,I_3}\cdots x^{i_p}{}_{,I_p}
\tilde x^{i_{p+1}}\cdots\tilde x^{i_{p+q}}
\vphantom{\sum_{p=0}^{2m}},&(7.2.7)\cr}
$$
where $F^{I_1\cdots I_p}_{i_1\cdots i_pi_{p+1}\cdots i_{p+q}}
=f^{I_1\cdots I_p}_{i_1\cdots i_pi_{p+1}\cdots i_{p+q}}\circ x$.
$\cal F$ has a natural grading corresponding to 
the total $\sans s$ degree of $x^i{}_{,I}$, $\tilde x^i$.

We define on $U_a\cap U_b\not=\emptyset$,
$$
Z_{ab}{}^i{}_j=\tilde D_{bj}X_a{}^i.
\eqno(7.2.8)
$$
It is easy to see that $Z$ is a $\GL(m,{\cal F})$ 1--cocycle on $M$.
$Z$ is called the fundamental 1 cocycle of the $N=2$ differential 
structure. One can introduce in standard fashion the sheaf 
${\cal F}_{r,s}:={\cal F}(Z^{\otimes r}\otimes Z^{\vee\otimes s})$ 
of germs of smooth $N=2$ sections of $Z^{\otimes r}\otimes 
Z^{\vee\otimes s}$. We denote by $\sans f_{\theta r,s}$ the real
vector space of sections of ${\cal F}_{r,s}$ on $M$. 

Notice that $\sans f_{\theta r,s}\simeq\sans f_{0 r,s}$, 
since, by (7.2.7), any $F\in{\cal F}(U)$ is completely determined 
by $f=F|_{\theta=0}$.

$z=Z|_{\theta=0}$ is nothing but the tangent bundle 1--cocycle of $M$. 
However, unlike the $N=1$ case, there is no simple geometrical 
interpretation of the spaces $\sans f_{0 r,s}{}^p$, 
$\sans f_{\theta r,s}{}^p$.

We are particularly interested in the space $\sans f_{\theta 0,0}$,
which is a graded algebra. 

We define 
$$
\matrix{~&~\cr
H_A=-X^i{}_{,A}D_i,\hfill&
H_{A,B}=X^i{}_{,A}D_{i,B}-\epsilon_{AB}\tilde X^iD_i\hfill\cr
~&~\cr
\tilde H_A=\tilde X^iD_{i,A}-X^i{}_{,A}\tilde D_i.\hfill&~\hfill\cr
~&~\cr}
\eqno(7.2.9a)-(7.2.9c)
$$
Using (7.2.5a), (7.2.5b), (7.2.6a)--(7.2.6c), it is easy to see that $H_A$,
$H_{A,B}$, $\tilde H_A$ are globally defined derivations 
on $\sans f_{\theta 0,0}$.

We set next
$$
\matrix{~\cr
\matrix{I(\xi)=C^i\xi D_i,\hfill&
I_{,A}(\xi)=X^j{}_{,A}\tilde D_jC^i\xi D_i+C^i\xi D_{i,A},\hfill\cr}\hfill\cr
~\cr
\tilde I(\xi)
=\Big[\tilde X^j\tilde D_jC^i\xi 
+\hbox{$1\over 2$}\epsilon^{KL}X^k{}_{,K}X^l{}_{,L}
\tilde D_k\tilde D_lC^i\xi\Big]D_i
+\epsilon^{KL}X^k{}_{,K}
\tilde D_kC^i\xi D_{i,L}+C^i\xi\tilde D_i,\hfill\cr
~\cr}
\eqno(7.2.10a)-(7.2.10c)
$$
where $C\xi$ is the element of $\sans f_{\theta 1,0}{}^0$ corresponding to 
$c\xi$ and is given explicitly by $C^i\xi=c^i\xi
+\theta^Kx^j{}_{,K}\partial_{xj}c^i\xi
+{1\over 2}\epsilon_{KL}\theta^K\theta^L
[\tilde x^j\partial_{xj}c^i\xi
+{1\over 2}\epsilon^{MN}x^j{}_{,M}x^k{}_{,N}
\partial_{xj}\partial_{xk}c^i\xi]$.
By (7.2.6a), (7.2.8), $I(\xi)$, $I_{,A}(\xi)$, $\tilde I(\xi)$ are globally 
defined derivations on $\sans f_{\theta 0,0}$.

Using the relation $D_iC^j\xi=0$, $D_{i,A}C^j\xi=0$, 
it is now straightforward to verify that $H_A$, $H_{A,B}$, $\tilde H_A$, 
$I$, $I_{,A}$, $\tilde I$ satisfy relations 
(2.2.2a)--(2.2.2f), (2.2.6a)--(2.2.6f), 
(2.2.9a)--(2.2.9i). In this way, $\sans f_{\theta 0,0}$ becomes
a $\Bbb Z$ graded real left module algebra of the $\Bbb Z$ graded real Lie
algebra $\sans t_\theta$ (cfr. sect 2.2).

Thus, $\sans f:=\sans f_{0 0,0}$ acquires the structure of $N=2$ $\goth g$ 
superoperation (cfr. def. 4.2.1), the relevant graded derivations being 
$$
\matrix{~&~\cr
h_A=-x^i{}_{,A}\tilde\partial_{xi},\hfill&
h_{A,B}=x^i{}_{,A}\epsilon_{BL}\partial_{xi}{}^{,L}
-\epsilon_{AB}\tilde x^i\tilde\partial_{xi}\hfill\cr
~&~\cr
\tilde h_A=\tilde x^i\epsilon_{AL}\partial_{xi}{}^{,L}
-x^i{}_{,A}\partial_{xi}.\hfill&~\hfill\cr
~&~\cr}
\eqno(7.2.11a)-(7.2.11c)
$$
$$
\matrix{~\cr
\matrix{i(\xi)=c^i\xi\tilde\partial_{xi},\hfill&
i_{,A}(\xi)=c^i\xi\epsilon_{AL}\partial_{xi}{}^{,L}
+x^j{}_{,A}\partial_{xj}c^i\xi\tilde\partial_{xi},\hfill\cr}\hfill\cr
~\cr
\tilde\imath(\xi)
=c^i\xi\partial_{xi}+x^j{}_{,K}\partial_{xj}c^i\xi\partial_{xi}{}^{,K}
+\Big[\tilde x^j\partial_{xj}c^i\xi
+\hbox{$1\over 2$}\epsilon^{KL}x^k{}_{,K}x^l{}_{,L}
\partial_{xk}\partial_{xl}c^i\xi\Big]\tilde\partial_{xi}.\hfill\cr
~\cr}
\eqno(7.2.12a)-(7.2.12c)
$$
This superoperation is canonically associated to the $N=2$ 
differential structure.

In spite of the fact that, in the $N=2$ case, $\sans f$ does not have any 
simple geometrical interpretation, unlike its $N=1$ counterpart,
the (basic) cohomology of $\sans f$ in the $N=2$ case has essentially the 
same content as that of the $N=1$ case and a theorem analogous to 
theor. 7.1.1 holds.

\proclaim Theorem 7.2.1. There is an isomorphism of the $N=2$ (basic) 
cohomology of $\sans f$ the de Rham (basic) cohomology of the
($G$) manifold $M$. Indeed, one has that $H^{n,p}(\sans f)=0$ 
($H_{\rm basic}^{n,p}(\sans f)=0$), except perhaps for $(n,p)=(1,0)$, 
$(r,r+1)$ with $1\leq r\leq m$, and 
$$
H^{1,0}(\sans f)\simeq H_{\rm dR}^0(M), \quad
H^{r,r+1}(\sans f)\simeq H_{\rm dR}^r(M)
\otimes\hbox{$\bigvee^{r-1}$}\Bbb R^2, \quad 
1\leq r\leq m, \vphantom{\Big]}
\eqno(7.2.13a),(7.2.13b)
$$
$$
H_{\rm basic}^{1,0}(\sans f)\simeq H_{\rm dR~basic}^0(M), \quad
H_{\rm basic}^{r,r+1}(\sans f)\simeq H_{\rm dR~basic}^r(M)
\otimes\hbox{$\bigvee^{r-1}$}\Bbb R^2, \quad 
1\leq r\leq m.\vphantom{\Big]}
\eqno(7.2.14a),(7.2.14b)
$$

{\it Proof}. By prop. 4.2.2, $H^{n,p}(\sans f)=0$
($H_{\rm basic}^{n,p}(\sans f)=0$) except perhaps for $p=\pm n+1$. 
On the other hand, from the definition of $\sans f$, given above, 
$\sans f^{n,p}=0$ for $p<0$. So, $H^{n,p}(\sans f)=0$
($H_{\rm basic}^{n,p}(\sans f)=0$) except perhaps for $(n,p)=(1,0)$, 
$(r,r+1)$ with $1\leq r$. Consider first the case where $(n,p)=(1,0)$. 
From (7.2.9b) and the representation theory of $\sans i=\goth s\goth l
(2,\Bbb R)\oplus\Bbb R$, it is immediate to see that $\sans f^{1,0}$ 
consists precisely of the $F$ of the form $F=\alpha$ for some smooth 
function $\alpha$ on $M$ and that $\sans f^{1,-2}=0$. 
Further, the conditions $d_AF=0$ is equivalent to 
$d_{\rm dR}\alpha=0$, hence to the local constance of $\alpha$.
We thus have a linear bijection $\nu: \sans f^{1,0}\cap\cap_{A=1,2}\ker d_A
\mapsto Z_{\rm dR}^0(M)$, where $Z_{\rm dR}^r(M)$ is the space of closed 
$r$ forms, given by $F\mapsto\alpha$. Being $\sans f^{1,-2}=0$, 
(7.2.13a) follows. (7.2.14a) also holds, as, clearly, 
$\sans f^{1,0}=\sans f_{\rm basic}^{1,0}$ and $Z_{\rm dR}^0(M)
=Z_{\rm dR~basic}^0(M)$. 
Consider next the case where $(n,p)=(r,r+1)$ with  
$1\leq r$. Let $F\in\sans f^{r,r+1}$. From (7.2.9b) and the representation 
theory of $\sans i=\goth s\goth l(2,\Bbb R)\oplus\Bbb R$, $F$ is locally of 
the form
$$
F=x^{i_1}{}_{,A_1}\cdots x^{i_{r-1}}{}_{,A_{r-1}}
\Big[\tilde x^{i_r}\alpha^{A_1\cdots A_{r-1}}_{i_1\cdots i_{r-1}i_r}
+\hbox{$1\over 2$}\epsilon^{MN}x^{i_r}{}_{,M}x^{i_{r+1}}{}_{,N}
\beta^{A_1\cdots A_{r-1}}_{i_1\cdots i_{r-1}i_ri_{r+1}}\Big]
\eqno(7.2.15)
$$
with $\alpha^{A_1\cdots A_{r-1}}_{i_1\cdots i_{r-1}i_r}$ a realvalued smooth 
map symmetric in $A_1,\cdots, A_{r-1}$ and antisymmetric in $i_1,\cdots,
i_{r-1}$ and $\beta^{A_1\cdots A_{r-1}}_{i_1\cdots i_{r-1}i_ri_{r+1}}$ a 
realvalued smooth map symmetric in $A_1,\cdots, A_{r-1}$, antisymmetric in 
$i_1,\cdots,i_{r-1}$ and symmetric in  $i_r,i_{r+1}$. 
Next, assume that $d_AF=0$. Substituting (7.2.15) into the relation 
$d_AF_{A_1\cdots A_{r-1}}=0$ and taking into account the fact that 
terms with different numbers of $x^{i}{}_{,I}$, $\tilde x^i$ 
are linearly independent and, thus, must vanish separately, one gets 
the following three identities 
$$
\eqalignno{
&x^{i_1}{}_{,A_1}\cdots x^{i_{r-2}}{}_{,A_{r-2}}\tilde x^{i_{r-1}}
\tilde x^{i_r}\alpha^{A_1\cdots A_{r-1}}_{i_1\cdots i_{r-1}i_r}=0,
\vphantom{\Big[}&(7.2.16a)\cr
&(r-1)\epsilon_{AA_{r-1}}x^{i_1}{}_{,A_1}\cdots x^{i_{r-2}}{}_{,A_{r-2}}
\hbox{$1\over 2$}\epsilon^{MN}x^{i_r}{}_{,M}x^{i_{r+1}}{}_{,N}
\tilde x^{i_{r-1}}\beta^{A_1\cdots A_{r-1}}_{i_1\cdots i_{r-1}i_ri_{r+1}}
\vphantom{\Big[}&\cr
&+x^{i_1}{}_{,A_1}\cdots x^{i_{r-1}}{}_{,A_{r-1}}\Big[
-x^{i_r}{}_{,A}\tilde x^{i_{r+1}}
\beta^{A_1\cdots A_{r-1}}_{i_1\cdots i_{r-1}i_ri_{r+1}}
+x^{i_{r+1}}{}_{,A}\tilde x^{i_r}\partial_{xi_{r+1}}
\alpha^{A_1\cdots A_{r-1}}_{i_1\cdots i_{r-1}i_r}\Big]=0,&\cr
&&(7.2.16b)\cr
&x^{i_1}{}_{,A_1}\cdots x^{i_{r-1}}{}_{,A_{r-1}}
\hbox{$1\over 2$}\epsilon^{MN}x^{i_r}{}_{,M}x^{i_{r+1}}{}_{,N}
x^{i_{r+2}}{}_{,A}\partial_{xi_{r+2}}
\beta^{A_1\cdots A_{r-1}}_{i_1\cdots i_{r-1}i_ri_{r+1}}=0.
\vphantom{\Big[}&(7.2.16c)\cr}
$$
From (7.2.16a), using the symmetry properties of 
$\alpha^{A_1\cdots A_{r-1}}_{i_1\cdots i_{r-1}i_r}$
and the fact that $x^i{}_{,A}$, $\tilde x^i$ are odd, even, respectively,  
it follows immediately that 
$\alpha^{A_1\cdots A_{r-1}}_{i_1\cdots i_{r-2}i_{r-1}i_r}
+\alpha^{A_1\cdots A_{r-1}}_{i_1\cdots i_{r-2}i_ri_{r-1}}=0$.
Since $\alpha^{A_1\cdots A_{r-1}}_{i_1\cdots i_{r-1}i_r}$ is already
antisymmetric in $i_1,\cdots,i_{r-1}$, 
$\alpha^{A_1\cdots A_{r-1}}_{i_1\cdots i_r}$ is antisymmetric in all
the indices $i_1,\cdots,i_r$. Thus, for fixed $A_1,\cdots,A_{r-1}$, the 
$\alpha^{A_1\cdots A_{r-1}}_{i_1\cdots i_r}$ 
are the coefficients of a local $r$ form $\alpha^{A_1\cdots A_{r-1}}$. 
Next, applying the derivation $u_B$ (cfr. eq. (7.2.11a)) to eq. (7.2.16b)
and contracting with $\epsilon^{BA}$, one gets 
$$
\eqalignno{
&x^{i_1}{}_{,A_1}\cdots x^{i_{r-1}}{}_{,A_{r-1}}
\hbox{$1\over 2$}\epsilon^{MN}x^{i_r}{}_{,M}x^{i_{r+1}}{}_{,N}
\beta^{A_1\cdots A_{r-1}}_{i_1\cdots i_{r-1}i_ri_{r+1}}&\cr
&={2\over r+1}x^{i_1}{}_{,A_1}\cdots x^{i_{r-1}}{}_{,A_{r-1}}
\hbox{$1\over 2$}\epsilon^{MN}x^{i_r}{}_{,M}x^{i_{r+1}}{}_{,N}
\partial_{xi_{r+1}}\alpha^{A_1\cdots A_{r-1}}_{i_1\cdots i_{r-1}i_r}.
&(7.2.17)\cr}
$$
Applying $d_A$ to this relation, one gets 
$$
\eqalignno{
&(r-1)\epsilon_{AA_{r-1}}x^{i_1}{}_{,A_1}\cdots x^{i_{r-2}}{}_{,A_{r-2}}
\hbox{$1\over 2$}\epsilon^{MN}x^{i_r}{}_{,M}x^{i_{r+1}}{}_{,N}
\tilde x^{i_{r-1}}\beta^{A_1\cdots A_{r-1}}_{i_1\cdots i_{r-1}i_ri_{r+1}}
\vphantom{r\over r}&\cr
&-x^{i_1}{}_{,A_1}\cdots x^{i_{r-1}}{}_{,A_{r-1}}x^{i_r}{}_{,A}
\tilde x^{i_{r+1}}\beta^{A_1\cdots A_{r-1}}_{i_1\cdots i_{r-1}i_ri_{r+1}}
\vphantom{r\over r}&\cr
&=2{r-1\over r+1}
\epsilon_{AA_{r-1}}x^{i_1}{}_{,A_1}\cdots x^{i_{r-2}}{}_{,A_{r-2}}
\hbox{$1\over 2$}\epsilon^{MN}x^{i_r}{}_{,M}x^{i_{r+1}}{}_{,N}
\tilde x^{i_{r-1}}\partial_{xi_{r+1}}
\alpha^{A_1\cdots A_{r-1}}_{i_1\cdots i_{r-1}i_r}&\cr
&-{1\over r+1}x^{i_1}{}_{,A_1}\cdots x^{i_{r-1}}{}_{,A_{r-1}}x^{i_r}{}_{,A}
\tilde x^{i_{r+1}}\big(
\partial_{xi_{r+1}}\alpha^{A_1\cdots A_{r-1}}_{i_1\cdots i_{r-1}i_r}
+\partial_{xi_r}\alpha^{A_1\cdots A_{r-1}}_{i_1\cdots i_{r-1}i_{r+1}}\big),&\cr
&&(7.2.18a)\cr
&x^{i_1}{}_{,A_1}\cdots x^{i_{r-1}}{}_{,A_{r-1}}
\hbox{$1\over 2$}\epsilon^{MN}x^{i_r}{}_{,M}x^{i_{r+1}}{}_{,N}
x^{i_{r+2}}{}_{,A}\partial_{xi_{r+2}}
\beta^{A_1\cdots A_{r-1}}_{i_1\cdots i_{r-1}i_ri_{r+1}}
\vphantom{r\over r}&\cr
&={2\over r+1}x^{i_1}{}_{,A_1}\cdots x^{i_{r-1}}{}_{,A_{r-1}}
\hbox{$1\over 2$}\epsilon^{MN}x^{i_r}{}_{,M}x^{i_{r+1}}{}_{,N}
x^{i_{r+2}}{}_{,A}\partial_{xi_{r+1}}
\partial_{xi_{r+2}}\alpha^{A_1\cdots A_{r-1}}_{i_1\cdots i_{r-1}i_r}.&\cr
&&(7.2.18b)\cr}
$$
Substituting (7.2.18a), (7.2.18b) into (7.2.16b), (7.2.16c), respectively, 
one obtains after a straightforward calculation the equations 
$$
\eqalignno{
&x^{i_1}{}_{,A_1}\cdots x^{i_{r-1}}{}_{,A_{r-1}}x^{i_{r+1}}{}_{,A_{r+1}}
\tilde x^{i_r}
\sum_{l=1}^{r+1}(-1)^{l-1}\partial_{xi_l}
\alpha^{A_1\cdots A_{r-1}}_{i_1\cdots i_{l-1}i_{l+1}\cdots i_{r+1}}=0,
&(7.2.19a)\cr
&x^{i_1}{}_{,A_1}\cdots x^{i_{r-1}}{}_{,A_{r-1}}
\hbox{$1\over 2$}\epsilon^{MN}x^{i_r}{}_{,M}x^{i_{r+1}}{}_{,N}
x^{i_{r+2}}{}_{,A}\partial_{xi_{r+1}}
\partial_{xi_{r+2}}\alpha^{A_1\cdots A_{r-1}}_{i_1\cdots i_{r-1}i_r}=0.~~~~
&(7.2.19b)\cr}
$$
Using the symmetry properties of 
$\alpha^{A_1\cdots A_{r-1}}_{i_1\cdots i_{r-1}i_r}$
and the fact that $x^i{}_{,A}$, $\tilde x^i$ are odd, even, respectively,
it is easy to see, that (7.2.19a) implies that 
$\sum_{l=1}^{r+1}(-1)^{l-1}\partial_{xi_l}
\alpha^{A_1\cdots A_{r-1}}_{i_1\cdots i_{l-1}i_{l+1}\cdots i_{r+1}}=0$ or
$d_{\rm dR}\alpha^{A_1\cdots A_{r-1}}=0$ so that the local $r$ form 
$\alpha^{A_1\cdots A_{r-1}}$ is closed and locally exact. 
By this reason and the fact that $x^i{}_{,I}\partial_{xi}
x^j{}_{,J}\partial_{xj}$ $x^k{}_{,K}\partial_{xk}=0$ by antisymmetry, 
one finds that eq. (7.2.19b) is automatically satisfied.
We note that, by (7.2.5a) and the global definition of $F$, 
the local exact $r$ form $\alpha^{A_1\cdots A_{r-1}}$ is the local 
restriction of a globally defined closed $r$ form, which will be 
denoted by the same symbol. To summarize, we have shown that 
(7.2.16a)--(7.2.16c) imply that, for fixed $A_1,\cdots,A_{r-1}$, 
$\alpha^{A_1\cdots A_{r-1}}$ is a closed $r$ form and that (7.2.17) holds. 
Conversely, assume that for fixed $A_1,\cdots,A_{r-1}$, 
$\alpha^{A_1\cdots A_{r-1}}$ is a closed $r$ form and that (7.2.17) 
holds. Using (7.2.5a), (7.2.5b), it is straightforward though tedious
to show that $F$, as given by (7.2.15), belongs to $\sans f^{r,r+1}$.
As shown above, (7.2.17) implies (7.2.18a), (7.2.18b) using which
eqs. (7.2.16b), (7.2.16c) become equivalent to eqs. (7.2.19a), (7.2.19b).
Eqs. (7.2.16a), (7.2.19a), (7.2.19b), are trivially satisfied by the closed
$r$ form $\alpha^{A_1\cdots A_{r-1}}$. 
Thus, (7.2.16a)--(7.2.16c) are satisfied as well implying that $d_AF=0$.
In conclusion, we have shown that $\sans f^{r,r+1}\cap\cap_{A=1,2}\ker d)_A$ 
consists precisely of the elements $F\in\sans f^{r,r+1}$ of the form 
$$
F=x^{i_1}{}_{,A_1}\cdots x^{i_{r-1}}{}_{,A_{r-1}}
\Big[\tilde x^{i_r}\alpha^{A_1\cdots A_{r-1}}_{i_1\cdots i_{r-1}i_r}
+{2\over r+1}\hbox{$1\over 2$}\epsilon^{MN}x^{i_r}{}_{,M}x^{i_{r+1}}{}_{,N}
\partial_{xi_{r+1}}\alpha^{A_1\cdots A_{r-1}}_{i_1\cdots i_{r-1}i_r}\Big]
\eqno(7.2.20)
$$
with $\alpha^{A_1\cdots A_{r-1}}$ an $r$ form symmetric in
$A_1,\cdots,A_{r-1}$ and such that $d_{\rm dR}\alpha^{A_1\cdots A_{r-1}}=0$. 
We thus have a linear bijection $\nu: 
\sans f^{r,r+1}\cap\cap_{A=1,2}\ker d_A\mapsto
Z_{\rm dR}^r(M)\otimes\bigvee^{r-1}\Bbb R^2$, where
$Z_{\rm dR}^r(M)$ is the space of closed $r$ forms,
given by $F\mapsto(\alpha^{A_1\cdots A_{r-1}})_{A_1,\cdots,A_{r-1}=1,2}$. 
Next, assume that $F\in{1\over 2}\epsilon^{KL}
d_Kd_L\sans f^{r,r-1}$. Then, $F={1\over 2}\epsilon^{KL}d_Kd_LG$
for some $G\in\sans f^{r,r-1}$. From (7.2.9b) and the representation 
theory of $\sans i=\goth s\goth l(2,\Bbb R)\oplus\Bbb R$, $G$ is of the form 
$$
G=x^{i_1}{}_{,A_1}\cdots x^{i_{r-1}}{}_{,A_{r-1}}
\gamma^{A_1\cdots A_{r-1}}_{i_1\cdots i_{r-1}},
\eqno(7.2.21)
$$
with $\gamma^{A_1\cdots A_{r-1}}_{i_1\cdots i_{r-1}}$ a realvalued smooth 
map symmetric in $A_1,\cdots,A_{r-1}$ and antisymmetric in 
$i_1,\cdots,i_{r-1}$. By a straightforward computation, one finds that
$$
\eqalignno{
\hbox{$1\over 2$}\epsilon^{KL}&d_Kd_LG
=(-1)^{r-1}x^{i_1}{}_{,A_1}\cdots x^{i_{r-1}}{}_{,A_{r-1}}
\Big[\tilde x^{i_r}\sum_{l=1}^r(-1)^{l-1}\partial_{xi_l}
\gamma^{A_1\cdots A_{r-1}}_{i_1\cdots i_{l-1}i_{l+1}\cdots i_{r-1}i_r}&\cr
&+{2\over r+1}\hbox{$1\over 2$}\epsilon^{MN}x^{i_r}{}_{,M}x^{i_{r+1}}{}_{,N}
\partial_{xi_{r+1}}\sum_{l=1}^r(-1)^{l-1}\partial_{xi_l}
\gamma^{A_1\cdots A_{r-1}}_{i_1\cdots i_{l-1}i_{l+1}\cdots i_{r-1}i_r}
\Big].&(7.2.22)\cr}
$$
Note that the $\gamma^{A_1\cdots A_{r-1}}_{i_1\cdots i_{r-1}}$ are 
the coefficients of a local $r-1$ form $\gamma^{A_1\cdots A_{r-1}}$. 
By (7.2.5a) and the global definition of $G$, $\gamma^{A_1\cdots A_{r-1}}$ 
is the restriction of a globally defined $r-1$ form, 
which we shall denote by the same symbol. As (7.2.22) indicates, 
the linear map $\nu$ maps cohomologically trivial elements of 
$\sans f^{r,r+1}\cap\cap_{A=1,2}\ker d_A$ into cohomologically trivial 
elements of $Z_{\rm dR}^r(M)\otimes\bigvee^{r-1}\Bbb R^2$. 
Thus, $\nu$ induces a linear bijection $\hat\nu:H^{r,r+1}(\sans f)
\mapsto H_{\rm dR}^r(M)\otimes\bigvee^{r-1}\Bbb R^2$. 
Next, assume that $F\in\sans f_{\rm basic}^{r,r+1}$ and that 
$d_AF=0$. In particular, $F$ is of the form (7.2.20) for some closed 
$r$ form $\alpha^{A_1\cdots A_{r-1}}$ symmetric in $A_1,\cdots,A_{r-1}$.
By (2.2.19d), (2.2.19e) and the relation $d_AF=0$, 
the basicity of $F$ is equivalent to the relation $j(\xi)F=0$, 
$\xi\in\goth g$, where $j(\xi)$, by (2.2.15a), is given in the present 
situation by (7.2.12a). A simple computation shows that this identity is 
equivalent to
$$
x^{i_1}{}_{,A_1}\cdots x^{i_{r-1}}{}_{,A_{r-1}}
c^{i_r}\xi\alpha^{A_1\cdots A_{r-1}}_{i_1\cdots i_{r-1}i_r}=0.
\eqno(7.2.23)
$$
As is straightforward to check, this relation entails that $c^{i_0}\xi
\alpha^{A_1\cdots A_{r-1}}_{i_0i_1\cdots i_{r-1}}=0$, so that 
$j_{\rm dR}(\xi)\alpha^{A_1\cdots A_{r-1}}=0$. 
As $l_{\rm dR}(\xi)=[d_{\rm dR},j_{\rm dR}(\xi)]$ and 
$d_{\rm dR}\alpha^{A_1\cdots A_{r-1}}=0$, the closed $r$ form 
$\alpha^{A_1\cdots A_{r-1}}$ is basic. Conversely, if 
$\alpha^{A_1\cdots A_{r-1}}$ is basic 
(7.2.23) obviously holds. So, the linear bijection $\nu$ introduced earlier 
maps $\sans f_{\rm basic}^{r,r+1}\cap\cap_{A=1,2}\ker d_A$ into 
$Z_{\rm dR~basic}^r(M)\otimes\bigvee^{r-1}\Bbb R^2$, where
$Z_{\rm dR~basic}^r(M)$ is the space of closed basic $r$ forms. Let 
$G\in\sans f_{\rm basic}^{r,r-1}$. Then, $G$ is of the form (7.2.21) and 
satisfies $j(\xi)G=0$, $j_A(\xi)G=0$, $l(\xi)G=0$, 
where $j(\xi)$, $j_A(\xi)$ and $l(\xi)$ are defined by
(2.2.15a)--(2.2.15c) and are given by (7.2.12a)--(7.2.12c).
It is straightforward to see that these identities 
yields the equations 
$$
\eqalignno{
&\epsilon_{AA_{r-1}}x^{i_1}{}_{,A_1}\cdots x^{i_{r-2}}{}_{,A_{r-2}}
c^{i_{r-1}}\xi\gamma^{A_1\cdots A_{r-1}}_{i_1\cdots i_{r-1}}=0,&(7.2.24a)\cr
&x^{i_1}{}_{,A_1}\cdots x^{i_{r-1}}{}_{,A_{r-1}}
\bigg[\sum_{l=1}^{r-1}\partial_{xi_l}c^{i_r}\xi
\gamma^{A_1\cdots A_{r-1}}_{i_1\cdots i_{l-1}i_ri_{l+1}\cdots i_{r-1}}
+c^{i_r}\xi\partial_{xi_r}\gamma^{A_1\cdots A_{r-1}}_{i_1\cdots i_{r-1}}\bigg]
=0.~~~~~~~~~&(7.2.24b)\cr}
$$
Thus, $c^{i_0}\xi\gamma^{A_1\cdots A_{r-1}}_{i_0i_1\cdots i_{r-2}}=0$, 
$\sum_{l=1}^{r-1}\partial_{xi_l}c^{i_r}\xi
\gamma^{A_1\cdots A_{r-1}}_{i_1\cdots i_{l-1}i_ri_{l+1}\cdots i_{r-1}}
+c^{i_r}\xi\partial_{xi_r}\gamma^{A_1\cdots A_{r-1}}_{i_1\cdots i_{r-1}}=0$, 
as is easy to see, so that $j_{\rm dR}(\xi)\gamma^{A_1\cdots A_{r-1}}=0$ and 
$l_{\rm dR}(\xi)\gamma^{A_1\cdots A_{r-1}}=0$ and $\gamma^{A_1\cdots A_{r-1}}$
is basic. Conversely the basicity of $\gamma^{A_1\cdots A_{r-1}}$ 
implies (7.2.24a), (7.2.24b). From (7.2.21), (7.2.22), we see that
$\nu$ maps cohomologically trivial elements of $\sans f_{\rm basic}^{r,r+1}
\cap\cap_{A=1,2}\ker d_A$ into cohomologically trivial elements of 
$Z_{\rm dR~basic}^r(M)\otimes\bigvee^{r-1}\Bbb R^2$.  
Thus, $\nu$ induces a linear bijection $\hat\nu: 
H_{\rm basic}^{r,r+1}(\sans f)\mapsto H_{\rm dR~basic}^r(M)\otimes
\bigvee^{r-1}\Bbb R^2$. \hfill $\square$

A theorem analogous to theor. 7.1.2 also holds.

\proclaim Theorem 7.2.2. One has
$$
\Conn(\sans f)\simeq\Conn(M)
\eqno(7.2.25)
$$
(cfr. def. 6.2.1).

{\it Proof}. 
From the representation theory of $\sans i=\goth s\goth l(2,\Bbb R)
\oplus\Bbb R$, any $a_A\in\sans f^{2,1}\otimes\goth g$ is locally of the 
form $a_A=a_i\tilde x^i{}_{,A}$, where $a_i$ is a $\goth g$ valued 
smooth map. Define $\lambda((a_A)_{A=1,2})=a_id_{\rm dR}x^i$. Then, 
from (7.2.12a)--(7.2.12c), it is easy to see that $\lambda((a_A)_{A=1,2})$ 
is a connection of $M$ if and only if $(a_A)_{A=1,2}$ is a connection 
of $\sans f$. The map $\lambda$ is clearly a bijection. 
\hfill $\square$

\vskip.2cm {\it 7.3. The relation between the $N=1$ and $N=2$ cohomologies 
of $\sans f$}

Let $\sans f(n)$ denote the superoperation $\sans f$ for $N=n$, $n=1,~2$, 
as defined in subsects. 7.1, 7.2. 

\proclaim Corollary 7.3.1. One has
$$
H^{n,\pm n+1}(\sans f(2))\simeq 
H^{\pm(n-{1\over 2})+{1\over 2}}(\sans f(1))\otimes
\hbox{$\bigvee^{n-1}$}\Bbb R^2,
\eqno(7.3.1)
$$
$$
H_{\rm basic}^{n,\pm n+1}(\sans f(2))\simeq 
H_{\rm basic}^{\pm(n-{1\over 2})+{1\over 2}}(\sans f(1))\otimes
\hbox{$\bigvee^{n-1}$}\Bbb R^2.
\eqno(7.3.2)
$$

{\it Proof}. Combine props. 7.1.1, 7.2.1. \hfill $\square$

Thus, the $N=1$ and $N=2$ cohomologies of $\sans f$ are closely related.
Note the analogy to relations (5.3.1), (5.3.2).

\proclaim Corollary 7.3.2. One has
$$
\Conn(\sans f(2))\simeq\Conn(\sans f(1)).
\eqno(7.3.3)
$$

{\it Proof}. Combine props. 7.1.2, 7.2.2. \hfill $\square$

Thus, the $N=1$ and $N=2$ connections of $\sans f$ are manifestations of
the same geometrical structure.
\par\vskip .6cm
{\bf 8. Concluding remarks}         
\vskip .4cm

There are a few fundamental questions which are still open and which are of 
considerable salience both in geometry and topological field theory. 

Cors. 5.3.1, 7.3.1 suggest that a relation formally analogous to (7.3.2) 
should hold also between the $N=1$ and $N=2$ equivariant cohomologies of 
$\sans f$ (cfr. sect. 6). Further, from (7.3.3), we expect 
that the range of the $N=1$ and $N=2$ Weil homomorphisms (cfr. subsects. 
6.1, 6.2) should have essentially the same content. This question is 
of fundamental importance to show conclusively that balanced topological gauge 
field theory does not contain new topological observables besides those 
coming from the underlying $N=1$ theory. We have not been able to 
either proove or disproove such assertions yet. 

There are other possible lines of inquiry. It is known that the $N=1$ 
Maurer--Cartan equations of a Lie algebra $\goth g$ can be obtained from 
the $N=1$ Weil algebra relation (3.1.11a) by formally setting $\phi=0$. 
By a similar procedure, one can obtain the $N=2$ Maurer--Cartan equations 
by formally setting $\phi_{AB}=0$, $\rho_A=0$ in the $N=2$ Weil algebra
relations (3.2.11e), (3.2.11g). Indeed, it is straigtforward to check that 
the basic relation $[d_A,d_B]=0$ still holds after this truncation.
This hints to a possible $N=2$ generalization of gauge fixing. 

Finally, note that, by obtaining the $N=2$ Weil algebra, we are in the 
position of formulating other models of equivariant cohomology in 
balanced topological field theory besides Cartan's used in \ref{31}, 
generalizing the $N=1$ intermediate or BRST model of \ref{7,8}. 

We leave these matters to future work \ref{34}.
\vskip.6cm
\par\noindent
{\bf Acknowledgements.} We are greatly indebt to R. Stora for providing 
his invaluable experience and relevant literature. 
\vskip.6cm
\centerline{\bf REFERENCES}

\item{\ref{1}} S. Cordes, G. Moore and S. Ramgoolam, 
{\it Lectures on 2--D Yang--Mills Theory, Equivariant Cohomology and 
Topological Field Theories}, presented at the 1994 Trieste Spring School on 
String Theory, Gauge Theory and Quantum Gravity, Trieste, Italy, 11 April --
22 April 1994, and at the NATO Advanced Study Institute, Les Houches Summer 
School, Session 62: Fluctuating Geometries in Statistical Mechanics and 
Field Theory, Les Houches, France, 2 August -- 9 September 1994, 
Nucl. Phys. Proc. Suppl. {\bf 41} (1995), 184, and Les Houches proceedings,
hep-th/9411210.

\item{\ref{2}} R. Dijkgraaf,
{\it Les Houches Lectures on Fields, Strings and Duality},
lectures given at NATO Advanced Study Institute at the 
Les Houches Summer School on Theoretical Physics, Session 64: 
Quantum Symmetries, Les Houches, France, 1 August - 8 September 1995, 
hep-th/9703136. 

\item{\ref{3}} J. M. F. Labastida and C. Lozano, 
{\it Lectures in Topological Quantum Field Theory},
talk given at La Plata Meeting on Trends in Theoretical Physics, La Plata,
Argentina, 28 April -- 6 May 1997, CERN-TH-97-250, hep-th/9709192.

\item{\ref{4}} E. Witten, 
{\it Introduction to Cohomological Field Theories},
Lectures at the Trieste Workshop on Topological Methods in Physics,
Trieste, Italy, June 1990, Int. J. Mod. Phys. {\bf A6} (1991), 2775.

\item{\ref{5}} L. Baulieu and I. M. Singer, 
{\it Topological Yang--Mills Symmetry}, 
Nucl. Phys. Proc. Suppl. {\bf 5B} (1988), 12.

\item{\ref{6}} L. Baulieu and I. M. Singer, 
{\it Conformally Invariant Gauge Fixed Actions for 2-D Topological Gravity}, 
Commun. Math. Phys. {\bf 135} (1991), 253.

\item{\ref{7}}  J. Kalkman, 
{\it BRST Model for Equivariant Cohomology and Representatives for the 
Equivariant Thom Class}, 
Commun. Math. Phys. {\bf 153} (1993), 447.

\item{\ref{8}} S. Ouvry, R. Stora and P. van Baal, 
{\it On the Algebraic Characterization of Witten Topological Yang--Mills 
Theory},
Phys. Lett. {\bf B220} (1989), 159.

\item{\ref{9}} R. Stora, F. Thuillier and J.-C. Wallet, 
{\it Algebraic Structure of Cohomological Field Theory Models 
and Equivariant Cohomology}, lectures presented at the First Carribean 
School of Mathematics and Theoretical Physics, Saint Francois,
Guadaloupe May--June 1993.

\item{\ref{10}} V. Mathai and D. Quillen, 
{\it Superconnections, Thom Classes and Equivariant Differential Forms}, 
Topology {\bf 25} (1986), 85.

\item{\ref{11}} M. F. Atiyah and L. Jeffrey, 
{\it Topological Lagrangians and Cohomology}, 
J. Geom. Phys. 7 (1990), 119.

\item{\ref{12}} M. Blau, 
{\it The Mathai--Quillen Formalism and Topological Field Theory}, 
J. Geom. Phys. {\bf 11} (1991), 129.

\item{\ref{13}} R. Stora, 
{\it Equivariant Cohomology and Topological Field Theories}, 
talk given at the International Symposium on BRS Symmetry
on the Occasion of its 20--th Anniversary, Kyoto Japan, 18 September --
22 September 1995, ENSLAPP-A-571-95;
{\it Exercises in Equivariant Cohomology},
talk given at NATO Advanced Study Institute on Quantum Fields and Quantum 
Space Time, Cargese, France, 22 July -- 3 August 1996. 
in {\it Cargese 1996, Quantum Fields and Quantum Space Time}, 265, 
hep-th/9611114;
{\it Exercises in Equivariant Cohomology and Topological Field Theories}, 
talk given at Symposium on the Mathematical Beauty of Physics, Gif-sur-Yvette,
France, 5 June -- 7 June 1996,
in {\it Saclay 1996, The Mathematical Beauty of Physics}, 51, 
hep-th/9611116.

\item{\ref{14}} J. P. Yamron,
{\it Topological Actions for Twisted Supersymmetric Theories},
Phys. Lett. {\bf B213} (1988), 325.

\item{\ref{15}} E. Witten,
{\it Topology Changing Amplitudes in $2+1$ Dimensional Gravity},
Nucl. Phys. {\bf B323} (1989), 113.

\item{\ref{16}} D. Birmingham, M. Blau and G. Thompson, 
{\it Geometry and Quantization of Topological Gauge Theories},
Int. J. Mod. Phys. {\bf A5} (1990), 4721.

\item{\ref{17}} M. Blau and G. Thompson,
{\it N=2 Topological Gauge Theory, the Euler Characteristic 
of Moduli Spaces  and the Casson Invariant},
Commun. Math. Phys. {\bf 152} (1993), 41, hep-th/9112012.

\item{\ref{18}} C. Vafa and E. Witten,
{\it A Strong Coupling Test of S Duality},
Nucl. Phys. {\bf B431} (1994), 3, hep-th/9408074.

\item{\ref{19}} M. Bershadsky, A. Johansen, V. Sadov and C. Vafa,
{\it Topological Reduction of 4-D SYM to 2-D Sigma Models},
Nucl. Phys. {\bf B448} (1995), 166, hep-th/9501096.

\item{\ref{20}} N. Marcus,
{\it The Other Topological Twisting of $N=4$ Yang--Mills}
Nucl. Phys. {\bf B452} (1995), 331, hep-th/9506002.

\item{\ref{21}}  M. Blau and G. Thompson,
{\it Aspects of $N_T\geq 2$ Topological Gauge Theories and D--Branes},
Nucl. Phys. {\bf B492} (1997), 545,  hep-th/9612143.

\item{\ref{22}} J. M. F. Labastida and C. Lozano,
{\it Mathai--Quillen Formulation of Twisted N=4 Supersymmetric Gauge 
Theories in Four Dimensions},
Nucl. Phys. {\bf B502} (1997), 741, hep-th/9702106.

\item{\ref{23}} J. M. F. Labastida and C. Lozano,
{\it Mass Perturbations in Twisted N=4 Supersymmetric Gauge Theories},
CERN-TH-97-316,  hep-th/9711132.

\item{\ref{24}} By R. Dijkgraaf, J. --S. Park and B. J. Schroers,
{\it N=4 Supersymmetric Yang--Mills Theory on a Kaehler Surface},
ITFA-97-09, hep-th/9801066.

\item{\ref{25}} M. Bershadsky, V. Sadov and C. Vafa, 
{\it D-Branes and Topological Field Theories},
Nucl. Phys. {\bf B463} (1996), 420, hep-th/9511222.

\item{\ref{26}} M. Blau and G. Thompson,
{\it Euclidean SYM Theories by Time Reduction and Special Holonomy 
Manifolds} Phys. Lett. {\bf B415} (1997), 242, hep-th/9706225.

\item{\ref{27}} B. S. Acharya, J. M. Figueroa--O'Farrill, B. Spence 
and M. O'Loughlin,
{\it Euclidean D--Branes and Higher Dimensional Gauge Theory}
QMW-PH-97-20, hep-th/9707118.

\item{\ref{28}} J. M. Figueroa--O'Farrill, A. Imaanpur and  J. McCarthy,
{\it Supersymmetry and Gauge Theory on Calabi--Yau Three--Folds},
QMW-PH-97-29, hep-th/9709178.

\item{\ref{29}} J. --S. Park,
{\it Monads and D--Instantons},
Nucl. Phys. {\bf B493} (1997), 198, hep-th/9612
096.

\item{\ref{30}} C. Hofman and J. --S. Park,
{\it Monads, Strings, and M Theory},
THU-97-14A, hep-th/9706130. 

\item{\ref{31}} R. Dijkgraaf and  G. Moore,
{\it Balanced Topological Field Theoreis}, 
Commun. Math. Phys. {\bf 185} (1997), 411, hep-th/9608169.

\item{\ref{32}} J. Manes, R. Stora and B. Zumino,
{\it Algebraic Study of Chiral Anomalies},
Commun. Math. Phys. {\bf 102} (1985), 157.

\item{\ref{33}} W. Grueb, S. Halperin and R. Vanstone,
{\it Connections, Curvature and Cohomology}, vol. III, Academic Press, 
New York 1973

\item{\ref{34}} R. Zucchini, im preparation. 

\bye